\documentclass[fleqn,usenatbib,twocolumn]{mnras}

\usepackage{newtxtext,newtxmath}

\usepackage[T1]{fontenc}

\DeclareRobustCommand{\VAN}[3]{#2}
\let\VANthebibliography\thebibliography
\def\thebibliography{\DeclareRobustCommand{\VAN}[3]{##3}\VANthebibliography}

\usepackage{graphicx}	
\usepackage{amsmath}	
\usepackage{bm}
\usepackage{tabularx}
\usepackage{pdflscape}
\usepackage{multicol}
\usepackage{multirow}
\usepackage{color}
\usepackage{times}
\usepackage{hyperref}
\usepackage{colortbl}
\usepackage{longtable}

\definecolor{Gray}{gray}{0.9}

\makeatletter

\newcommand{\HeII}{{\rm He\,{\scriptstyle II}}}

\newcommand{\CII}{{\rm C\,{\scriptstyle II}}}
\newcommand{\CIII}{{\rm C\,{\scriptstyle III}}}
\newcommand{\CIV}{{\rm C\,{\scriptstyle IV}}}
\newcommand{\SIV}{{\rm S\,{\scriptstyle IV}}}
\newcommand{\FeII}{{\rm Fe\,{\scriptstyle II}}}
\newcommand{\SiII}{{\rm Si\,{\scriptstyle II}}}
\newcommand{\SiIII}{{\rm Si\,{\scriptstyle III}}}

\newcommand{\NII}{{\rm N\,{\scriptstyle II}}}
\newcommand{\NI}{{\rm N\,{\scriptstyle I}}}

\newcommand{\OII}{{\rm O\,{\scriptstyle II}}}
\newcommand{\OIII}{{\rm O\,{\scriptstyle III}}}

\newcommand{\Rmnum}[1]{\expandafter\@slowromancap\romannumeral #1@}

\newcommand{\Muv}{M_{\mbox{\tiny UV}}}

\newcommand{\A}{\mbox{\AA}}
\newcommand{\TIGM}{T_{\rm IGM}}
\newcommand{\TIGMi}{T_{{\rm IGM},i}}

\title[LAE $\times$ IGM tomography]{
Photometric IGM tomography with Subaru/HSC: the large-scale structure of Ly$\bm{\alpha}$ emitters and IGM transmission in the COSMOS field at $\bm{z\sim5}$}

\author[K. Kakiichi et al.]
{Koki Kakiichi,$^{1}$\thanks{E-mail: kakiichi@ucsb.edu (KK)}
Joseph F. Hennawi,$^{1,2}$
Yoshiaki Ono,$^{3}$
Akio K.\ Inoue,$^{4,5}$
Masami Ouchi,$^{6,3,7}$  \newauthor
Richard S. Ellis,$^{8}$
Romain A. Meyer,$^{9}$ and
Sarah I. Bosman$^{9}$
\\
$^{1}$Department of Physics, Broida Hall, University of California, Santa Barbara, CA 93106-9530, USA\\
$^{2}$Leiden Observatory, Leiden University, Niels Bohrweg 2, 2333 CA Leiden, Netherlands \\
$^{3}$Institute for Cosmic Ray Research, The University of Tokyo, 5-1-5 Kashiwanoha, Kashiwa, Chiba 277-8582, Japan\\
$^{4}$Waseda Research Institute for Science and Engineering, Faculty of Science and Engineering, Waseda University, 3-4-1, Okubo, Shinjuku, Tokyo 169-8555, Japan \\
$^{5}$Department of Physics, School of Advanced Science and Engineering, Faculty of Science and Engineering,  Waseda University, 3-4-1, Okubo, Shinjuku, Tokyo \\ 169-8555, Japan \\ 
$^{6}$National Astronomical Observatory of Japan, 2-21-1 Osawa, Mitaka, Tokyo 181-8588, Japan \\
$^{7}$Kavli Institute for the Physics and Mathematics of the Universe (WPI), University of Tokyo, Kashiwa, Chiba 277-8583, Japan \\
$^{8}$Department of Physics and Astronomy, University College London, Gower Street, London WC1E 6BT, UK \\ 
$^{9}$Max Planck Institut für Astronomie, K\"{o}nigstuhl 17, D-69117, Heidelberg, Germany
}

\date{Accepted XXX. Received YYY; in original form ZZZ}

\pubyear{2015}

\begin{document}
\label{firstpage}
\pagerange{\pageref{firstpage}--\pageref{lastpage}}
\maketitle

\begin{abstract}
We present a novel technique called ``photometric IGM tomography'' to map the intergalactic medium (IGM) at $z\simeq4.9$ in the COSMOS field. It utilizes deep narrow-band (NB) imaging to photometrically detect faint Ly$\alpha$ forest transmission in background galaxies across the Subaru/Hyper-Suprime Cam (HSC)'s $1.8\rm\,sq.\,deg$ field of view and locate Ly$\alpha$ emitters (LAEs) in the same cosmic volume. Using ultra-deep HSC images and Bayesian spectral energy distribution fitting, we measure the Ly$\alpha$ forest transmission at $z\simeq4.9$ along a large number ($140$) of background galaxies selected from the DEIMOS10k spectroscopic catalogue at $4.98<z<5.89$ and the SILVERRUSH LAEs at $z\simeq5.7$. We photometrically measure the mean Ly$\alpha$ forest transmission and achieve a result consistent with previous measurements based on quasar spectra. We also measure the angular LAE-Ly$\alpha$ forest cross-correlation and Ly$\alpha$ forest auto-correlation functions and place an observational constraint on the large-scale fluctuations of the IGM around LAEs at $z\simeq4.9$. Finally, we present the reconstructed 2D tomographic map of the IGM, co-spatial with the large-scale structure of LAEs, at a transverse resolution of $11 \,h^{-1}\rm cMpc$ across $140\,h^{-1}\rm cMpc$ in the COSMOS field at $z\simeq4.9$. We discuss the observational requirements and the potential applications of this new technique for understanding the sources of reionization, quasar radiative history, and galaxy-IGM correlations across $z\sim3-6$. Our results represent the first proof-of-concept of photometric IGM tomography, offering a new route to examining early galaxy evolution in the context of the large-scale cosmic web from the epoch of reionization to cosmic noon.
\end{abstract}




\section{Introduction}

Cosmography, i.e. the science of mapping the Universe, is a fundamental pillar of astronomy. Mapping the distribution of objects on the sky has been pivotal for the discovery of the large-scale structure of the Universe. Our modern cosmological model is largely based on the maps of cosmic microwave background fluctuations \citep[e.g][]{Planck2020}, the large-scale distribution of galaxies \citep[e.g. eBOSS Collaboration:][]{eBOSS2021}, and gravitational lensing \citep[e.g DES Collaboration:][]{DES2022}. Mapping the structure of the intergalactic medium (IGM) with 21-cm tomography has great promise in advancing our understanding of the epoch of reionization and cosmic dawn \citep[e.g.][]{Pritchard2012}. However, while significant progress has been made in searching for the 21-cm power spectrum \citep{Mertens2020,Trott2020,HERA2022} and the global signal \citep{Bowman2018,Singh2022}, there are still many challenges to overcome before IGM tomography can be achieved with the 21-cm line.

Meanwhile, the Ly$\alpha$ forest remains the best probe of the IGM available to date (e.g \citealt{Becker2015review, McQuinn2016} for reviews). Recent measurements of effective optical depth indicate an ending of reionization as late as $z\sim5.3-5.7$ \citep{Becker2015, Eilers2018, Bosman2022}. The presence of long Gunn-Peterson troughs extending $\sim10-100\,h^{-1}$ comoving Mpc (cMpc) \citep{Becker2015,Zhu2022} and transmission spikes \citep{Barnett2017,Yang2020} in the same redshift range indicates large spatial variation in the IGM opacity at the tail end of reionization. Simulations suggest the large-scale fluctuations of Ly$\alpha$ forest transmission could be caused by the UV background fluctuations from galaxies \citep{Becker2015,DAloisio2018,Davies2018} or luminous rare sources such as AGN \citep{Chardin2015,Chardin2017,Meiksin2020}, thermal fluctuations in the IGM \citep{DAloisio2015,Keating2018}, islands of neutral hydrogen due to the late end of reionization \citep{Kulkarni2019,Keating2020,Nasir2020}, and/or the spatially-varying distribution of self-shielded absorbers which modulate the mean free path of the ionizing radiation \citep{Davies2016,DAloisio2018}. However, without directly observing the sources of ionizing radiation, it is difficult to understand how the interplay of these various physical processes and how they shape the physical state of the IGM during the reionization process. 

Establishing the direct spatial correlation between galaxy populations and Ly$\alpha$ forest transmission of the IGM is key to understanding how reionization proceeded. Previous ground-based surveys have mapped the distribution of galaxies both photometrically \citep{Becker2018,Kashino2020,Christenson2021,Ishimoto2022} and spectroscopically \citep{Kakiichi2018,Meyer2019,Meyer2020,Bosman2020} along sightlines to luminous $z\gtrsim6$ quasars where high quality Ly$\alpha$ forest spectra are available. By using Subaru/Hyper-Suprime Cam (HSC) narrow-band (NB) imaging in a total of six quasar fields, \citet{Becker2018,Christenson2021,Ishimoto2022} find that $\sim20\,h^{-1}\rm cMpc$ scale galaxy underdensities (overdensities) at $z\simeq5.7$ correlate with opaque (transmissive) regions of the IGM on $\sim50-100\,h^{-1}\rm cMpc$ scale. These observations favour the scenario where ionizing radiation from galaxies drives the large-scale UV background fluctuations at the tail end of reionization and/or completely neutral islands still exist in the IGM at $z<5.7$. The spectroscopic survey using Keck/DEIMOS and VLT/MUSE \citep{Kakiichi2018,Meyer2020} supports a similar picture. Surveying a total of eight quasar fields, \citet{Meyer2020} find a large scale excess transmission of Ly$\alpha$ forest around $z\simeq5.8$ galaxies on scales of $\sim10-40\,h^{-1}\rm cMpc$ at $\sim2-3\sigma$. By modelling the observed galaxy-Ly$\alpha$ forest cross-correlation signal, they interpreted that this excess transmission is caused by the UV background fluctuations driven by the faint unseen population of galaxies clustered around luminous galaxies, with an average Lyman continuum (LyC) escape fraction of $\langle f_{\rm esc}\rangle\simeq14\,\%$ at $z\simeq5.8$. 

Recent fully-coupled cosmological radiation hydrodynamic simulations also show the excess transmission in the large-scale galaxy-Ly$\alpha$ forest cross-correlation and suggest that the cross-correlation signal contains important information about the timing of reionization \citep{Garaldi2022}. The large-scale UV background fluctuations in the Ly$\alpha$ forest have long been recognised to contain valuable information about the nature of LyC sources (e.g. host halo mass) through their clustering properties \citep{Pontzen2014,Gontcho2014,Meiksin2019,Wolfson2022}. In summary, establishing the spatial correlation between galaxies and Ly$\alpha$ forest at $5<z<7$ offers a smoking gun test of the reionization process and is one of the key science goals of ongoing JWST quasar field surveys (ID 2078, PI: \citealt{Wang2021}; ID 1243, PI: \citealt{Lilly2017}, see also \citealt{Kashino2022}). However, due to the rarity of high-redshift quasars, the connection between galaxies and the IGM in these surveys will remain limited to one-dimensional skewers.

Ultimately we seek to map both galaxies and the IGM in three dimensions. At intermediate redshifts $z\sim2-3$, deep spectroscopic samples of background galaxies have enabled the construction of 3D Ly$\alpha$ forest tomographic maps of the IGM \citep{Lee2014a,Lee2014b,Lee2018,Newman2020,Horowitz2021} and the measurements of the galaxy-Ly$\alpha$ forest cross-correlation \citep{Steidel2010,Chen2020}. However, extending this approach to higher redshifts is extremely challenging because the diminishing Ly$\alpha$ forest transmission demands a much larger investment of telescope time. Spectroscopically detecting the UV continua and Ly$\alpha$ forest transmission of $z\simeq5-6$ galaxies would require 30-m class telescopes such as the Thirty-Meter Telescope (TMT), Giant Magellan Telescope (GMT), and Extremely Large Telescope (ELT) \citep{Japelj2019}. 

An alternative approach for IGM tomography is to utilize ultra-deep NB imaging to detect the Ly$\alpha$ forest transmission at a fixed redshift against the backdrop of more distant galaxies. As the throughput of an imager is much higher than a typical spectrograph, this photometric measurement of the Ly$\alpha$ forest transmission can be more sensitive than one based on spectroscopy. For the $\sim60\,\%$ throughput of the HSC imager (cf. $\sim10-20\,\%$ throughput of a typical spectrograph), the NB measurement of the Ly$\alpha$ forest with the Subaru 8.2-m can be comparable to undertaking a spectroscopic IGM survey with a 14-20 m telescope. The typical NB filter width is $\simeq100\,\A$ which corresponds to a line-of-sight distance of $\sim30\,h^{-1}\rm cMpc$ at $z\sim5-6$. This matches the scale of fluctuations in the Ly$\alpha$ forest transmission seen by previous quasar field surveys \citep{Becker2018,Kakiichi2018,Kashino2019,Meyer2019,Meyer2020,Christenson2021,Ishimoto2022}. Since up to a few hundred background galaxies can be identified using extant spectroscopic catalogues and NB-selected Ly$\alpha$ emitters (LAEs) in well-studied extragalactic fields, photometric IGM tomography can provide a $\times$ 100 increase in the number of galaxy-Ly$\alpha$ sightline pairs compared to  quasar surveys - a huge boost in statistical power. In an earlier article, \citet{Kakiichi2022} outlined the strategy for photometric IGM tomography which provides a path forward to map to map the Ly$\alpha$ forest transmission of the IGM and Ly$\alpha$ emitting galaxies in the same cosmic volume using
only imaging data. 

In this paper, we apply this photometric IGM tomography technique to the well-studied extragalactic COSMOS field and present  measurements of the LAE-Ly$\alpha$ forest cross-correlation and the auto-correlation of the Ly$\alpha$ forest at $z\simeq4.9$. The wealth of deep multi-wavelength imaging and spectroscopic data makes COSMOS an ideal field to demonstrate the method. We provide the first large-scale 2D tomographic map of the IGM at $z\simeq4.9$ across the $1.8\rm\,deg^2$ ($\sim140\,h^{-1}\rm cMpc$ in diameter) field of view of Subaru/HSC, enabling us to directly visualise the spatial connection between galaxies and the IGM. After the reionization process is complete, we expect that the galaxy-Ly$\alpha$ forest cross-correlation will evolve from positive \citep{Kakiichi2018,Meyer2020,Garaldi2019} owing to the large-scale UV background fluctuations and/or ionized bubbles to a negative (i.e. anti-correlation) signal \citep[e.g.][]{Turner2017,Nagamine2021} due to the increasing impact of gas overdensities around galaxies at lower redshifts. Our study at $z\simeq4.9$ will provide a clue for how the cross-correlation evolves across cosmic time.

In Section \ref{sec:data} we describe the imaging data used in this analysis. Section \ref{sec:catalog} describes the galaxy catalogues and the selection for the photometric IGM tomography. Section \ref{sec:measurement} presents the method to estimate the Ly$\alpha$ forest transmission along background galaxies using a Bayesian spectral energy distribution (SED) fitting framework. In Sections \ref{sec:analysis}-\ref{sec:map}, we present our main results including the measurements of mean Ly$\alpha$ forest transmission (Section \ref{sec:analysis}), LAE-Ly$\alpha$ forest cross-correlation (Section \ref{sec:cross}), auto-correlation of the Ly$\alpha$ forest (Section \ref{sec:auto}), and the reconstruction of the 2D IGM tomographic map (Section \ref{sec:map}). In Section \ref{sec:discussions} we discuss the requirement to improve the photometric IGM tomography and various science applications. We summarise our results and conclusions in Section \ref{sec:conclusion}. Throughput this paper we assume cosmological parameters $(\Omega_m, \Omega_\Lambda, \Omega_b, h, \sigma_8,n_s)=(0.3089,0.6911,0.0486,0.6774,0.8159,0.9667)$ \citep{Planck2016}. We use cMpc (pMpc) to indicate distances in comoving (proper) units. All magnitudes in this paper are quoted in the AB system \citep{Oke1983}.

\section{Data}\label{sec:data}

\begin{figure*}
  \centering
    \includegraphics[width=0.33\textwidth]{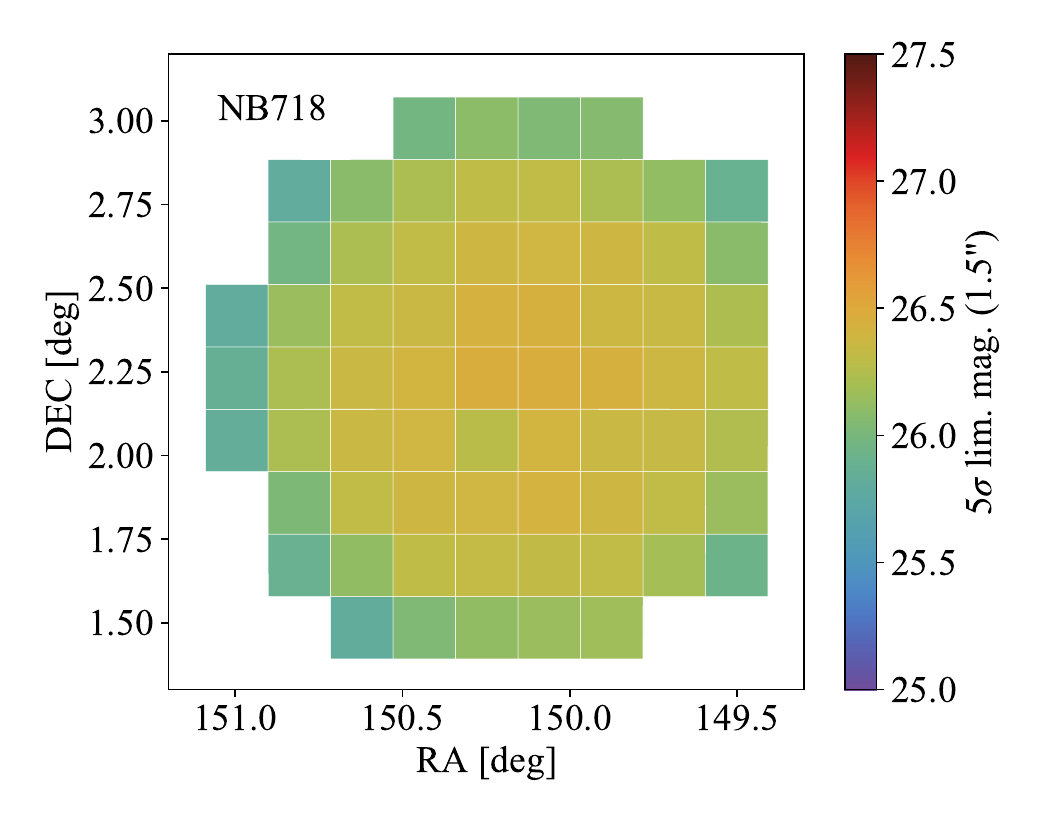}
    \includegraphics[width=0.33\textwidth]{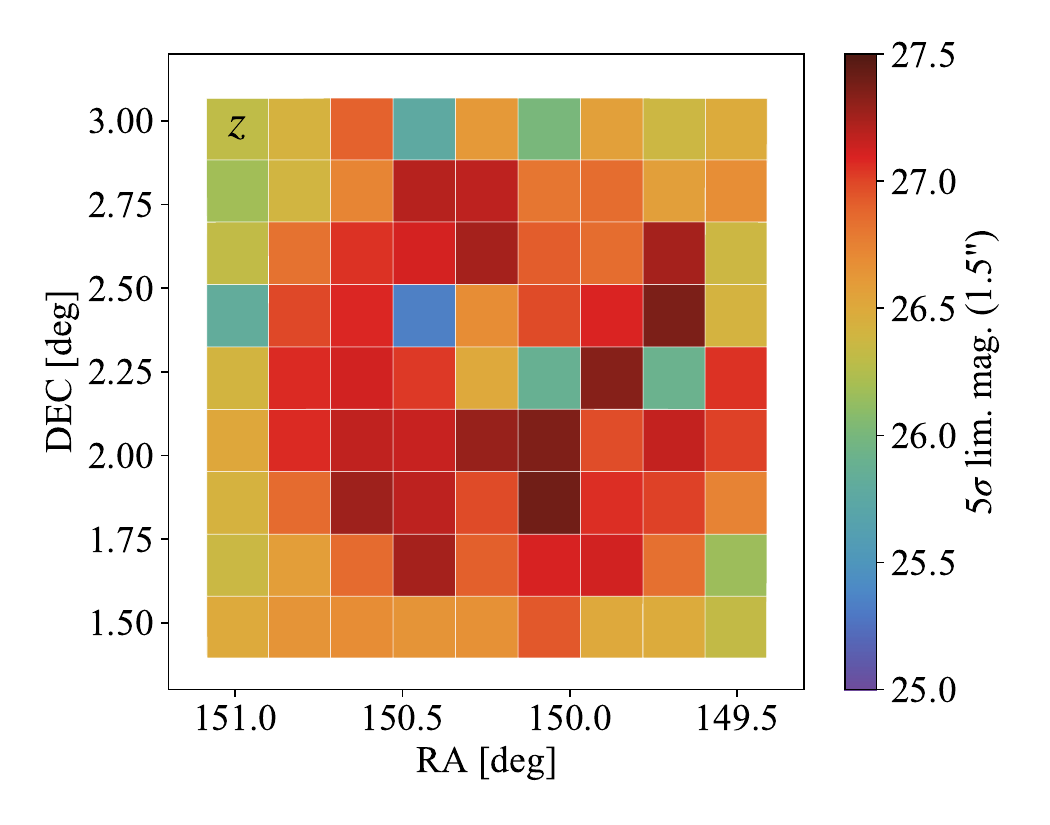}
    \includegraphics[width=0.33\textwidth]{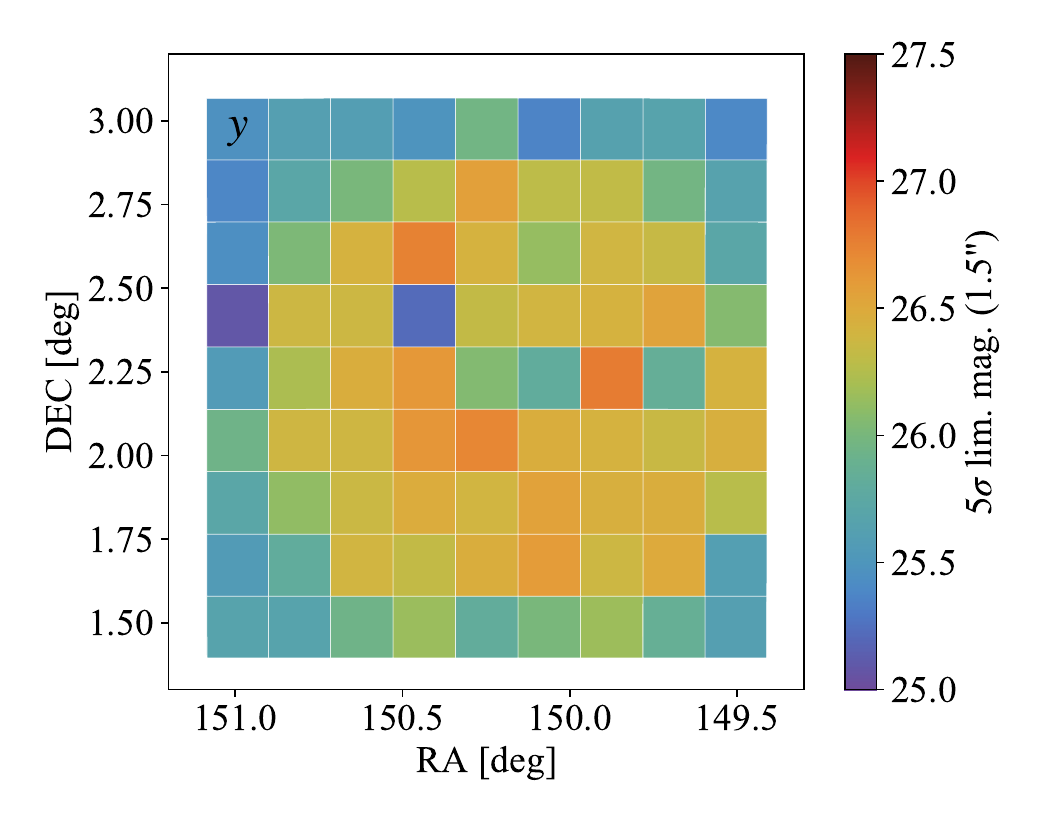}
    \vspace{-0.5cm}
      \caption{$5\sigma$ limiting magnitudes of the NB718, $z$, and $y$-band images with fixed $1.5''$ aperture in the COSMOS field (tract 9813).}
      \label{fig:limmag}
  \end{figure*}
  
  \begin{table*}
  \centering
  \caption{Summary of the foreground NB images}\label{table:filter_summary}
  \begin{tabular}{lllllll}
  \hline\hline
  Filter      & Ly$\alpha$ redshift & bkg. source redshift & $5\sigma$ depth (1.5\arcsec)  & $3\sigma$ depth (1.5\arcsec)  & $1\sigma$ depth (1.5\arcsec) & Ref \\
        &  &  & {[}AB mag{]} &  {[}AB mag{]} &  {[}AB mag{]} & \\
  \hline
  NB718 & 4.90 $(4.85\,<\,z_{\rm Ly\alpha}\,<\,4.94)$ &   $4.98<z<5.89$      & 26.30                    & 26.86    & 28.05 & Inoue et al (2020)                \\
  \hline\hline
  \multicolumn{7}{l}{$^b$ computed from the median of random sky objects in masked region (i.e. excluding near bright stars and artefacts) in each patch of tract 9813.}
  \end{tabular}
  \end{table*}
  
  \begin{table}
  \centering
  \caption{Summary of the BB images from HSC-SSP DR3 in the UD-COSMOS field (tract 9813)}\label{table:BB_filter_summary}
  \begin{tabular}{cccccc}
  \hline\hline
  \multicolumn{5}{c}{Median $5\sigma$ depth (1.5\arcsec)  {[}mag{]}}  & Reference \\
  $g$   & $r$ & $i$    & $z$ & $y$ & \\
  \hline
  27.85         & 27.39         & 27.22         & 26.86         & 26.23  & Aihara et al (2022)      \\
  \hline\hline
  \end{tabular}
  \end{table}

We use the public release of a Subaru HSC NB718 image from CHORUS DR1 \citep{Inoue2020} and broad-band (BB) $grizy$ and NB816 images from HSC-SSP DR3 \citep{Aihara2022} in the ultra-deep layer of COSMOS field (tract 9813). These images cover an area of approximately $1.67\times1.67\rm\,deg^2$ centred at $\rm(RA,DEC)=(10h01m00s, +2d14m00s)$. Co-added images are retrieved from the public data release website.\footnote{HSC-SSP DR3: \url{https://hsc-release.mtk.nao.ac.jp/doc/index.php/data-access__pdr3/}}$^{,}$\footnote{CHORUS DR1: \url{https://hsc-release.mtk.nao.ac.jp/doc/index.php/chorus/} The ancillary data including PSF FWHM and limiting magnitudes for each patch is also downloaded from here.} We use NB718 as a foreground NB filter to measure the Ly$\alpha$ forest transmission at $z\simeq4.9$ (the central wavelength of $7170.5\,\A$ corresponds to Ly$\alpha$ redshift of $z_{\rm Ly\alpha}=4.898$) covering the range of $4.85<z<4.94$ and $z$- and $y$-bands to measure the UV continua of the background galaxies.

We mask regions contaminated by artefacts (bright star haloes, ghosts, blooming, channel-stop, dip). Since we measure the residual transmitted fluxes in the NB718 image, particular care is needed for this filter as artefacts could potentially cause false positive detections. To address this we first flag all the masked pixels reported by CHORUS PDR1 \citep{Inoue2020}. This includes the pixels with the \texttt{hscPipe} flags: \texttt{pixelflags\_bright\_object=True} (pixels affected by bright objects) or \texttt{pixelflags\_saturatedcenter=True} (pixels affected by count saturation). Pixels affected by the haloes of bright stars are also masked. After visual inspection of the NB718 image with the CHORUS PDR1 mask overlaid, we find that there are still some regions affected by the outer ghosts of bright stars, which extend to approximately $\sim270\rm\,arcsec$ in radius. To mask these, we conservatively follow the procedure adopted by HSC-SSP DR3 \citep{Aihara2022}. We select bright stars with $G<10\rm\,mag$ from GAIA DR3 catalogue\footnote{\url{https://gea.esac.esa.int/archive/}} in the footprint of tract 9813. We then mask regions inside 320 arcsec radius around these Gaia stars. 320 arcsec corresponds to the size of the outermost ghost identified by \citet{Aihara2022}. They find that the outer ghost is significant for a star brighter than $\sim7\rm\,mag$ and the inner ghost is dominant at $\sim7-9\rm\,mag$. We apply the same mask for the broad-band images. 

The limiting magnitudes of the NB and BB images are estimated using synthetic apertures randomly  
distributed in the blank sky regions of the image. For NB718 we use the published limiting magnitude map
from the CHORUS PDR1. For the BB images, we retrieve the synthetic apertures located in the empty regions of the sky (hereafter sky objects) from HSC database and perform photometry with a fixed  $1.5\arcsec$ aperture.
As the sensitivity varies across the field of view, we estimate the limiting magnitudes for each patch. In each patch, there are typically $\sim40$ sky objects and we compute the limiting magnitude from the standard deviation of the photometric measurements of the fluxes for the sky objects. 
Figure \ref{fig:limmag} shows the limiting magnitudes of NB718, $z$, and $y$-bands for each patch in the field. The NB limiting magnitudes vary by about $\sim0.18$ mag. The median depths of the NB and BB images are summarised in Tables~\ref{table:filter_summary} and \ref{table:BB_filter_summary}.

A rule-of-thumb for the required depth for IGM tomography is given by \citet{Kakiichi2022}. Assuming the flat UV continuum slope of a background galaxy, the NB magnitude needs to reach
\begin{equation}
m_{\rm NB}=m_{\rm BB}-2.5\log_{10}e^{-\tau_{\rm eff}(z)}\approx m_{\rm BB}+\tau_{\rm eff}(z),
\end{equation}
to detect the Ly$\alpha$ forest transmission with an effective optical depth $\tau_{\rm eff}(z)$. Here, the NB magnitude corresponds to NB718 and the BB magnitude corresponds to $z$-band which covers the UV continuum of a background galaxy. Assuming the effective optical depth of $\tau_{\rm eff}(z)=1.5$ at $z=4.9$ \citep{Becker2013,Eilers2018,Bosman2022}, the required NB718 depth for a $z=25.0\rm\,mag$ background source is $26.5\,\rm mag$. The existing NB718 depth meets this requirement at $>3\sigma$ (Table~\ref{table:filter_summary}). For a fainter background source with $z=26.5\rm\,mag$, the existing NB718 depth still has a $\sim1\sigma$ sensitivity to the mean Ly$\alpha$ forest transmission. The HSC imaging of the COSMOS field thus has sufficient sensitivity for photometric IGM tomography.

\subsection{Photometry}

We measure the $grizy$ and NB718 photometry from HSC-SSP DR3 and CHORUS PDR1 data using a fixed $1.5\arcsec$ aperture for the background sources. The zero points of all photometric bands is $27\,\rm mag/DN$ according to the HSC-SSP data release. We ignore a few percent level correction arising from aperture corrections during the photometric calibration stage. This is negligible compared with the other photometric errors described below. We assign the photometric error of an object based on the limiting magnitude of the patch where the object is located.
    
\section{Catalogues}\label{sec:catalog}

The redshift range for the background sources is chosen such that the Ly$\alpha$ forest range between Ly$\beta$ ($\lambda_\beta=1026$\,\AA) and Ly$\alpha$ ($\lambda_\alpha=1216$\,\AA) lines is covered by the NB718 filter. 
The lower and upper redshifts are set by $z_{\rm min}=\lambda_{\rm NB,max}/\lambda_\alpha-1$ and $z_{\rm max}=\lambda_{\rm NB,min}/\lambda_\beta-1$ where $\lambda_{\rm NB,max}$ $\lambda_{\rm NB,min}$ are the maximum and minimum wavelengths of the filter. We define the minimum and maximum wavelengths as a range where the NB718 filter transmission is $>50\,\%$. The appropriate background source redshift for $z=4.9$ IGM tomography is thus $4.98<z<5.89$. 

To locate foreground galaxies at $z=4.9$ in the same redshift slice corresponding to that for which our IGM transmission is being measured, we employ the SILVERRUSH catalogue of $z\simeq4.9$ LAEs (\textsc{ver20210224}, \citealt{Ono2021}). 

To identify background galaxies at $4.98<z<5.89$, we use both the SILVERRUSH catalogue of $z\simeq5.7$ LAEs \citep{Ono2021} and the spectroscopic redshift (spec-z) catalogue compiled with HSC-SSP DR3 \citep{Aihara2022}. For the latter, we find that DEIMOS10k \citep{Hasinger2018} was the primary source of the background galaxies as we will describe below. Thus in the remainder of the paper, we refer the background galaxies derived from the catalogues to as the LAE and DEIMOS10k samples, respectively.

\subsection{Spectroscopic redshift catalogue}

The spec-z catalogue associated with HSC-SSP DR3 is a compilation of public spectroscopic redshifts from numerous previous redshift surveys including 2dFGRS \citep{Colless2003}, 3D-HST \citep{Skelton2014,Momcheva2016}, 6dFGRS \citep{Jones2004,Jones2009}, C3R2 DR2 \citep{Masters2017,Masters2019}, DEEP2 DR4 \citep{Davis2003,Newman2013}, DEEP3 \citep{Cooper2011,Cooper2012}, DEIMOS10k \citep{Hasinger2018}, FMOS-COSMOS \citep{Silverman2015,Kashino2019}, GAMA DR2 \citep{Liske2015}, LEGA-C DR2 \citep{Straatman2018}, PRIMUS DR1 \citep{Coil2011,Cool2013}, SDSS DR16 \citep{Ahumada2020}, SDSS IV QSO catalog \citep{Paris2018}, UDSz \citep{Bradshaw2013,McLure2013}, VANDELS DR1 \citep{Pentericci2018}, VIPERS PDR1 \citep{Garilli2014}, VVDS \citep{LeFevre2013}, WiggleZ DR1 \citep{Drinkwater2010}, and zCOSMOS DR3 \citep{Lilly2009}. 

Spectroscopic sources are matched to the HSC photometry by position, thus the catalogue only includes the objects detected by HSC-SSP DR3. Using the CAS search, we retrieve the HSC-SSP DR3 spec-z catalogue after it was cross-matched using the \texttt{object\_id}. We re-measure the fixed $1.5''$ aperature photometry in $grizy$ and NB718 for all the objects to ensure consistent measurements of the fluxes across all bands.


\begin{figure*}
  \centering
    \includegraphics[width=\textwidth]{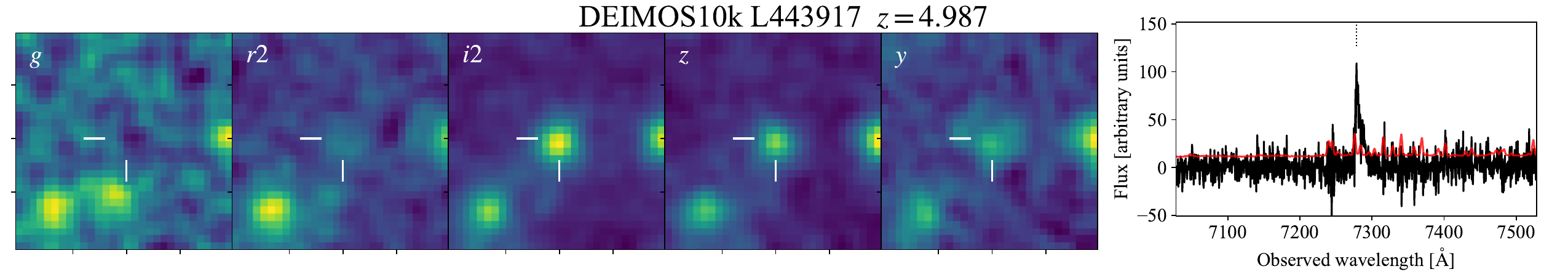}
    \vspace{-0.5cm}
      \caption{An example background galaxy from the DEIMOS10k catalogue. ({\bf Left}): Postage stamps show $5\times5$ arcsec cutouts of $grizY$ images around the object marked with a white crosshairs. ({\bf Right}): DEIMOS 1D spectrum (flux: black, red: noise). The vertical dotted line indicates the Ly$\alpha$ line.}
      \label{fig:bkg_source_DEIMOS10k}
  \end{figure*}
  
  \begin{figure*}
  \centering
    \includegraphics[width=\textwidth]{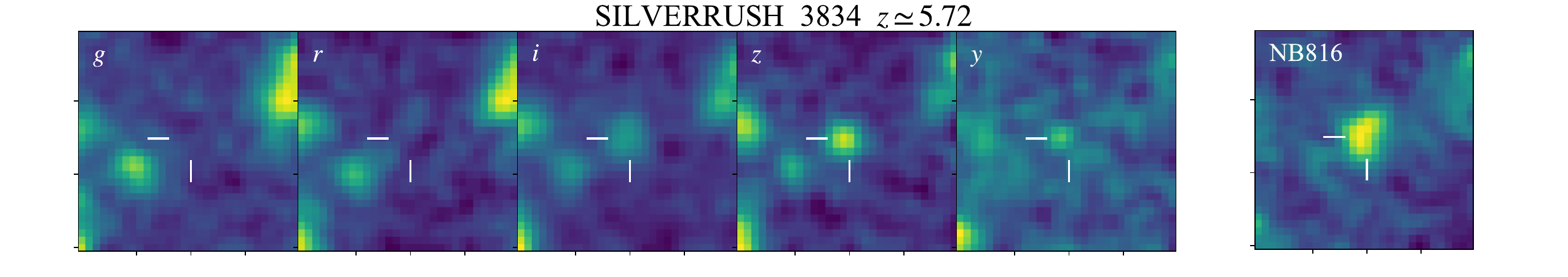}
    \vspace{-0.5cm}
    \caption{An example background LAE from the SILVERRUSH catalogue. ({\bf Left}): Postage stamp showing $5\times5$ arcsec cutouts of $grizY$ images with the object marked with a white cross. ({\bf Right}): NB816 image. The NB colour excess is clearly detected.}
    \label{fig:bkg_source_SILVERRUSH}
  \end{figure*}

Using the spec-z catalogue, we search for background source candidates at $4.98<z<5.89$ in the UD-COSMOS field (tract 9813). The catalogue contains a spec-z flag (\texttt{specz\_flag\_homogeneous=True} for secure and \texttt{False} for insecure) after homogenising the quality flags\footnote{\url{https://hsc-release.mtk.nao.ac.jp/doc/index.php/catalog-of-spectroscopic-redshifts__pdr3/}} of spectroscopic redshifts from the above surveys. Selecting only objects with \texttt{specz\_flag\_homogeneous=True}, we find 236 candidates in the required redshift range, of which 60 belong to DEIMOS10k \citep{Hasinger2018} and 176 belong to 3D-HST (v4.1.5, \citealt{Momcheva2016}). The majority of bright candidates with $z\lesssim25.5$ comes from DEIMOS10k whereas fainter candidates are mostly from 3D-HST.

In order to visually confirm the spectroscopic redshifts, we downloaded the original DEIMOS10k spectra from NASA/IPAC Infrared Science Archive (IRSA)\footnote{\url{https://irsa.ipac.caltech.edu/data/COSMOS/overview.html}}, and the 3D-HST spectra from MAST archive\footnote{\url{https://archive.stsci.edu/prepds/3d-hst/}}. For 54 of the 60 DEIMOS10k sources, we confirmed an emission line feature (mostly single Ly$\alpha$ line, one Lyman break only, one quasar). We rejected 6 objects because either (1) no published spectrum is available (DEIMOS10k ID: L234173) or (2) we could not visually confirm the reported redshift (L420065, L430951, L378903, C563716, L442206). 

Out of the remaining 54 DEIMOS10k sources, we removed 11 residing in the masked regions. Furthermore, in order to secure a reliable UV continuum detection in each background sources, we applied a $5\sigma$ detection cut in the $z$-band using the limiting magnitude appropriate for the relevant patch. 3 candidates fail to meet this criterion in the $z$-band of HSC-SSP DR3. We also require a $3\sigma$ non-detection in $g$-band in order to reject low-redshift interlopers. This leads to the removal of a further 4  sources. As a result we finally have 36 DEIMOS10k background sources for our IGM tomography. We present the postage stamp image and DEIMOS spectrum of a representative background source in Figure \ref{fig:bkg_source_DEIMOS10k}.

For the 3D-HST sources, the catalogued redshifts are determined from either photometric and/or grism spectroscopic data. The ACS/G800L grism spectra cover Ly$\alpha$ in our desired redshift range. We downloaded the 3D-HST catalogue and find that all the relevant sources have only photometric redshifts; we could not find any sources with a convincing Ly$\alpha$ line or Lyman break in the grism spectra. In principle, we can use background objects with photo-z's whose 95\% confidence interval (i.e. \textsc{z\_best\_l95, z\_best\_u95} from \textsc{cosmos\_3dhst.v4.1.5.zbest.dat}) lie within our required range. This would ensure that their Ly$\alpha$ forest region is appropriately covered by the NB718 filter. However, since it is unclear how catastrophic photo-z errors might affect the quality of our IGM tomography, we decided to remove all the 3D-HST objects from our final background sources in this paper. Nonetheless, in future work, it will be interesting to examine the utility of the photo-z background sources for IGM tomography.

\begin{figure}
  \centering
    \includegraphics[width=0.9\columnwidth]{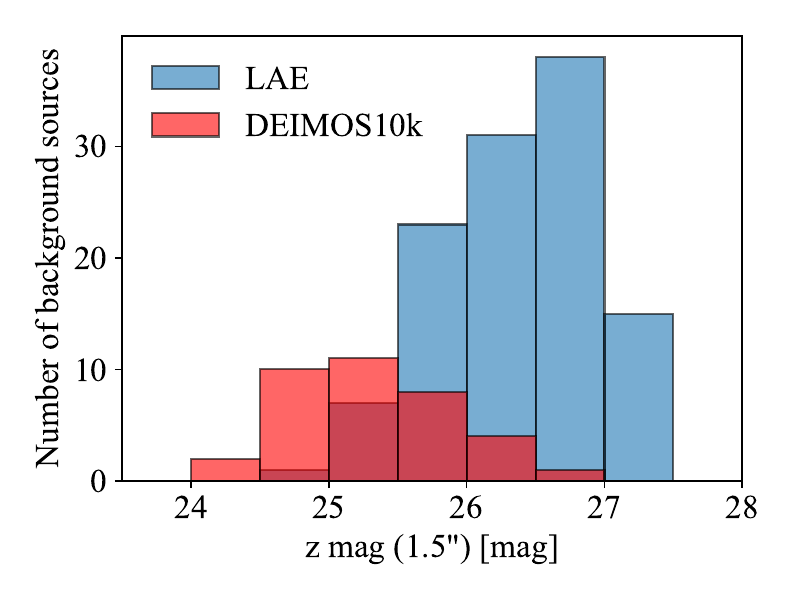}
    \vspace{-0.3cm}
      \caption{Magnitude distribution of background sources for NB718 IGM tomography at $z\simeq4.9$. Our final background source catalogue contains 151 sources in total (red: 36 spec-z objects from DEIMOS10k, blue: 115 LAE objects from SILVERRUSH).}
      \label{fig:bkg_source_hist}
\end{figure}

\begin{figure}
\centering
  \includegraphics[width=\columnwidth]{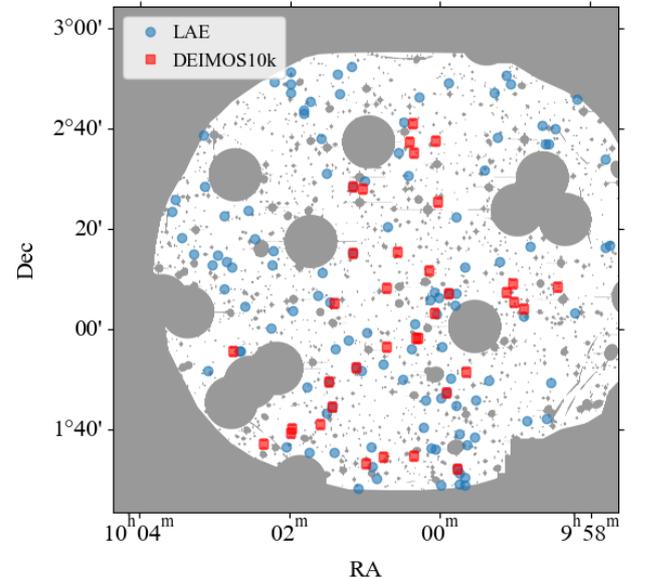}
  \vspace{-0.8cm}
    \caption{Sky distribution of background sources for NB718 IGM tomography (blue circles: $z=5.7$ LAEs from SILVERRUSH, red squares: spec-z from DEIMOS 10k). Masked regions and those outside the field-of-view are indicated by the gray shaded regions.}
    \label{fig:bkg_source_map}
\end{figure}

To summarise, our final catalogue of background spec-z sources contains 36 objects from DEIMOS10k. Their $z$-band magnitudes and distribution in the UD-COSMOS field are shown in Figures \ref{fig:bkg_source_hist} and \ref{fig:bkg_source_map}, respectively. The $z$-band magnitudes range from 24.2 to 26.9 with the median SNR $\approx 19.5$. Because the previous DEIMOS spectroscopic campaigns are focused near the central region of the COSMOS field, their distribution reflects their survey footprints.

\subsection{LAE catalogue}

We use the LAE catalogue from \citet{Ono2021} constructed as part of the SILVERRUSH programme. The catalogue is based on the data from CHORUS survey \citep{Inoue2020} and from the HSC-SSP internal data release of S18A which is basically identical to the Public Data Release 2 \citep{Aihara2019}. To ensure homogeneous photometric measurements for the IGM tomography, we re-measure the fixed $1.5''$ aperture photometry at the coordinates of the SILVERRUSH LAEs using HSC-SSP DR3 $grizy$ and CHORUS NB718 images.
We use the SILVERRUSH catalogue to select both background and foreground LAEs at $z=5.7$ and $z=4.9$ located by NB816 and NB718 colour excess, respectively. 

\subsubsection{Background LAE selection: $z\simeq5.7$}

To select $z\simeq5.7$ background LAEs, we draw a sample from the SILVERRUSH catalogue \citep{Ono2021} which applies a NB colour excess $i-{\rm NB816}\ge1.2$ and $5\sigma$ detection in NB816. Such a NB816 colour excess can locate LAEs at a redshift $z=5.726$ with an $\Delta z\simeq0.1$ accuracy \citep{Ono2021} sufficient to ensure that the Ly$\alpha$ forest transmission is covered by the NB718 filter. Details of the LAE catalogue construction are described in \citet{Ono2021}. The SILVERRUSH catalogue contains 378 $z\simeq5.7$ LAEs in the UD-COSMOS field. This double NB technique \citep{Kakiichi2022} allows us to efficiently assemble a large number of background sources for IGM tomography.

In addition to the standard NB selection, we require a $5\sigma$ detection in $z$-band. This criterion is met by 176 objects out of 378 LAEs. We also removed 35 further objects which reside in the masked regions. While \citet{Ono2021} already applied masks in constructing the original catalogue, our revised mask in NB718 is more conservative. As before, we also require a $3\sigma$ non-detection in the $g$-band to avoid possible low-redshift interlopers; this removes a further 26 objects. 

Thus, our final background LAE catalogue contains 115 objects. We visually inspected all of these objects in HSC DR3 $grizy$ and NB816 images. The $z$-band magnitudes and the spatial distribution of the background LAEs are shown in Figures \ref{fig:bkg_source_hist} and \ref{fig:bkg_source_map}. The $z$-band magnitudes range from 24.7 to 27.2 with the median SNR $\approx 7.4$. For comparison, the $i$-band magnitudes of the background LAEs, which cover the Ly$\alpha$ forest flux, Ly$\alpha$ emission line, and UV continuum rewards of Ly$\alpha$ line, are much fainter than the $z$-band magnitudes, ranging from 28.0 to 29.7 with the median SNR $\approx6.8$. The background LAEs are typically fainter than the DEIMOS10k sample, but distributed more evenly across the entire UD-COSMOS field as they are selected homogeneously via NB816 colour excess. We present an example postage stamp of a background LAE in Figure \ref{fig:bkg_source_SILVERRUSH}.

\subsubsection{Foreground LAE selection: $z\simeq4.9$}

In order to cross-correlate foreground LAEs with the Ly$\alpha$ forest transmission, we also use the HSC NB718 data to select LAEs at $z\simeq4.9$. The SILVERRUSH catalogue applies the selection criteria: $ri-{\rm NB718}>0.7$ and $r-i>0.8$ and $ri-{\rm NB718}>(ri-{\rm NB718})_{3\sigma}$ and $g>g_{2\sigma}$ where $ri$ is calculated by the linear combination of the fluxes in $r$- and $i$-bands,  $f_r$ and $f_i$, following $f_{ri}=0.3f_r+0.7f_i$ and the $2\sigma$ and $3\sigma$ subscripts denote $2\sigma$ and $3\sigma$ limiting magnitudes \citep{Ono2021}. Further detail is described in \citet{Ono2021}. This gives 280 $z\simeq4.9$ LAEs in the UD-COSMOS field. We remove 17 objects lying in our updated masked regions. For these foreground LAEs, unlike the background galaxies, we do not apply any $z$-band detection cut and use all 263 NB718-selected LAEs for our subsequent analysis. The average luminosities of the foreground LAEs are summarised in Table \ref{table:LAEs}.

\begin{table}
\centering
\caption{Average physical properties of foreground LAEs.}\label{table:LAEs}
\begin{tabular}{lll}
\hline\hline
Redshift & $\log_{10}\langle L_{\alpha}\rangle$ (2.0\arcsec) & $\langle M_{\rm UV}\rangle$ (2.0\arcsec)  \\
       & {[}$\rm erg\,s^{-1}${]}    & {[}AB mag{]}  \\
\hline
 $z=4.89$  & 42.62 & $-20.09$  \\
\hline\hline
\end{tabular}
\end{table}

\section{IGM Ly$\alpha$ forest transmission}\label{sec:measurement}

A measurement of the IGM Ly$\alpha$ forest transmission $T_{\rm IGM}$ using the foreground NB718 filter requires us to infer the intrinsic spectral energy distribution (SED) of each background galaxy in the absence of any IGM absorption. The NB-integrated Ly$\alpha$ forest transmission is then determined by the ratio between the observed and intrinsic NB fluxes,
\begin{equation}
    T_{\rm IGM}=\frac{f^{\rm obs}_{\rm NB}}{f^{\rm intr}_{\rm NB}}.
\end{equation}

There are several ways to perform this measurement. One obvious way to estimate $f_{\rm NB}^{\rm intr}$, similar to the approach employed by \citet{Mawatari2017}, is to first fit a SED to each background galaxy using broad-band photometry redward of the Ly$\alpha$ emission line $>1216$~\AA~, and then extrapolate the continuum to the relevant rest-frame range of the Ly$\alpha$ forest between $1026$\,\AA~and $1216$\,\AA~covered by the foreground NB718 filter. While intuitive, it is difficult to rigorously propagate the photometric errors and systematic uncertainties
of the SED modelling into the final measurement of $T_{\rm IGM}$. Also, it is hard to quantify likely degeneracies between the SED parameters (e.g. UV continuum slope, or age and dust attenuation law) and the Ly$\alpha$ forest transmission.

A better way is to {\it simultaneously fit both the galaxy SED and the Ly$\alpha$ forest transmission of the IGM in a fully Bayesian framework}. This allows a rigorous propagation of photometric and systematic errors 
in the $T_{\rm IGM}$ estimate for each background galaxy and characterises the full posteriors including the degeneracy with the SED model parameters. 

\begin{figure*}
\hspace{0.1cm}
	\includegraphics[width=1.1\columnwidth]{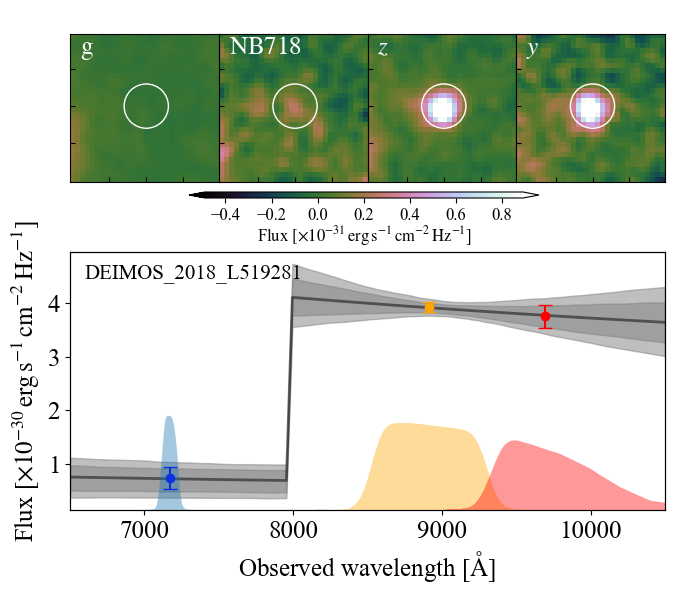}
  \includegraphics[width=0.95\columnwidth]{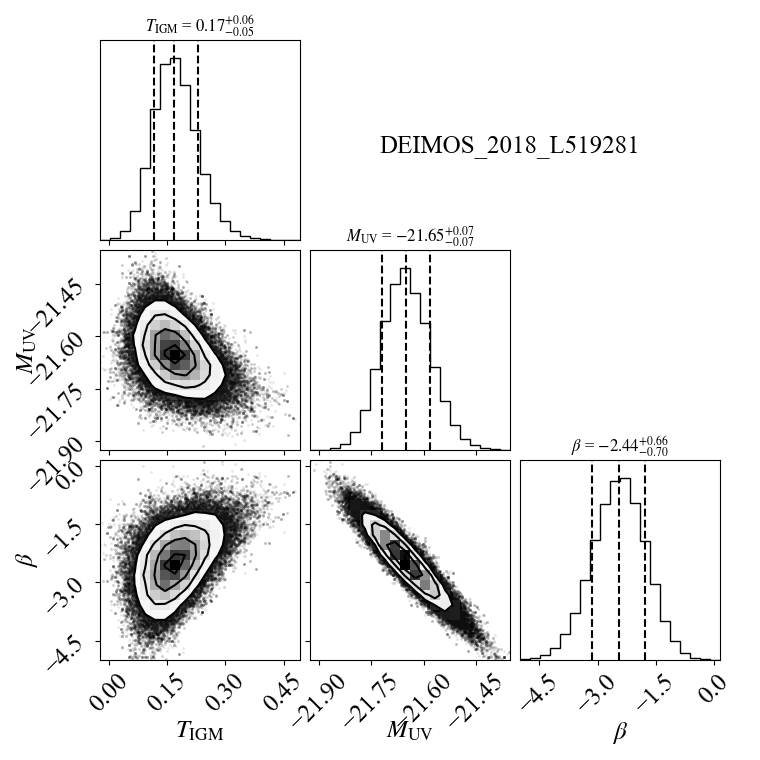}
\hspace{0.1cm}
  \includegraphics[width=1.1\columnwidth]{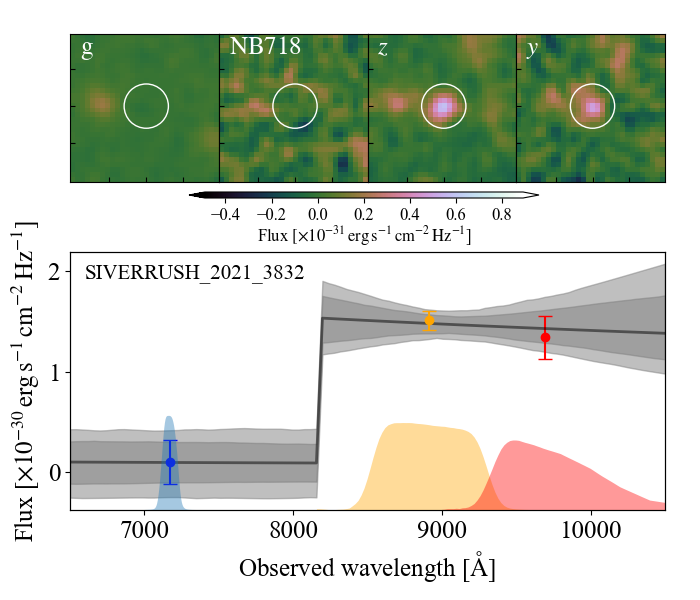}
	\includegraphics[width=0.95\columnwidth]{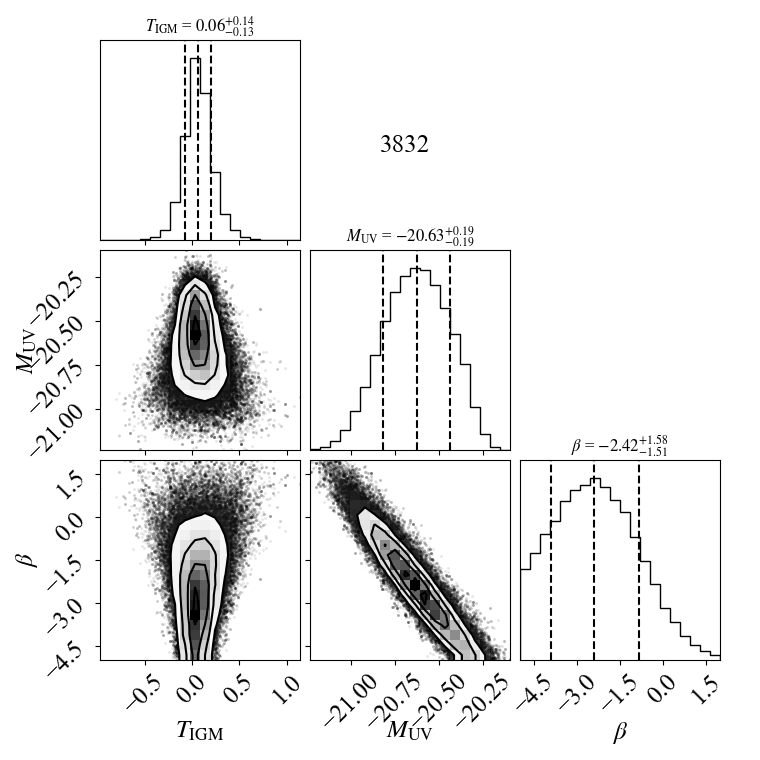}
    \caption{Representative examples illustrating results from the Bayesian SED fitting framework for the DEIMOS10k (top panels) and LAE (bottom panels) samples. The examples show a case for the detection of the transmitted Ly$\alpha$ forest flux in NB718 (e.g. DEIMOS\_2018\_519281) and one for a non-detection (e.g. SILVERRUSH\_2021\_7140). ({\bf Left}): The best-fit power-law SEDs (black solid) with the $14-86\%$ and $5-95\%$ confidence intervals (dark and light grey shaded regions) are overlaid on the measured NB718 (blue), $z$ (yellow), and $y$ (red) band fluxes of the background source. The wavelength coverage of each NB718, $z$ and $y$-band filter is indicated by the transparent filled curves. $5''\times5''$ postage stamps show the images of $g$, NB718, $z$, and $y$ fluxes smoothed by a Gaussian kernel with a standard deviation of one pixel. The colourbar scales are the same for all images. ({\bf Right}): MCMC corner plots of the key parameters ($\TIGM,\Muv,\beta$).}
    \label{fig:MCMC_result}
\end{figure*}

\subsection{Bayesian SED fitting framework}

We apply a Bayesian SED fitting framework to measure $T_{\rm IGM}$. We forward model the observed photometric fluxes in narrow- and broad-band filters using realistic HSC filter transmission curves $t_{\rm NB}(\nu)$ and $t_{\rm BB}(\nu)$ including the CCD quantum efficiency, the transmittance of the dewar window and the Primary Focus Unit of the HSC. For our IGM tomography, $\rm NB=NB718$ and ${\rm BB}=z,y$ since we use NB718 to measure the Ly$\alpha$ forest transmission and $z$- and $y$-bands to constrain the intrinsic galaxy SED.

We denote a model SED of a background galaxy by $L_\nu(\nu_{\rm e}|\boldsymbol{\Theta})$ (in unit of $\rm erg\,s^{-1}\,Hz^{-1}$) where $\nu_e$ is the emitted frequency at the rest-frame of the galaxy and $\boldsymbol{\Theta}$ is a set of the SED parameters. We assume a power-law SED with $L_\nu=L_\nu(1500\A)(\nu_e/\nu_{1500})^{-(2+\beta)}$ where $L_\nu(1500\A)$ and $\nu_{1500}$ represent the luminosity and frequency at $1500\,\A$ respectively, and $\beta$ is the UV continuum slope. Thus our SED parameters are $\boldsymbol{\Theta}=\{M_{\rm UV},\beta\}$. At the known redshift of the object (either from spectroscopy or NB detection of Ly$\alpha$), the observed flux is $f_\nu(\nu|\boldsymbol{\Theta})=(1+z)L_\nu[\nu_e=\nu(1+z)|\boldsymbol{\Theta}]/(4\pi D_L(z)^2)$ where $\nu$ is the observed frequency and $D_L(z)$ is the luminosity distance.

As the foreground NB718 filter covers a portion of Ly$\alpha$ forest of a background galaxy, the observed NB718 flux is attenuated by $e^{-\tau_\alpha}$ where $\tau_\alpha$ is the the Ly$\alpha$ optical depth of the IGM, therefore,
\begin{equation}
f_{\rm NB}(T_{\rm IGM},\boldsymbol{\Theta})=\frac{\int e^{-\tau_\alpha}f_\nu(\nu|\boldsymbol{\Theta})\,t_{\rm NB}(\nu)d\nu}{\int t_{\rm NB}(\nu)d\nu}\approx T_{\rm IGM }f_{\rm NB}^{\rm intr}(\boldsymbol{\Theta}).
\end{equation}
We define the NB-integrated Ly$\alpha$ forest transmission of the IGM as $T_{\rm IGM}=\left.\int e^{-\tau_\alpha}\,t_{\rm NB}(\nu)d\nu\right/\int t_{\rm NB}(\nu)d\nu$. In the absence of the IGM, the NB718 flux is $f_{\rm NB}^{\rm intr}(\boldsymbol{\Theta})=\left.\int f_\nu(\nu|\boldsymbol{\Theta})\,t_{\rm NB}(\nu)d\nu\right/\int t_{\rm NB}(\nu)d\nu$. The BB fluxes redward of Ly$\alpha$ are not affected by the IGM. They are thus modelled as
\begin{equation}
f_{\rm BB}(\boldsymbol{\Theta})=\frac{\int f_\nu(\nu|\boldsymbol{\Theta})\,t_{\rm BB}(\nu)d\nu}{\int t_{\rm BB}(\nu)d\nu}.
\end{equation}

We assume the observed photometric noise follows a Gaussian distribution and that noise levels in the various filters do not correlate with one another. Therefore, the likelihood can be written as the sum of Gaussian likelihoods,
\begin{equation}
    \ln\mathcal{L}=-\frac{1}{2}\left[\frac{f_{\rm NB}^{\rm obs}-f_{\rm NB}(T_{\rm IGM},\bm{\Theta})}{\sigma_{\rm NB}}\right]^2-\frac{1}{2}\sum_{{\rm BB}= z,y}\left[\frac{f_{\rm BB}^{\rm obs}-f_{\rm BB}(\bm{\Theta})}{\sigma_{\rm BB}}\right]^2.
\end{equation}

Using Bayes theorem, we can express the posterior as a product of prior $P(T_{\rm IGM},\boldsymbol{\Theta})$ and likelihood $\mathcal{L}$,
\begin{equation}
P(T_{\rm IGM},\boldsymbol{\Theta}|f_{\rm NB}^{\rm obs}, \boldsymbol{f}_{\rm BB}^{\rm obs})\propto P(T_{\rm IGM},\boldsymbol{\Theta})\mathcal{L}(f_{\rm NB}^{\rm obs}, \boldsymbol{f}_{\rm BB}^{\rm obs}|T_{\rm IGM},\boldsymbol{\Theta}).
\end{equation}
Thus the measurement of the Ly$\alpha$ forest transmission $T_{\rm IGM}$ along each background galaxy is given by the marginalized posterior over the SED parameters $\boldsymbol{\Theta}$,
\begin{equation}
P(T_{\rm IGM}|f_{\rm NB}^{\rm obs}, \boldsymbol{f}_{\rm BB}^{\rm obs})=\int P(T_{\rm IGM},\boldsymbol{\Theta}|f_{\rm NB}^{\rm obs}, \boldsymbol{f}_{\rm BB}^{\rm obs})d\boldsymbol{\Theta}.\label{eq:posterior}
\end{equation}

We implement this Bayesian SED fitting framework using a Markov Chain Monte Carlo method \textsc{emcee} \citep{Foreman-Mackey2013}. We use flat priors in the range of $-100<\TIGM<100$, $-30<\Muv<-15$, and $-5<\beta<2$ as our default. We justify the very wide range of flat priors (instead of imposing a physical range between 0 and 1 for individual measurements for $\TIGM$) in Section \ref{sec:analysis}.

When fitting the observed SEDs, we find occasional cases (13\,\% in the background LAE sample and 0\,\% in the DEIMOS10k sample) where the marginalized posterior peaks at an unphysically large value $\TIGM>1$. This indicates there may be an unaccounted systematic error in our data. In order to assess and remove such objects, we compute the probability that the estimated $\TIGM$ is greater than 1 using the posterior, i.e. $P(T_{\rm IGM}>1|f_{\rm NB}^{\rm obs}, \boldsymbol{f}_{\rm BB}^{\rm obs})=\int_1^\infty P(T_{\rm IGM}|f_{\rm NB}^{\rm obs}, \boldsymbol{f}_{\rm BB}^{\rm obs})d\TIGM$. We then flag objects with $P(T_{\rm IGM}>1|f_{\rm NB}^{\rm obs}, \boldsymbol{f}_{\rm BB}^{\rm obs})>50\,\%$. This choice is motivated by the fact that a Gaussian posterior centred at $T_{\rm IGM}>1$ gives $>50\,\%$ probability that $\TIGM$ is greater than 1, suggesting an unmodelled systematic error while SED fitting. We then visually check the original images and confirm such cases originate from systematic errors such as the under-subtraction of the sky background in NB718 and/or contamination from nearby objects in the photometric aperture. We apply this validation procedure\footnote{Note that while one can similarly flag objects with $T_{\rm IGM}<0$ to avoid systematics due to over-subtracted sky background, non-detection of NB flux from faint background sources will also result in the value centred on $T_{\rm IGM}=0$. This means that removing these objects will unphysically bias the result to a larger mean Ly$\alpha$ forest transmission. We thus decided not to remove these objects from our analysis.} in both the DEIMOS10k and LAE samples. We find no such cases (out of 36) in the DEIMOS10k catalogue. However, 11 cases (out of 115) are flagged in the LAE sample. Here, our visual inspection confirms likely contamination from nearby objects or a diffuse NB718 image compared to that in the BB $z$-band, suggesting that the region may be affected by the faint halo or ghost of a nearby bright star or by incorrect sky subtraction. Thus these 11 objects are removed and we use the remaining 104 background LAEs for our subsequent analysis.

\subsection{Individual IGM Ly$\alpha$ forest transmission measurements}\label{sec:individual_TIGM}

\begin{figure}
  \centering
    \includegraphics[width=\columnwidth]{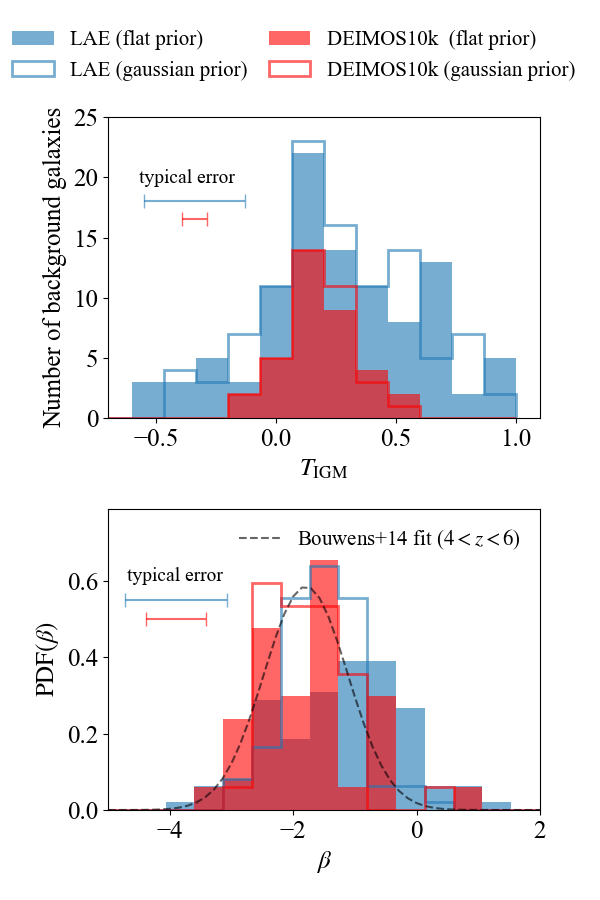}
    \vspace{-0.8cm}
    \caption{({\bf Top}): Distribution of estimated $\TIGM$ values from the mean of each posterior of individual background galaxies (blue: LAEs, red: DEIMOS10k) using a flat prior (filled histogram) and Gaussian prior (step histogram) on the $\beta$ UV slope. The median standard deviation of the posteriors is indicated as the typical error of the measurement. ({\bf Bottom}): As above but for the UV slope $\beta$. The Gaussian fit to the distribution of $\beta$ slopes from \citet{Bouwens2014} is indicated by the dashed line.}
      \label{fig:TIGM_histogram}
  \end{figure}

\begin{figure*}
  \centering
    \includegraphics[width=\columnwidth]{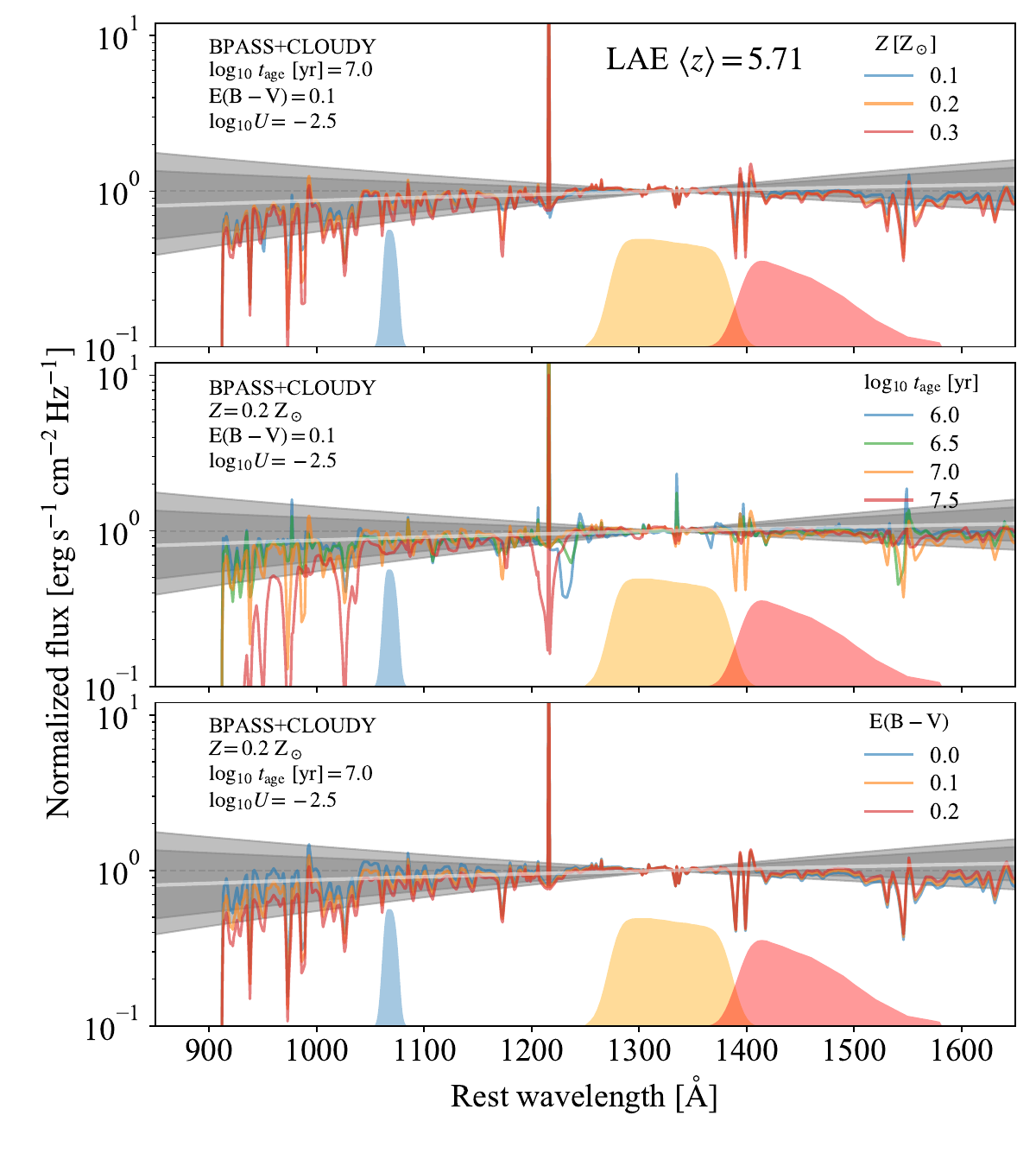}
    \includegraphics[width=\columnwidth]{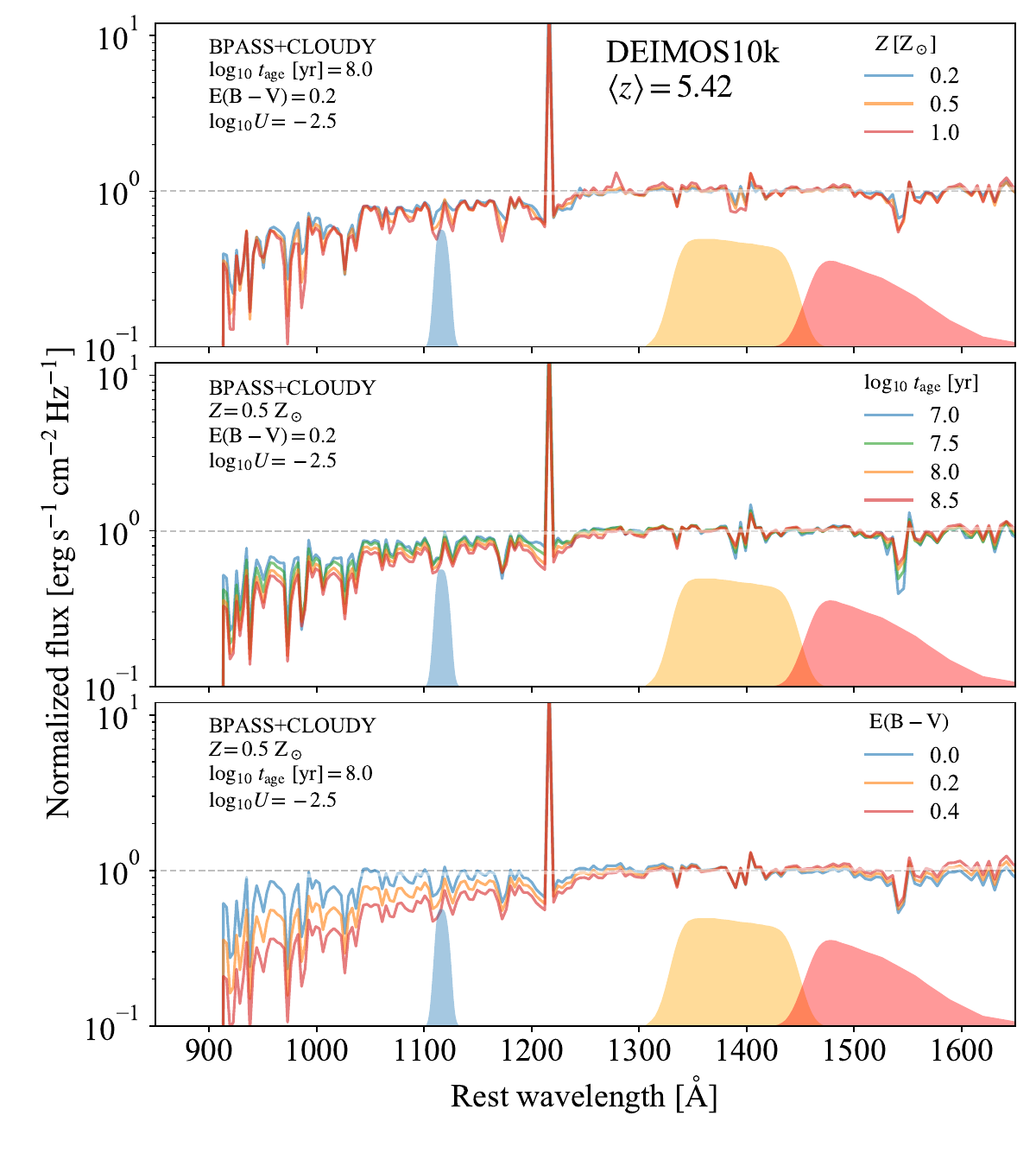}
    \caption{Comparison of the intrinsic galaxy SED without the IGM absorption from BPASS+CLOUDY and power-law fits to the LAE (left) and DEIMOS10k (right) samples. The dark (light) gray regions indicate the $16-84$ ($5-95$) percentiles of the range of the best-fit power-law models with the white line indicating the population-averaged $\beta$ slope. The default parameters of the BPASS+CLOUDY model are stellar age $t_{\rm age}=10\rm\,Myr$, metallicity $Z=0.20Z_\odot$, ionization parameter $\log_{10}U=-2.5$ and \citet{Calzetti2001} dust attenuation $E(B-V)=0.1$. The filter transmission curves for NB718 (blue), $z$ (yellow), and $y$ (red) bands are indicated by the shaded regions.}
    \label{fig:BPASS_SED}
\end{figure*}

By applying the Bayesian SED fitting framework, we measure the Ly$\alpha$ forest transmission of the IGM along sightlines to individual background galaxies. The method simultaneously returns the constraints on the Ly$\alpha$ forest transmission $T_{\rm IGM}$ and the SED parameters ($\Muv$ and $\beta$) for each source. In Figure \ref{fig:MCMC_result} we show two representative examples of the derived constraints on these physical parameters for the DEIMOS10k and LAE samples. When the transmitted Ly$\alpha$ forest flux is detected in NB718 as demonstrated by the DEIMOS10k sample (e.g. DEIMOS\_2018\_L519281), we can clearly measure the Ly$\alpha$ forest transmission. In the case of a non-detection in NB718 (e.g. SILVERRUSH\_2021\_3832) the derived constraint on $\TIGM$ is consistent with zero within the photometric uncertainty. The table of the MCMC results for the full DEIMOS10k and LAE samples is available as the supplementary online material.

Figure \ref{fig:TIGM_histogram} shows the distribution of the estimated Ly$\alpha$ forest transmission. The SED fitting provides $|\delta\TIGM/\TIGM|\sim49\%$ and $76\%$ determinations of the Ly$\alpha$ forest transmission for the DEIMOS10k and LAE samples with median $\delta\TIGM\sim0.10$ and $0.42$, where $\delta\TIGM$ and $\TIGM$ are given by the standard deviation and mean of the posterior. This includes errors from both photometric noise and the uncertainty in the intrinsic UV continuum. Defining the signal-to-noise ratio (SNR) to be the inverse of the relative error ${\rm SNR}=|\delta\TIGM/\TIGM|^{-1}$, the median SNR of the individual $\TIGM$ measurements is thus ${\rm SNR}\simeq2.0$ and $1.3$ for DEIMOS10k and LAE samples, respectively.

The uncertainty in $\TIGM$ for the LAEs is larger than the DEIMOS10k sample. Since the LAE sample is typically fainter than the DEIMOS10k sample, it is more severely affected by photometric noise because of ({\it i}) the reduced contrast between the UV continuum level ($z$-band) and the Ly$\alpha$ forest flux (NB718) and ({\it ii}) a less precise determination of the UV continuum slope ($z-y$ colour). 

The uncertainty in the UV continuum slope introduces a degeneracy in the estimate of the Ly$\alpha$ forest transmission. As discussed in \citet{Kakiichi2022}, for power-law spectra the uncertainty in the continuum slope enters as $\TIGM^{\rm estimated}=(\lambda_{\rm NB}/\lambda_{\rm BB})^{\beta_{\rm true}-\beta_{\rm temp}}\TIGM^{\rm true}$ where $\beta_{\rm true}$ and $\beta_{\rm temp}$ are the continuum slopes of true and template spectra and the central wavelengths of the filters are $\lambda_{\rm NB}=7170.5\,\A$ for NB718 and $\lambda_{\rm BB}=8912.6\,\A$ for the $z$-band. The typical relative error $\delta\epsilon_{\rm cont}\equiv |T_{\rm IGM}^{\rm true}-T_{\rm IGM}^{\rm estimated}|/T_{\rm IGM}^{\rm true}$ is then estimated as:
\begin{equation}
\langle\delta\epsilon_{\rm cont}\rangle=\int |1-(\lambda_{\rm NB}/\lambda_{\rm BB})^{\beta_{\rm true}-\beta_{\rm temp}}|P(\beta_{\rm true}|\bar{\beta},\sigma_\beta)d\beta_{\rm true},
\end{equation}
resulting in $\langle\delta\epsilon_{\rm cont}\rangle\approx11\,\%$ ($21\,\%$) error for $\beta_{\rm temp}=-1.8$ ($-1.8\pm1$) and assuming that true $\beta$ slopes follows a Gaussian PDF with mean $\bar{\beta}=-1.8$ and $\sigma_\beta=0.7$ consistent with \citet{Bouwens2014}. In comparison, the relative error due to the photometric noise in the Ly$\alpha$ forest transmission can be estimated by approximating $\TIGM\approx f_{\rm NB}/f_{\rm z}$ as
\begin{equation}
  \delta\epsilon_{\rm phot}=\sqrt{(\delta f_{\rm NB718}/f_{\rm NB718})^2+(\delta f_{\rm z}/f_{\rm z})^2},
\end{equation}
where $\delta f_{\rm NB718}$ and $\delta f_{\rm z}$ are the limiting fluxes in NB718 and $z$. 
The median relative error of the DEIMOS10k and LAE samples are $\sim46\,\%$ and $\sim89\,\%$. These estimates indicate that for our current HSC depth, the photometric noise dominates the continuum slope uncertainty. 

This point is further reinforced by the fact that the choice of flat or Gaussian priors on the continuum slope has little impact on the resulting distribution of individual $\TIGM$ (Figure \ref{fig:TIGM_histogram}). 
A two-sample Kolmogorov-Smirnov test indicates the difference is not statistically significant with p-values of $p=0.392$ and $0.999$ for the LAE and DEIMOS10k samples respectively. 

\subsubsection{Population synthesis vs power-law SEDs}\label{sec:SED}

We now test whether a power-law spectrum is sufficient to accurately represent the intrinsic galaxy spectrum for the purposes of IGM tomography. We compare the power-law spectrum with the intrinsic galaxy spectrum model without the Ly$\alpha$ forest absorption from the stellar population synthesis code BPASS (v2.2.1, \citealt{Eldrige2017,Stanway2018}) processed with the photionization code CLOUDY \citep[c17.01][]{Ferland2017}. We generate a grid of model spectra with metallicities, $Z=0.1,0.2,0.3,0.5,1.0\rm\,Z_\odot$ and stellar ages, $\log_{10}t_{\rm age}/{\rm yr}=6.0,6.5,7.0,7.5,8.0,8.5$, with a Salpeter initial mass function with the upper mass limit set to $300\,\rm M_\odot$ including binary stars. We assume an instantanous starburst to model the LAEs and a continuous star formation history to model the Lyman-break galaxies (LBGs) in the DEIMOS10k sample. We assume the stellar population is spherically surrounded by gas with an electron density $n_e=200\rm\,cm^{-3}$ and ionization parameter $\log_{10}U=-2.5$ \citep[e.g.][]{Davis2021,Reddy2023}. We then apply the \citet{Calzetti2001} dust attenuation curve with $E(B-V)=0.0,0.1,0.2,0.4$ to the BPASS+CLOUDY outputs to model the intrinsic galaxy spectra.

In Figure \ref{fig:BPASS_SED} (left) we compare the LAE power-law spectra with a range of the best-fit UV continuum slopes with intrinsic spectra calculate from the BPASS+CLOUDY model with varying metallicities, ages, and dust extinctions. All the spectra are normalized at $\sim1330\,\A$ corresponding to the rest-frame wavelength coverage of the $z$-band. For a typical range of metallicities $Z\sim0.1-0.3\,Z_\odot$, stellar ages $t_{\rm age}\sim1-30\rm\,Myr$, and dust extinction $E(B-V)\sim0.0-0.3$ for LAEs (e.g. \citealt{Ono2010,Guaita2011, Nakajima2012, Hagen2014, Trainor2016}, see also reviews by \citealt{Hayes2019,Ouchi2020}), the power-law template approximates the continuum shape of the BPASS+CLOUDY spectra at $\sim1000-1600\,\A$ very well. The relative error in the estimated $\TIGM$ due to adopting different SEDs, $\delta\epsilon_{\rm SED}\equiv|(\TIGM^{\rm power-law}-\TIGM^{\rm BPASS+CLOUDY})/\TIGM^{\rm BPASS+CLOUDY}|$, is given by
\begin{equation}
  \delta\epsilon_{\rm SED}=|1-f_{\rm NB}^{\rm intr, BPASS+CLOUDY}/f_{\rm NB}^{\rm intr, power-law}|.
\end{equation}
Comparing the power-law template with median $\beta$ slope and the BPASS+CLOUDY spectrum with typical LAE parameters of $Z=0.20\rm\,Z_\odot$, $\log_{10} t_{\rm age}/{\rm yr}=7.0$, and $E(B-V)=0.10$, the relative error is $\delta\epsilon_{\rm SED}\sim 6\,\%$ ($\sim 15$ and $17\,\%$ for $E(B-V)=0.2$ and $\log_{10} t_{\rm age}/{\rm yr}=7.5$ respectively). This is much smaller than the error from photometric noise and be captured by the continuum slope error budget of the power-law template. We conclude the power-law SED fit is sufficient to predict the intrinsic flux at the Ly$\alpha$ forest region for the current HSC depth.

We expect a similar uncertainty for the DEIMOS10k sample, which should primarily consist of LBGs. Figure \ref{fig:BPASS_SED} (right) shows the same comparison normalized at $\sim1400\,\A$ corresponding to the $z$-band coverage at the mean redshift of the DEIMOS10k sample. LBGs typically span a range of dust extinction $E(B-V)\sim0.0-0.4$ \citep{deBarros2014,Reddy2016}, metallicities $Z\sim0.2-1.0\rm\,Z_\odot$ \citep{Steidel2014}, and stellar ages $t_{\rm age}\sim50-500\rm\,Myr$ \citep{Stark2009,Curtis-Lake2013,deBarros2014}. Being more mature star-forming galaxies than LAEs, their older stellar ages or higher dust extinctions introduce a downward trend towards shorter wavelengths. For typical LBG parameters of $Z=0.5\rm\,Z_\odot$, $\log_{10}\,t_{\rm age}/{\rm yr}=8.0$, and $E(B-V)=0.2$, the impact on the estimated $T_{\rm IGM}$ between the BPASS+CLOUDY and power-law SEDs is $\delta\epsilon_{\rm SED}\sim27\rm\,\%$ ($\sim39$ and $31\rm\,\%$ for $E(B-V)=0.4$ and $\log_{10}\,t_{\rm age}/{\rm yr}=8.5$ respectively). While larger than for the LAEs, this is still within the photometric error for the individual measurements of $T_{\rm IGM}$. However, if this range of stellar ages and dust extinctions is representative of the true intrinsic spectra of background spec-z LBGs, it could introduce a systematic bias in stacked measurements. The use of power-law SEDs for background spec-z sample could systematically underestimate the measured Ly$\alpha$ forest transmission because it predicts an intrinsic UV continuum level larger than the true value. To quantify and eliminate the possible impact of this limitation, we would need to extend our analysis (which currently only uses $z$- and $y$-bands) including near-infrared data to better constrain the ages and dust extinctions of background galaxies.
We discuss this strategy further in Section \ref{sec:discussions}, but in the following analysis we only discuss the effect of this potential bias.

Another possible systematic arising from the assumption of a power-law spectrum is the presence of stellar photospheric and interstellar absorption lines in the Ly$\alpha$ forest region of a background galaxy. Prominant absorption lines between Ly$\beta$ and Ly$\alpha$ lines ($1026-1216\,\A$) include $\CII\,\lambda1036$, $\SIV/\FeII\,\lambda1063$, $\NII\,\lambda1084$, $\NI\,\lambda1134$, $\CIII\,\lambda1176$, $\SiII\,\lambda\lambda1190,\!1193$, $\SiIII\,\lambda1207$ \citep{Reddy2016}. Previous spectroscopic IGM tomographic surveys mitigated this issue by masking the $\pm2-5\,\A$ regions around each absorption line \citep{Lee2014b,Lee2018,Newman2020}. For $z\sim5.7$ background LAEs, the NB718 filter covers the rest-frame wavelength between 1050 and 1082\,\A, which coincides with the $\SIV/\FeII\,\lambda1063$ line. According to a stacked galaxy spectrum \citep{Newman2020}, the typical rest-frame equivalent width of the absorption line is ${\rm EW}_{\rm abs}\sim0.8\,\A$.  The effect of an absorption line on the NB718 flux is thus $(1+z){\rm EW}_{\rm abs}/\Delta\lambda_{\rm NB718}\approx4.8\,\%$ reduction in flux integrated over the NB718 filter. Accordingly, the resulting bias in the estimated $\TIGM$ is only
\begin{equation}
  \delta\epsilon_{\rm abs}=(1+z){\rm EW}_{\rm abs}/\Delta\lambda_{\rm NB718},
\end{equation}
i.e. $4.8\,\%$, which is negligible compared to the other sources of error discussed above. For the DEIMOS10k sample, the spectroscopic redshifts span the range of $4.98<z<5.89$. As the contamination in NB718 flux by the absorption lines is expected to be randomized, the effect of absorption lines on the overall DEIMOS10k sample should be negligible.

\section{Mean Ly$\alpha$ forest transmission}\label{sec:analysis}
\subsection{Estimating mean Ly$\alpha$ forest transmission}  

The mean Ly$\alpha$ forest transmission is simply the mean in a representative volume $V$, $\overline{T}_{\rm IGM}=V^{-1}\int\TIGM(\boldsymbol{x})dV$ where $\boldsymbol{x}$ is a 3D spatial position in the Universe. For IGM tomography, we can only sample $T_{\rm IGM}(\boldsymbol{x})$ along sightlines to background galaxies. For each background galaxy at a location $\boldsymbol{\theta}_i$, we sample $T_{\rm IGM}(\boldsymbol{\theta}_i)=T_{{\rm IGM},i}$ where $T_{{\rm IGM},i}$ is the measured Ly$\alpha$ forest transmission integrated along the line of sight to an $i$th galaxy over the width of the NB filter. Since the background galaxies are uncorrelated with the foreground IGM structure, providing random sampling of $\TIGM(\boldsymbol{x})$, this is equivalent to performing the Monte Carlo integration of the mean Ly$\alpha$ forest transmission, i.e.
\begin{equation}
\overline{T}_{\rm IGM}=\frac{1}{N_{\rm bg}}\sum_{i=1}^{N_{\rm bg}}T_{{\rm IGM},i}.
\end{equation}
Since we have noisy measurements of $T_{{\rm IGM},i}$ characterised by the posterior $P(T_{{\rm IGM},i}|f_{{\rm NB},i}^{\rm obs}, \boldsymbol{f}_{{\rm BB},i}^{\rm obs})$ (equation \ref{eq:posterior}) for a set of background galaxies $i=1,\dots,N_{\rm bg}$, our estimate of the mean Ly$\alpha$ forest transmission is also noisy. Thus, to characterise the uncertainty in the mean Ly$\alpha$ forest transmission, we need to know the posterior probability of $\overline{T}_{\rm IGM}$, that is, $P(\overline{T}_{\rm IGM}|\{f_{{\rm NB},i}^{\rm obs}, \boldsymbol{f}_{{\rm BB},i}^{\rm obs}\}_{i=1,\dots,N_{\rm bg}})$. Assuming individual measurements of $\TIGMi$ are statistically independent, the joint posterior probability is simply the product of all the individual posteriors, i.e. $P(\{T_{{\rm IGM},i}\}_{i=1,\dots,N_{\rm bg}}|\{f_{{\rm NB},i}^{\rm obs}, \boldsymbol{f}_{{\rm BB},i}^{\rm obs}\}_{i=1,\dots,N_{\rm bg}})=\prod_{i=1}^{N_{\rm bg}} P(T_{{\rm IGM},i}|f_{{\rm NB},i}^{\rm obs}, \boldsymbol{f}_{{\rm BB},i}^{\rm obs})$. Formally, the posterior probability of the mean Ly$\alpha$ forest transmission can then be written in terms of the convolution of the individual posteriors of $\TIGMi$ of all background galaxies \citep[e.g.][Section 3.6]{Sivia2006},
\begin{align}
&P(\overline{T}_{\rm IGM}|\{f_{{\rm NB},i}^{\rm obs}, \boldsymbol{f}_{{\rm BB},i}^{\rm obs}\}_{i=1,\dots,N_{\rm bg}})= \nonumber \\
&\int \prod_{i=1}^{N_{\rm bg}} dT_{{\rm IGM},i} P(T_{{\rm IGM},i}|f_{{\rm NB},i}^{\rm obs}, \boldsymbol{f}_{{\rm BB},i}^{\rm obs}) \delta_{\rm D}\left(\overline{T}_{\rm IGM}-\frac{1}{N_{\rm bg}}\sum_{i=1}^{N_{\rm bg}}T_{{\rm IGM},i}\right), \label{eq:full_posterior}
\end{align}
where $\delta_{\rm D}(x)$ is the Dirac delta function. Numerically, it is easy to generate the posterior of $\overline{T}_{\rm IGM}$, i.e. $P(\overline{T}_{\rm IGM}|\{f_{{\rm NB},i}^{\rm obs}, \boldsymbol{f}_{{\rm BB},i}^{\rm obs}\}_{i=1,\dots,N_{\rm bg}})$, by computing the histogram of many $\overline{T}_{\rm IGM}$'s using random draws of $\TIGMi$ from $P(T_{{\rm IGM},i}|f_{{\rm NB},i}^{\rm obs}, \boldsymbol{f}_{{\rm BB},i}^{\rm obs})$.

Equation (\ref{eq:full_posterior}) illustrates an important, but subtle point on the choice of prior on $\TIGMi$. When the data has no constraining power, i.e. $P(T_{{\rm IGM},i}|f_{{\rm NB},i}^{\rm obs}, \boldsymbol{f}_{{\rm BB},i}^{\rm obs})\rightarrow P(T_{{\rm IGM},i})$, the posterior of $\overline{T}_{\rm IGM}$ becomes a convolution of multiple priors of $\TIGMi$. Assuming any prior with mean $\overline{T}_{\rm prior}$ and standard deviation $\sigma_{\rm prior}$ for all the individual measurements, by the virtue of central limit theorem, the posterior of $\overline{T}_{\rm IGM}$ approaches a Gaussian distribution with mean $\overline{T}_{\rm prior}$ and standard deviation $\sigma_{\overline{T}_{\rm IGM}}=\sigma_{\rm prior}/\sqrt{N_{\rm bg}}$. In our case, if we were to impose a flat prior between 0 and 1 (i.e. $\overline{T}_{\rm prior}=0.5$ and $\sigma_{\rm prior}=0.29$) for the 104 individual measurements, the end result would be a Gaussian posterior on $\overline{T}_{\rm IGM}$ with mean $\overline{T}_{\rm prior}=0.5$ and $\sigma_{\overline{T}_{\rm IGM}}=0.29/\sqrt{104}=0.028$ even for a completely uninformative dataset. This demonstrates that a reasonable prior on individual measurements propagates into an unreasonably tight constraint on the mean Ly$\alpha$ forest transmission. While imposing a flat prior with $0<\TIGM<1$ seems an innocent assumption, when we are interested in the mean we should use a maximally non-informative prior on the individual measurement to avoid an artificial constraint on the final estimate of $\overline{T}_{\rm IGM}$. We thus use a flat prior on individual measurement allowing a very wide range between $-100<\TIGM<100$.

Given the full posterior probability of $\overline{T}_{\rm IGM}$, it is natural to take our best estimate of the mean Ly$\alpha$ forest transmission as the expectation value $\langle \overline{T}_{\rm IGM}\rangle=\int \overline{T}_{\rm IGM} P(\overline{T}_{\rm IGM}|\{f_{{\rm NB},i}^{\rm obs}, \boldsymbol{f}_{{\rm BB},i}^{\rm obs}\}_{i=1,\dots,N_{\rm bg}})d\overline{T}_{\rm IGM}$, which simplifies as the average of the expectation values of individual measurements of $\TIGMi$ (see Appendix \ref{app:derivation} for derivation),
\begin{equation}
\langle \overline{T}_{\rm IGM}\rangle=\frac{1}{N_{\rm bg}}\sum_{i=1}^{N_{\rm bg}} \langle T_{{\rm IGM},i}\rangle,\label{eq:expectation_value}
\end{equation}
where $\langle T_{{\rm IGM},i}\rangle=\int T_{{\rm IGM},i}P(T_{{\rm IGM},i}|f_{{\rm NB},i}^{\rm obs}, \boldsymbol{f}_{{\rm BB},i}^{\rm obs})dT_{{\rm IGM},i}$. Similarly, the variance of the mean Ly$\alpha$ forest transmission, $\sigma^2_{\overline{T}_{\rm IGM}}=\int (\overline{T}_{\rm IGM}-\langle\overline{T}_{\rm IGM}\rangle)^2 P(\overline{T}_{\rm IGM}|\{f_{{\rm NB},i}^{\rm obs}, \boldsymbol{f}_{{\rm BB},i}^{\rm obs}\}_{i=1,\dots,N_{\rm bg}})d\overline{T}_{\rm IGM}$, is given by
\begin{align}
\sigma^2_{\overline{T}_{\rm IGM}}=\frac{1}{N_{\rm bg}^2}\sum_{i=1}^{N_{\rm bg}} {\rm Var}[\TIGMi].\label{eq:variance}
\end{align}
This again simply follows from the sum of the variances of individual measurements, ${\rm Var}[T_{{\rm IGM},i}]=\int (T_{{\rm IGM},i}-\langle T_{{\rm IGM},i}\rangle)^2P(T_{{\rm IGM},i}|f_{{\rm NB},i}^{\rm obs}, \boldsymbol{f}_{{\rm BB},i}^{\rm obs})dT_{{\rm IGM},i}$. Note that because the final variance $\sigma^2_{\overline{T}_{\rm IGM}}$ scales like $1/N_{\rm bg}$ times the average variance of the individual measurement, the final error scales as $\sigma_{\overline{T}_{\rm IGM}}\propto 1/\sqrt{N_{\rm bg}}$. All these quantities are easy to compute using the MCMC sample of $\TIGMi$ of the individual posteriors from the Bayesian SED fitting framework.

\begin{figure}
  \centering
    \includegraphics[width=\columnwidth]{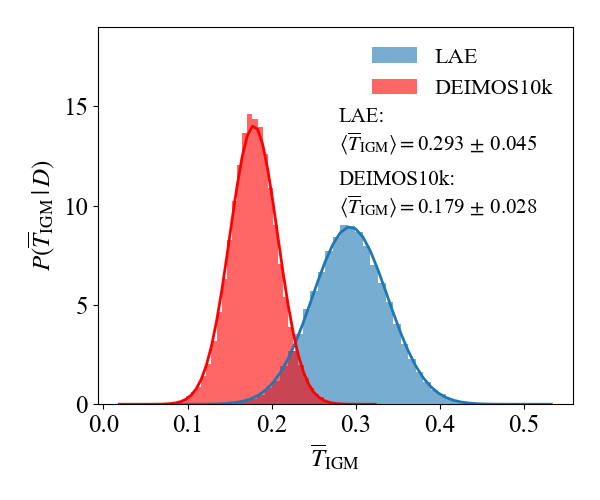}
    \vspace{-0.8cm}
    \caption{Full posterior of the mean Ly$\alpha$ forest transmission $\overline{T}_{\rm IGM}$, $P(\overline{T}_{\rm IGM}|\{f_{{\rm NB},i}^{\rm obs}, \boldsymbol{f}_{{\rm BB},i}^{\rm obs}\}_{i=1,\dots,N_{\rm bg}})$, for the LAE (blue histogram) and DEIMOS10k (red histogram) samples. The histogram is computed using the mean Ly$\alpha$ forest transmission from 1,000,000 random realizations of a set of $\TIGMi$ from the individual posteriors. The blue and red solid curves are the Gaussian distributions with the expectation value and variance computed via (\ref{eq:expectation_value}) and (\ref{eq:variance}). Note that both the expectation value and the maximum a posteriori estimate of $\overline{T}_{\rm IGM}$ are equivalent.}
    \label{fig:TIGM_mean}
  \end{figure}

The central limit theorem guarantees that for a large number of background galaxies $P(\overline{T}_{\rm IGM}|\{f_{{\rm NB},i}^{\rm obs}, \boldsymbol{f}_{{\rm BB},i}^{\rm obs}\}_{i=1,\dots,N_{\rm bg}})$ approaches a Gaussian distribution. Thus the expectation value $\langle \overline{T}_{\rm IGM}\rangle$ is equivalent to the the maximum {\it a posteriori} estimate of the mean Ly$\alpha$ forest transmission, $\overline{T}_{\rm IGM}^{\rm MP}$, which is given by
\begin{equation}
  \overline{T}_{\rm IGM}^{\rm MP}=\underset{\scriptscriptstyle{\overline{T}}_{\rm IGM} }{\rm arg\,max}~P(\overline{T}_{\rm IGM}|\{f_{{\rm NB},i}^{\rm obs}, \boldsymbol{f}_{{\rm BB},i}^{\rm obs}\}_{i=1,\dots,N_{\rm bg}}).
\end{equation}
Figure \ref{fig:TIGM_mean} shows the full posterior of $\overline{T}_{\rm IGM}$ directly computed from random realizations of individual measurements, which explicitly confirms the equivalence between the expectation value and the maximum {\it a posteriori} estimate of $\overline{T}_{\rm IGM}$, i.e. $\overline{T}_{\rm IGM}^{\rm MP}=\langle \overline{T}_{\rm IGM}\rangle$. As the full posterior is Gaussian, the variance (equation \ref{eq:variance}) completely characterises the total error in the estimated mean Ly$\alpha$ froest transmission as $\langle\bar{T}_{\rm IGM}\rangle\pm\sigma_{\overline{T}_{\rm IGM}}$ at one sigma level. This error is fully propagated, including both photometric noise and UV continuum uncertainties, from the Bayesian SED fitting procedure for the individual measurements of $T_{{\rm IGM},i}$.

\subsection{Results}\label{sec:mean_TIGM_result}

Using the DEIMOS10k and LAE samples, the mean Ly$\alpha$ forest transmission at $z\simeq4.9$ are estimated to be
\begin{equation}
\langle \overline{T}_{\rm IGM}\rangle=0.179\pm0.028~~~~\mbox{(for DEIMOS10k)}
\end{equation}
and
\begin{equation}
\langle \overline{T}_{\rm IGM}\rangle=0.293\pm0.045~~~~\mbox{(for LAE)}
\end{equation}
This represents the first $\sim15\,\%$ photometric measurement of the mean Ly$\alpha$ forest transmission of the IGM using background galaxies. The statistical precision is comparable to the $\sim8\,\%$ determinations based using quasar spectra \citep{Becker2013,Becker2015,Eilers2018,Yang2020,Bosman2018,Bosman2022}. Our precision arises from a large number of background galaxies roughly consistent with the $\propto 1/\sqrt{N_{\rm bg}}$ scaling law, from which we expect $\delta \TIGM/\TIGM\sim0.49/\sqrt{36}$ $(0.76/\sqrt{104})$ $=8\,\%$ $(7\,\%)$ error on the mean transmission for the DEIMOS10k (LAE) sample. Note that the error budget quoted above using equation (\ref{eq:variance}) takes into account only the photometric noise and UV continuum uncertainties, but not the error from cosmic or patch-to-patch variance. To quantify its impact, we empirically estimate the total error using Jackknife resampling (described in Section \ref{sec:cross}). We find that the Jackknife errors are $\sigma_{\rm JK}=0.024$ and $0.047$ for the DEIMOS10k and LAE samples respectively, comparable to our analytic estimate of the (photometric noise + continuum) error. Thus, the additional error from cosmic variance is not significant.

\begin{figure}
  \centering
    \includegraphics[width=\columnwidth]{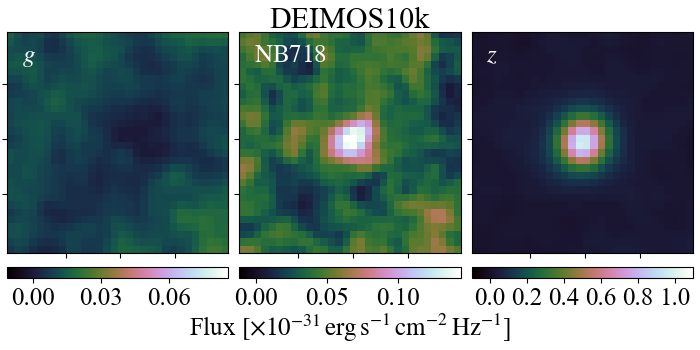} 	
    \includegraphics[width=\columnwidth]{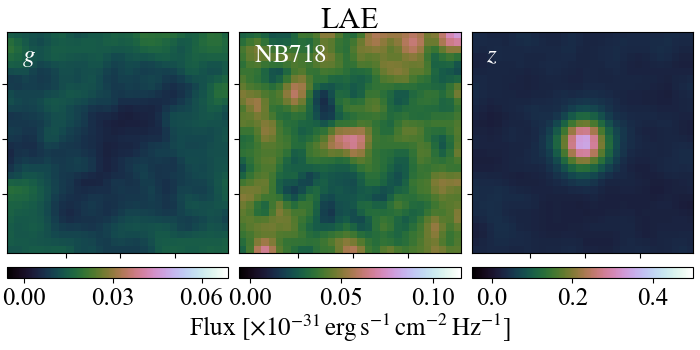}    
	\vspace{-0.5cm}    
    \caption{$5\arcsec\times5\arcsec$ cutout images of the sigma-clipped mean stack of $g$ (left), NB718 (middle), $z$ (right) images for the DEIMOS10k (top) and LAE (bottom) samples.}\label{fig:stack}
  \end{figure}

To visually confirm that the measured Ly$\alpha$ forest transmission is truly representative of the physical value at $z\simeq4.9$, we applied a sigma-clipped mean stacking of the NB718 and BB images of the DEIMOS10k and LAE sample in Figure \ref{fig:stack}. The signal is clearly detected in NB718 as well as via a clear UV continuum detection in $z$. The signal is also detected in the median stack. The non-detection in mean $g$ stack reinforces that the detected signal originates from the Ly$\alpha$ forest transmission towards the background galaxies and not due to low-redshift interlopers.

\subsubsection{Comparison with the literature}

\begin{figure}
  \hspace*{-0.6cm}
	\includegraphics[width=1.1\columnwidth]{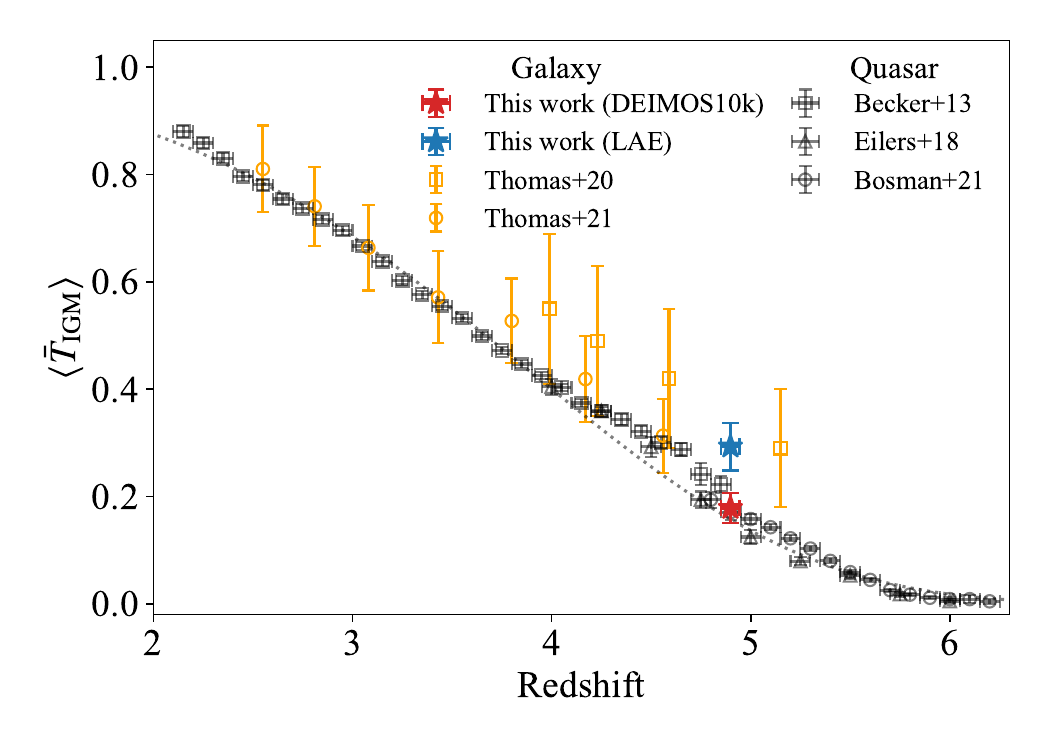}
  \vspace{-0.7cm}
  \caption{Comparison of the mean Ly$\alpha$ forest transmission measured in this work using the DEIMOS10k (red) and LAE (blue) samples with the previous measurements based on quasar \citep{Becker2013,Eilers2018,Bosman2022} and galaxy spectra \citep{Thomas2020,Thomas2021}.}
    \label{fig:TIGM}
\end{figure}

In Figure \ref{fig:TIGM} we compare our mean Ly$\alpha$ forest transmission with the measurements using quasars \citep{Becker2013,Eilers2018,Bosman2022} and galaxy spectra \citep{Thomas2017,Thomas2020,Thomas2021} in the literature. Using high signal-to-noise quasar spectra, the former measures the mean Ly$\alpha$ forest transmission in bins of $50\,h^{-1}\rm cMpc$ length. This is comparable to the line-of-sight comoving length $34\,h^{-1}\rm cMpc$ of the NB718 filter. \citet{Thomas2017,Thomas2020,Thomas2021} used a large spectroscopic sample of galaxies from the VANDELS and VUDS surveys and measured the Ly$\alpha$ forest transmission from the rest-frame $1070-1170\,\A$ region of background galaxies by using various IGM templates in their spectral fitting method (see also \citet{Monzon2020} who used a stacking method).

Our measurement using the DEIMOS10k sample is in excellent agreement with the quasar studies, in particular with the latest high signal-to-noise measurement of $\langle \overline{T}_{\rm IGM}\rangle=0.171\pm0.014$ based on the XQR-30 quasar sample \citep{Bosman2022}. Although our measurement using background LAEs is slightly higher, it is still broadly in agreement with the previous values from the quasar- and galaxy-based measurements. 

Our conclusion disagrees with the claim by \citet{Thomas2020} who argued that photometric data is insufficient to constrain the Ly$\alpha$ forest transmission. This is because their SED fitting used broad-band photometry to determine the Ly$\alpha$ forest transmission which is contaminated by the Ly$\alpha$ emission line and UV continuum and covers a region below the Ly$\beta$ line depending on the redshift of a background galaxy. This renders the resulting measurement of the Ly$\alpha$ forest transmission uncertain and is a likely source of their $\sim20\,\%$ discrepancy between their photometric and spectroscopic measurements. In contrast, our method uses a NB filter precisely covering the appropriate Ly$\alpha$ forest region enabling a clean photometric measurement of the transmitted Ly$\alpha$ forest flux. We argue that there is no fundamental limitation to the photometric approach when a carefully chosen combination of a NB filter and background galaxy redshifts is used.

\subsection{Systematics: low-redshift interlopers and SEDs}\label{sec:systematics}

There is a $\sim2\sigma$ tension in our measurements of the mean Ly$\alpha$ forest transmission between the DEIMOS10k and LAE samples. The statistical significance is calculated from the difference between the posterior means divided by the quadrature sum of the errors \citep{Lemos2021}, $(0.293-0.179)/\sqrt{0.045^2+0.028^2}=2.15\sigma$. While the discrepancy is statistically insignificant, it could indicate systematic effects outside our statistical error budget.

\subsubsection{Low-redshift interlopers}

The first possibility is contamination by the low-redshift interlopers in the background $z\simeq5.7$ LAE sample. A low-redshift interloper could bias the result by introducing a fictitious transmissive sightline. Possible interlopers in the LAE sample selected by a NB816 excess include low-redshift galaxies with strong emission lines from $z\simeq0.25$ H$\alpha$, $z\simeq0.63$ $[\OIII]\,\lambda5008$, and $z\simeq1.19$ $[\OII]\,\lambda\lambda3727,3729$, as well as slightly lower redshift AGN at $z\simeq4.28$ with $\CIV\,\lambda1549$ emission \citep[e.g.][]{Ouchi2008,Shibuya2018,Sobral2018}. Their rest-frame optical or UV continua could be mistaken in the NB718 filter as the transmitted Ly$\alpha$ forest flux at $z\simeq4.9$, artificially increasing the estimated mean Ly$\alpha$ forest transmission. The effect of a low-redshift interloper can be written as
\begin{equation}
    \langle\overline{T}_{\rm IGM}\rangle=(1-f_{\rm bg.int})\langle\overline{T}_{\rm IGM}\rangle^{\rm true}+f_{\rm bg.int}\langle\overline{T}_{\rm IGM}\rangle^{\rm bg.int},
\end{equation}
where $f_{\rm bg.int}$ is the contamination rate by low-redshift interlopers in the background LAE sample and $\langle\overline{T}_{\rm IGM}\rangle^{\rm bg.int}$ is the fictitious mean Ly$\alpha$ forest transmission along the sightlines of the interlopers. The interloper fraction in the LAE selection is typically $\sim20\%$ \citep{Shibuya2018}. Assuming that $\langle\overline{T}_{\rm IGM}\rangle^{\rm bg.int}=0.7$ and the measured mean Ly$\alpha$ forest transmission from DEIMOS10k sample is the 
true value  $\langle\overline{T}_{\rm IGM}\rangle^{\rm true}=0.179$, the observed value using NB-selected background LAEs will become $\langle\overline{T}_{\rm IGM}\rangle=(1-0.2)\times0.179+0.2\times0.7=0.283$, being consistent with the measurement from our LAE sample. This would resolve the tension between DEIMOS10k and LAE samples.

In order to explore this further, we cross-matched our background LAE catalogue with the spectroscopic catalogue compiled with HSC-SSP DR3 \citep{Aihara2022} containing 70,358 objects in the UD-COSMOS field (tract 9813). We find no interloper in our background LAE catalogue while 10 objects are spectroscopically-confirmed to be at $z\simeq5.7$. We also applied a stricter $g$-band non-detection cut compared to our default $3\sigma$ threshold and find that for a $1\sigma$ ($2\sigma$) threshold the estimated mean Ly$\alpha$ forest transmission becomes $\langle \overline{T}_{\rm IGM}\rangle=0.277\pm0.055$ $(0.298\pm0.048)$ for LAE sample; the values are consistent with our result from the $3\sigma$ threshold within $1\sigma$ error. Thus we find no obvious evidence for low-redshift interlopers in our LAE sample.

Note that as we require $>5\sigma$ detection in the $z$-band ($<26.86\rm\,mag$), potential interlopers, if any, need to have red $g-z\gtrsim2.0$ colours and be fainter than $29.60-28.84$ mag $(1-2\sigma)$ in $g$-band. Such interlopers could be low-redshift dusty red galaxies or Balmer break galaxies with H$\alpha$, $[\OIII]$, or $[\OII]$ doublet emission lines at $z\simeq0.25,0.63,1.19$. Ultimately, spectroscopic follow-up of the background LAEs would be necessary to fully reject systematic bias from low-redshift interlopers. Following up a random subset would determine the interloper fraction $f_{\rm bg.int}$ and by measuring the mean Ly$\alpha$ forest transmission along the interlopers, one can also determine $\langle\overline{T}_{\rm IGM}\rangle^{\rm bg.int}$. Then the observed Ly$\alpha$ forest transmission using the parent photometric background LAE sample, $\langle\overline{T}_{\rm IGM}\rangle^{\rm obs}$, could be statistically corrected via
\begin{equation}
    \langle\overline{T}_{\rm IGM}\rangle^{\rm corrected}=\frac{\langle\overline{T}_{\rm IGM}\rangle^{\rm obs}-f_{\rm bg.int}\langle\overline{T}_{\rm IGM}\rangle^{\rm bg.int}}{1-f_{\rm bg.int}}.
\end{equation}

\subsubsection{SED templates}

As discussed in Section \ref{sec:SED}, model galaxy SEDs may introduce a systematic error in $\TIGM$. The effect is expected to be larger in the DEIMOS10k sample which contains more mature galaxies with dust or more complex stellar populations for which the UV continuum cannot precisely be determined by only $z$ and $y$-band photometry. If the intrinsic continua were systematically overestimated by $\sim30\,\%$ (for example, because of an underestimated $E(B-V)$), then a more flexible galaxy SED template would lead to $\langle \overline{T}_{\rm IGM}\rangle=0.233\pm0.036$ relaxing the tension between the DEIMOS10k and LAE samples to $\sim1\sigma$. 
However, this explanation would weaken the agreement between both our measures and those determined using quasars.

Our choice of the Ly$\alpha$ forest wavelength range $(1026-1216\,\A)$ is more generous compared to the $1040-1190\,\A$  range commonly used for spectroscopic IGM tomography \citep{Lee2014b,Lee2018,Newman2020}. Although the effect of absorption lines is small (Section \ref{sec:SED}), the wider range means that a power-law continuum might neglect the effect of Ly$\beta$+$\CII\,\lambda1036$ and Ly$\alpha$+$\SiIII\,\lambda1207$ absorption lines. Also, neutral gas in the circumgalactic medium of a background galaxy could contribute to the Ly$\alpha$ absorption blueward of the line centre \citep{Rudie2013,Kakiichi2018,Bassett2021}. We can test this effect by adopting a more restrictive range of $1040-1190\,\A$ and limiting the redshift range of background galaxies to $5.072<z<5.841$ for the DEIMOS10k sample. We find $\langle\overline{T}_{\rm IGM}\rangle=0.174\pm0.035$, 
consistent with our main result. Thus contamination by absorption lines is negligible and cannot explain the tension.

\subsubsection{Cosmic variance}
Finally, cosmic variance may cause the tension between our two subsamples. Although the Jackknife resampling error should include the effects of cosmic variance, the small sample size may underestimate the effect. 
However, since both the DEIMOS10k and LAE samples probe similar regions of the sky in the COSMOS field, we consider cosmic variance an unlikely source of the tension.

\vspace{0.2cm}
In conclusion, we believe that the combination of a modest contribution from low-redshift interlopers and some variation in the SED templates is the mostly likely source of the $2\sigma$ tension between the DEIMOS10k and LAE samples. This can only be tested and corrected by spectroscopic follow-up of the background LAEs and inclusion of near-infrared photometry to better constrain the background galaxy SEDs.

\section{LAE-Ly$\alpha$ forest cross-correlation}\label{sec:cross}

\begin{figure}
 \centering
	\includegraphics[width=\columnwidth]{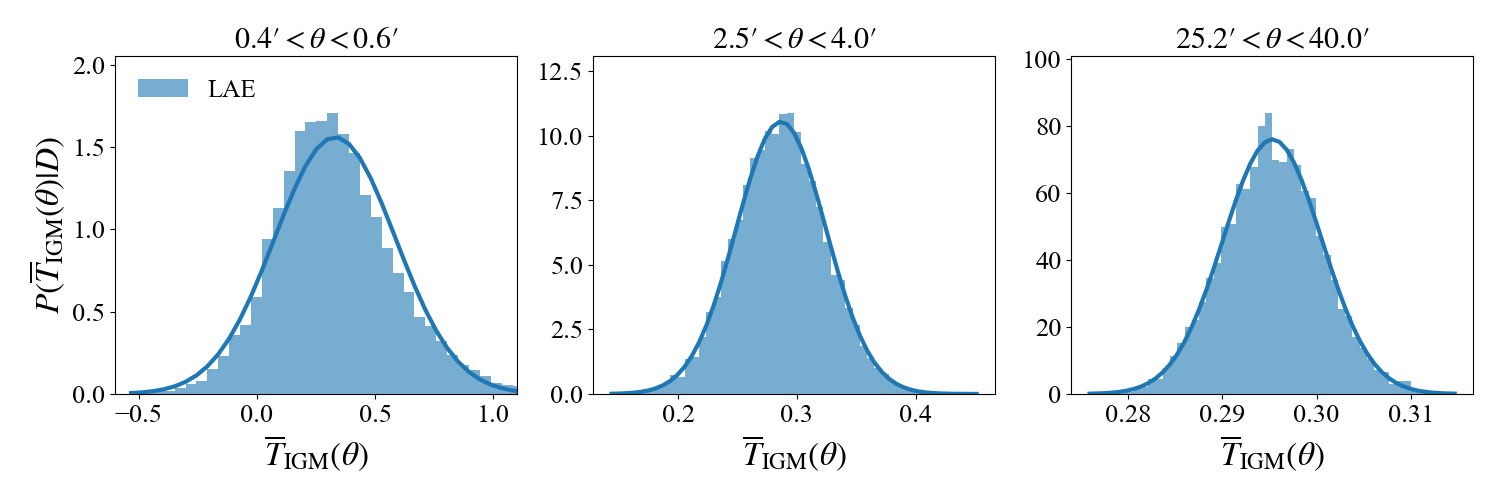}
  \includegraphics[width=\columnwidth]{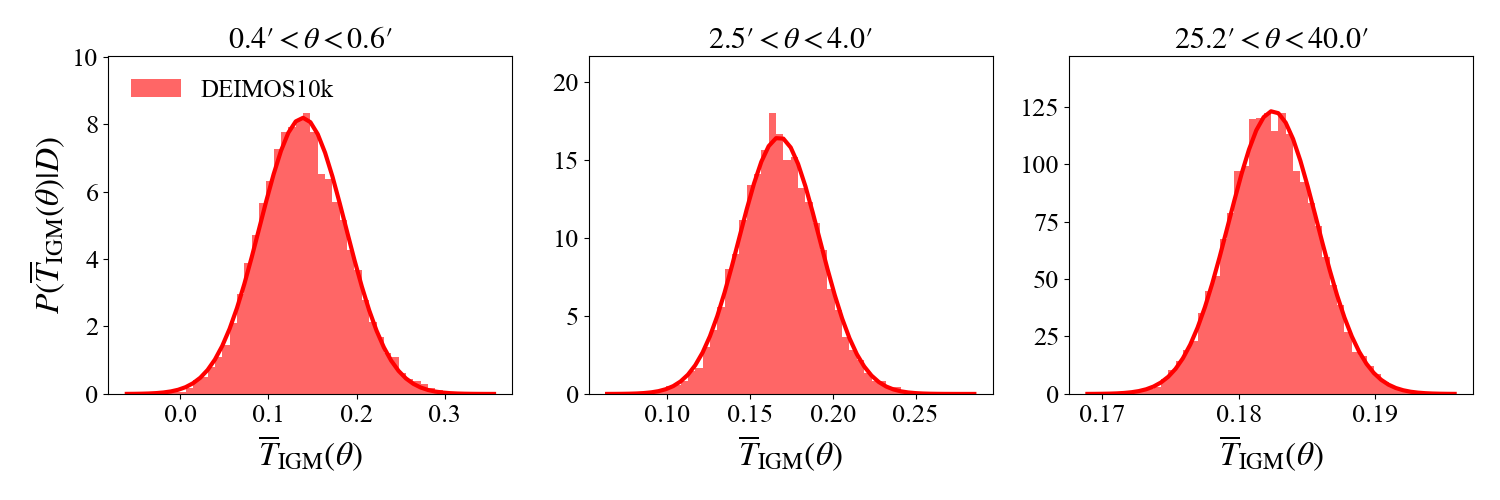}
  \vspace{-0.4cm}
  \caption{The full posterior of the mean Ly$\alpha$ forest transmission around $z\simeq4.9$ LAEs, $P(\overline{T}_{\rm IGM}(\theta)|\{f_{{\rm NB},i}^{\rm obs}, \boldsymbol{f}_{{\rm BB},i}^{\rm obs}\}_{i=1,\dots,N_{\rm bg}})$, using background $z\simeq5.7$ LAE (blue histogram) and DEIMOS10k (red histogram) samples. The histogram is computed via $\overline{T}_{\rm IGM}(\theta)$ from 10,000 random realizations of a set of $\TIGMi$ from the individual posteriors. The innermost (left), middle (middle), outermost (rightl) angular bins are shown. The blue and red solid curves are the Gaussian distributions with the expectation value and variance computed via equations (\ref{eq:expectation_value_xcorr}) and (\ref{eq:variance_xcorr}). The equivalence between the expectation value and the maximum {\it a posteriori} estimate of the mean Ly$\alpha$ forest transmission around LAEs is verified.}
  \label{fig:full_posterior_cross-correlation}
\end{figure}

\subsection{Estimating the mean Ly$\alpha$ forest transmission around LAEs} 

The NB718 dataset can be used to reveal $z\simeq4.9$ LAEs in the same redshift slice as the IGM Ly$\alpha$ forest transmission and thus how the transmission varies as a function of angular separation $\theta$ from the foreground LAEs. As before, we can use a Monte Carlo sampling to evaluate the angular mean Ly$\alpha$ forest transmission around LAEs, $\overline{T}_{\rm IGM}(\theta)=V(\theta)^{-1}\int \TIGM(\boldsymbol{x}) dV(\theta)$. By angular averaging the transmission for all pairs of foreground LAEs and background galaxy sightlines, we obtain
\begin{equation}
\overline{T}_{\rm IGM}(\theta)=\frac{1}{N_{\rm pair}(\theta)}\sum_{j=1}^{N_{\rm fg}}\sum_{i=1}^{N_{\rm bg}}\TIGMi \mathcal{I}(|\theta-\theta_{ij}|),
\end{equation}
where $N_{\rm pair}(\theta)=\sum_{j=1}^{N_{\rm fg}}\sum_{i=1}^{N_{\rm bg}}\mathcal{I}(|\theta-\theta_{ij}|)$ is the number of pairs in each angular bin and $\mathcal{I}(|\theta-\theta_{ij}|)$ is an indicator function equal to unity when the angular separation $\theta_{ij}=|\boldsymbol{\theta}_i-\boldsymbol{\theta}_j|$ between $i$ and $j$ is in the angular bin specified by $\theta$ with width $\Delta \theta$, i.e. $\mathcal{I}(|\theta-\theta_{ij}|)=1$ if  $|\theta-\theta_{ij}|<\Delta \theta$, and zero otherwise. The sums run over all foreground LAEs $j=1,\dots,N_{\rm fg}$ and all background galaxy sightlines $i=1,\dots,N_{\rm bg}$.

Again as each background galaxy sightline provides a noisy measurement of $\TIGMi$, analogous to the argument made for the mean Ly$\alpha$ forest transmission, the full posterior probablity of the mean Ly$\alpha$ forest transmission around LAEs, $P(\overline{T}_{\rm IGM}(\theta)|\{f_{{\rm NB},i}^{\rm obs}, \boldsymbol{f}_{{\rm BB},i}^{\rm obs}\}_{i=1,\dots,N_{\rm bg}})$,  can be expressed in terms of the posteriors of individual $\TIGMi$ measurements, which can be numerically computed by randomly sampling the individual posteriors. 
The expectation value of the angular mean Ly$\alpha$ forest transmission around LAEs is given by (see Appendix \ref{app:derivation})
\begin{align}
\langle\overline{T}_{\rm IGM}(\theta)\rangle=
\frac{1}{N_{\rm pair}(\theta)}\sum_{j=1}^{N_{\rm fg}}\sum_{i=1}^{N_{\rm bg}}\langle\TIGMi\rangle\mathcal{I}(|\theta-\theta_{ij}|).\label{eq:expectation_value_xcorr}
\end{align}
The variance of the estimated mean Ly$\alpha$ forest transmission around LAEs at each angular bin is given by
\begin{align}
{\rm Var}[\overline{T}_{\rm IGM}(\theta)]=\frac{1}{N_{\rm pair}(\theta)^2}\sum_{j=1}^{N_{\rm fg}}\sum_{i=1}^{N_{\rm bg}}{\rm Var}[\TIGMi]\mathcal{I}(|\theta-\theta_{ij}|).\label{eq:variance_xcorr}
\end{align}

Figure \ref{fig:full_posterior_cross-correlation} verifies that the full posterior $P(\overline{T}_{\rm IGM}(\theta)|\{f_{{\rm NB},i}^{\rm obs}, \boldsymbol{f}_{{\rm BB},i}^{\rm obs}\}_{i=1,\dots,N_{\rm bg}})$ follows a Gaussian distribution, which can be fully characterised by an expectation value (\ref{eq:expectation_value_xcorr}) and variance (\ref{eq:variance_xcorr}). The maximum {\it a posteriori} estimation,
\begin{equation}
  \overline{T}^{\rm MP}_{\rm IGM}(\theta)=\underset{\overline{T}_{\rm \scriptscriptstyle IGM}(\theta) }{\rm arg\,max}~P(\overline{T}_{\rm IGM}(\theta)|\{f_{{\rm NB},i}^{\rm obs}, \boldsymbol{f}_{{\rm BB},i}^{\rm obs}\}_{i=1,\dots,N_{\rm bg}}),
\end{equation}
is therefore equivalent to the expectation value $\langle\overline{T}_{\rm IGM}(\theta)\rangle$. The error estimated by the square root of the variance (\ref{eq:variance_xcorr}) in the mean Ly$\alpha$ forest transmission around LAEs scales as $\propto 1/\sqrt{N_{\rm pair}(\theta)}$ and includes the full uncertainties from photometric noise and continuum error in our individual $\TIGM$ measurements from the Bayesian SED fitting framework.

\subsection{The LAE-Ly$\alpha$ forest cross-correlation function}

While the ``mean Ly$\alpha$ forest transmission around LAEs''  is well defined given the observed distribution of LAEs, in order to examine the angular ``cross-correlation'', which is given by,
\begin{equation}
\omega_{\rm g\alpha}(\theta)=\langle\overline{T}_{\rm IGM}(\theta)\rangle/\langle\overline{T}_{\rm IGM}\rangle-1,
\end{equation}
we must quantify whether the excess probability of finding galaxies in the environment of high or low Ly$\alpha$ forest transmission is statistically significant. 
An additional uncertainty arises from the Poisson sampling of foreground galaxies in the survey footprint, including the effect of mask regions and the edge of the field-of-view.

To understand the scatter, we generate a random galaxy catalogue.  We populate $N_{\rm rand}$ objects at random locations $\{\boldsymbol{\theta}_i^{\rm rand}\}_{i=1,\dots,N_{\rm rand}}$ within the survey footprint excluding the actual masked regions. We then repeat the measurement of the mean Ly$\alpha$ forest transmission around the random objects using the actual background sightlines,
\begin{equation}
  \langle\overline{T}_{\rm IGM}^{\rm random}(\theta)\rangle=
  \frac{1}{N_{\rm pair}(\theta)}\sum_{j=1}^{N_{\rm rand}}\sum_{i=1}^{N_{\rm bg}}\langle\TIGMi\rangle\mathcal{I}(|\theta-\theta_{ij}^{\rm rand}|),
\end{equation}
where $\theta_{ij}^{\rm rand}$ is the angular separation between $i$-th random galaxy position and the observed location of $j$-th background galaxy sightline, $\theta_{ij}^{\rm rand}=|\boldsymbol{\theta}_i^{\rm rand}-\boldsymbol{\theta}_j|$.
This de-correlates the real cross-correlation signal between LAEs and Ly$\alpha$ forest transmission and should converge to the mean Ly$\alpha$ forest transmission, $\langle\overline{T}^{\rm random}_{\rm IGM}(\theta)\rangle\approx\langle\overline{T}_{\rm IGM}\rangle$. 

One can also de-correlate by randomly shuffling the observed values of $\TIGMi$ among the background galaxy sightlines but keeping their angular locations fixed,
\begin{equation}
\langle\overline{T}_{\rm IGM}^{\rm shuffle}(\theta)\rangle=
\frac{1}{N_{\rm pair}(\theta)}\sum_{j=1}^{N_{\rm fg}}\sum_{i=1}^{N_{\rm bg}}\langle\TIGMi^{\rm shuffle}\rangle\mathcal{I}(|\theta-\theta_{ij}|),
\end{equation}
where $\langle\TIGMi^{\rm shuffle}\rangle$ is a random draw from a set of real measurements $\{\langle T_{\rm IGM,1}\rangle,\langle T_{\rm IGM,2}\rangle,\dots,\langle T_{{\rm IGM},N_{\rm bg}}\rangle\}$ without replacement. The shuffled approach is convenient as it does not require us to model the galaxy selection function. 
This serves to verify that the observed LAE-Ly$\alpha$ forest cross-correlation is uncontaminated by the particular distribution of background galaxies on the sky. 
Both the shuffled and random measurements should be statistically identical, $\langle\overline{T}_{\rm IGM}^{\rm shuffle}(\theta)\rangle\approx\langle\overline{T}_{\rm IGM}^{\rm random}(\theta)\rangle$, and should converge to the mean Ly$\alpha$ forest transmission $\langle\overline{T}_{\rm IGM}\rangle$ within the statistical error.

\begin{figure}
  \centering
	\includegraphics[width=\columnwidth]{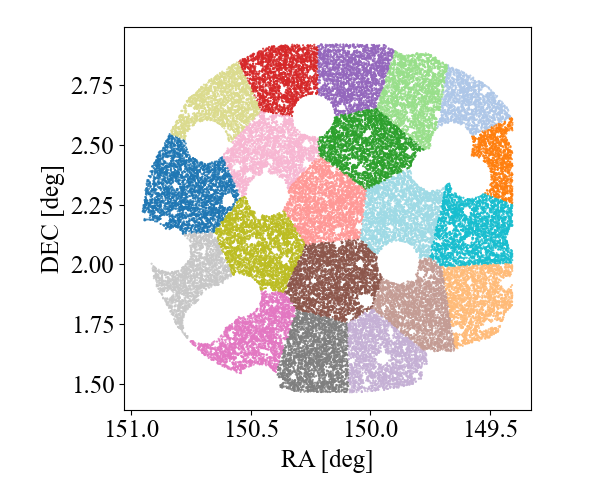}
  \vspace{-0.8cm}
  \caption{Angular regions (coloured patches) used to compute the Jackknife covariance matrix. Our survey footprint of the UD-COSMOS field are subdivided into $N_{\rm JK}=20$ regions.}
  \label{fig:JK}
\end{figure}
\begin{figure}
  \vspace{-0.4cm}
  \hspace{-0.5cm}
	\includegraphics[width=1.1\columnwidth]{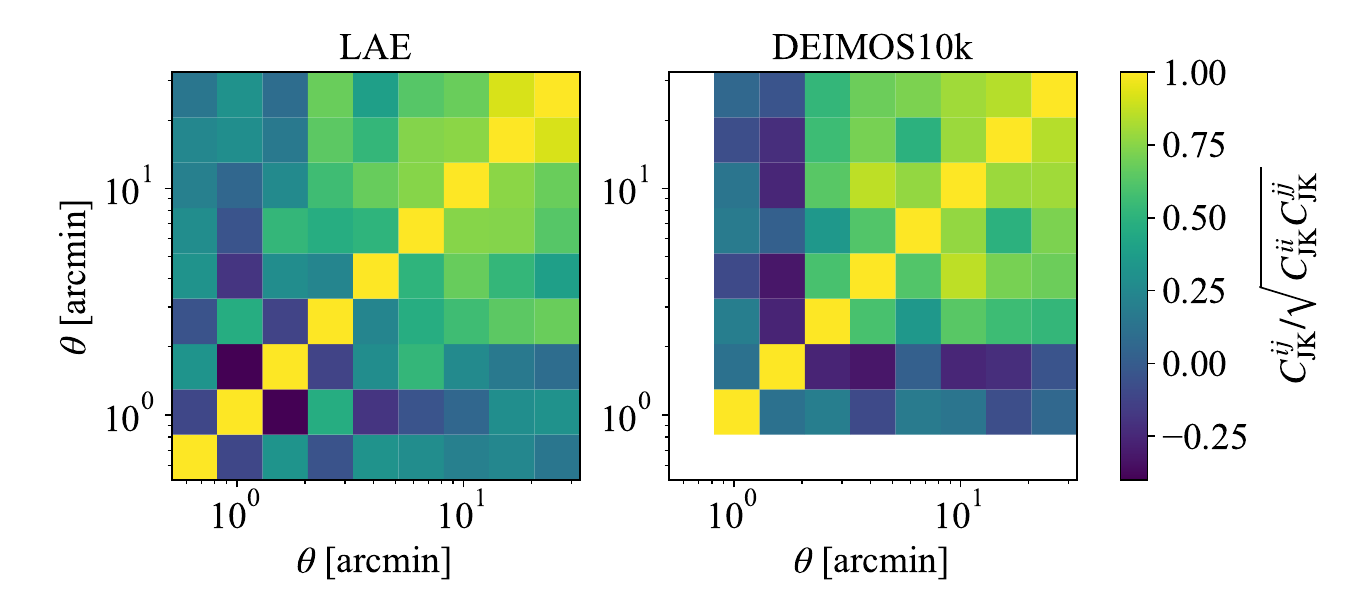}
  \vspace{-0.5cm}
  \caption{Correlation coefficients of the Jackknife covariance matrix for the LAE (left) and DEIMOS10k (right) samples. The covariance matrix at the innermost bin for the DEIMOS10k sample is not determined due to the small sample size in that bin. }
  \label{fig:Cov_JK}
\end{figure}

\begin{figure*}
  \hspace*{-0.75cm}
\includegraphics[width=1.1\columnwidth]{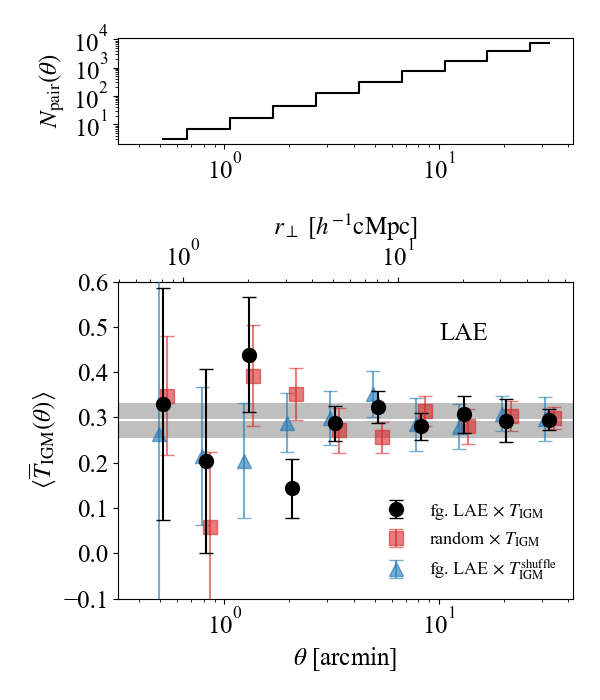}
  \hspace*{-0.5cm}
\includegraphics[width=1.1\columnwidth]{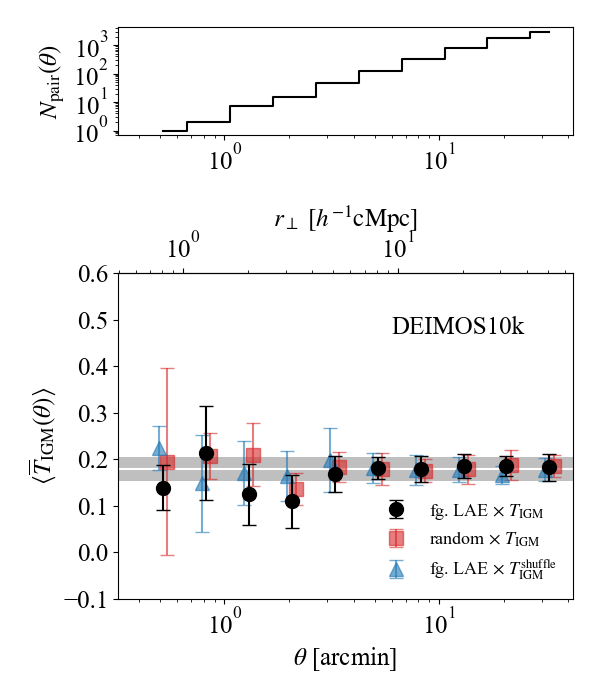}
\vspace{-0.5cm}
\caption{Mean Ly$\alpha$ forest transmission profiles around $z=4.9$ LAEs (black circles) using the $z=5.7$ LAE (left) and DEIMOS10k (right) samples as background sources. The same profiles around random foreground objects (red squares) and around foreground LAEs but using shuffled $\TIGM$ along the background sources (blue triangles) are offset by $\pm0.02$ dex offset along the x-axis for clarity. The gray shaded region indicates the mean Ly$\alpha$ forest transmission and its $1\sigma$ error. Top panels indicate the number of foreground LAE - background sightline pairs for the LAE and DEIMOS10k samples respectively.}
  \label{fig:LAE-TIGM}
\end{figure*}

We estimate the error on the cross-correlation using the Jackknife estimator \citep[e.g.][]{Norberg2009}. The Jackknife covariance matrix is given by
\begin{align}
&{\rm Cov}_{\rm JK}\left[\omega_{\rm g\alpha}(\theta),\omega_{\rm g\alpha}(\theta')\right]=\nonumber \\
&~~~~~~~~~\frac{N_{\rm JK}-1}{N_{\rm JK}}\sum_{k=1}^{N_{\rm JK}}\left[\omega^k_{\rm g\alpha}(\theta)-\overline{\omega}_{\rm g\alpha}^{\rm JK}(\theta)\right]\left[\omega_{\rm g\alpha}^k(\theta')-\overline{\omega}_{\rm g\alpha}^{\rm JK}(\theta')\right],
\end{align}
where
\begin{equation}
\overline{\omega}_{\rm g\alpha}^{\rm JK}(\theta)=\frac{1}{N_{\rm JK}}\sum_{k=1}^{N_{\rm JK}}\omega^k_{\rm g\alpha}(\theta),
\end{equation}
is the average of the cross-correlation functions from Jackknife resampling. Jackknife regions are obtained using the k-means clustering algorithm\footnote{\url{https://github.com/esheldon/kmeans_radec}} \citep{Kwan2017} on a random galaxy catalogue with $N_{\rm rand}=50,000$. This algorithm subdivides the observed survey area into $N_{\rm JK}$ regions of a roughly equal area as shown in Figure \ref{fig:JK}. To compute the Jackknife covariance, we omit foreground LAEs and $\TIGM$ along background galaxy sightlines located in each Jackknife region at a time and compute $N_{\rm JK}$ Jackknife re-sampled cross-correlation functions $\omega^k_{\rm g\alpha}(\theta)$, $k=1,\dots,N_{\rm JK}$, using the remaining objects. We use the same procedure to compute the Jackknife error for the other summary statistics (the mean Ly$\alpha$ forest transmission and Ly$\alpha$ forest auto-correlation function) in this paper.

The correlation coefficients of the Jackknife covariance matrix are shown in Figure \ref{fig:Cov_JK}. The covariance matrix of the innermost bins of the DEIMOS10k sample could not be determined due to the small sample size. For both LAE and DEIMOS10k samples, there are significant off-diagonal correlations between angular bins as the same sightlines contribute multiple foreground LAE-background sightline pairs.

\subsection{Result}

Figure \ref{fig:LAE-TIGM} shows the angular mean Ly$\alpha$ forest transmission around LAEs at $z\simeq4.9$. We find no excess transmission or absorption in the NB-integrated Ly$\alpha$ forest around the LAEs. The result is consistent with the global mean $\langle\overline{T}_{\rm IGM}\rangle$ within the $2\sigma$ error both for the LAE and DEIMOS10k samples. We compare our result with the random and shuffled measurements using the same number of foreground LAEs and background galaxies in the real data. Both measurements show similar fluctuations with the observed values, confirming that our result is consistent with no spatial correlation. 

\subsubsection{Correcting for contamination by low-redshift interlopers}

Figure~\ref{fig:LAE-TIGM} shows a mean offset between $\langle\overline{T}_{\rm IGM}(\theta)\rangle$ measured using the LAE and DEIMOS10k samples. As discussed in Section~\ref{sec:systematics}, this is likely caused by the low-redshift interlopers in both the foreground and background LAE samples \citep{Grasshorn-Gebhardt2019,Farrow2021}. Since the distribution of any low-redshift interlopers would be random relative to structures in the tomographic slice of interest, the interlopers will dilute the observed cross-correlation. Assuming foreground and background contamination fractions $f_{\rm fg.int}$ and $f_{\rm bg.int}$, the observed angular mean Ly$\alpha$ forest transmission around the foreground LAEs can be expressed as (see Appendix \ref{app:interloper})
\begin{align}
\langle\overline{T}_{\rm IGM}(\theta)\rangle=&\,(1-f_{\rm fg.int})(1-f_{\rm bg.int})\langle\overline{T}_{\rm IGM}(\theta)\rangle^{\rm true} \nonumber \\
&+f_{\rm fg.int}(1-f_{\rm bg.int})\langle\overline{T}_{\rm IGM}\rangle^{\rm true}
+f_{\rm bg.int}\langle\overline{T}_{\rm IGM}\rangle^{\rm bg.int},
\end{align}
where $\langle\overline{T}_{\rm IGM}\rangle^{\rm true}$ is the true mean Ly$\alpha$ forest transmission and $\langle\overline{T}_{\rm IGM}\rangle^{\rm bg.int}$ is the fictitious mean Ly$\alpha$ forest transmission measured along the low-redshift interlopers in the background LAE sample. The second and third terms indicate contaminations from the low-redshifts interlopers in the foreground and background LAE samples. 

Following our definition of the observed LAE-Ly$\alpha$ forest cross-correlation $\omega_{\rm g\alpha}(\theta)=\langle\overline{T}_{\rm IGM}(\theta)\rangle/\langle\overline{T}_{\rm IGM}\rangle-1$, we can similarly find that low-redshift interlopers dilute the cross-correlation amplitude by
\begin{align}
\omega_{\rm g\alpha}(\theta)=\frac{(1-f_{\rm fg.int})(1-f_{\rm bg.int})\langle \overline{T}_{\rm IGM}\rangle^{\rm true}}{(1-f_{\rm bg.int})\langle \overline{T}_{\rm IGM}\rangle^{\rm true}+f_{\rm bg.int}\langle \overline{T}_{\rm IGM}\rangle^{\rm bg.int}}\omega^{\rm true}_{\rm g\alpha}(\theta).\label{eq:xcorr_interloper}
\end{align}

The offset in the mean Ly$\alpha$ forest transmission around LAEs between the background LAE and DEIMOS10k samples can be explained by this effect. 
As in Section \ref{sec:systematics}, we set $f_{\rm fg.int}=f_{\rm bg.int}=0.2$ both for foreground and background LAE samples. The contamination fraction for the DEIMOS10k sample is $f_{\rm bg.int}=0.0$ as all are confirmed spectroscopically. We assume that the fictitious mean Ly$\alpha$ forest transmission along the interlopers in the background LAE sample is $\langle\overline{T}_{\rm IGM}\rangle^{\rm bg.int}=0.7$. As before, LAE interlopers can explain the offset between the background LAE and DEIMOS10k samples. 

These interlopers depress the observed LAE-Ly$\alpha$ forest cross-correlation by $\omega_{\rm g\alpha}(\theta)\approx0.40\,\omega^{\rm true}_{\rm g\alpha}(\theta)$ and $\omega_{\rm g\alpha}(\theta)\approx0.80\,\omega^{\rm true}_{\rm g\alpha}(\theta)$ for background LAE and DEIMOS10k samples, respectively. The cross-correlation measurement using the DEIMOS10k sample is also affected because of the interloper contamination in the foreground LAEs. 

All the terms in the damping pre-factor in Equation \ref{eq:xcorr_interloper} can be determined and statistically corrected {\it a posteriori} by spectroscopic follow up of a random subset of the foreground and background LAE samples as described in Section \ref{sec:systematics}.

\subsubsection{Limit on the IGM fluctuations around LAEs}

\begin{figure}
\hspace*{-0.75cm}
\includegraphics[width=1.1\columnwidth]{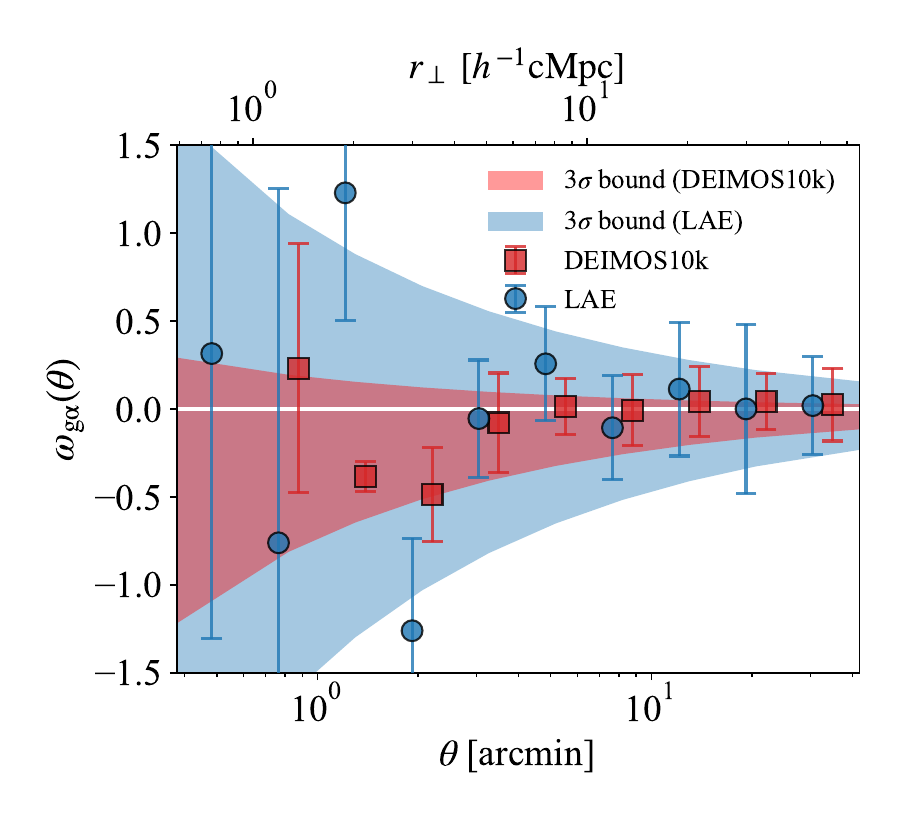}
\vspace{-1.0cm}
\caption{Observed LAE-Ly$\alpha$ forest cross-correlation function for the DEIMOS10k (red squares) and LAE (blue circles) samples after correcting for likely interloper contamination. The derived $3\sigma$ lower and upper limits of the cross-correlation assuming the power-law with slope $\gamma=0.5$ are shown with the red and blue shaded regions for the DEIMOS10k and LAE sample, respectively. The Jackknife covariance matrices scaled by the interloper correction factors are used to estimate the error.}
  \label{fig:cross-correlation}
\end{figure}

Figure \ref{fig:cross-correlation} shows the observed LAE-Ly$\alpha$ forest cross-correlation at $z\simeq4.9$ after correcting for possible low-redshift interloper contamination. To place an empirical constraint, we assume a simple power-law form, 
\begin{equation}
  \omega_{\rm g\alpha}^{\rm model}(\theta)=A_0(\theta/\theta_0)^{-\gamma},
\end{equation}
where the fluctuations are characterised by the amplitude $A_0$ at angular distance $\theta_0$ and the power-law slope $\gamma$. We assume a Gaussian likelihood with the measured Jackknife covariance matrix  and a fixed slope of $\gamma=0.5$. 
The resulting $3\sigma$ bounds 
are shown in Figure \ref{fig:cross-correlation}. The $3\sigma$ lower and upper limits are
\begin{equation}
  -0.29\left(\frac{r_\perp}{10\,h^{-1}\rm cMpc}\right)^{-0.5}<\omega_{\rm g\alpha}^{\rm model}<0.07\left(\frac{r_\perp}{10\,h^{-1}\rm cMpc}\right)^{-0.5}
  \end{equation}
for the DEIMOS10k sample, and  
\begin{equation}
-0.58\left(\frac{r_\perp}{10\,h^{-1}\rm cMpc}\right)^{-0.5}<\omega_{\rm g\alpha}^{\rm model}<0.40\left(\frac{r_\perp}{10\,h^{-1}\rm cMpc}\right)^{-0.5}
\end{equation}
for the LAE sample. The derived lower and upper limits are consistent for both the LAE and DEIMOS10k samples. While the bound from the DEIMOS10k sample is slightly shifted to the negative cross-correlation, this is likely due to an underestimated Jackknife error at small angular bins. The directly propagated error (equation \ref{eq:variance_xcorr}) from the Bayesian SED fitting framework indicate the error from photometric noise and UV continuum uncertainty in the DEIMOS10k sample at the inner bins should be larger than the empirical estimate from the Jackknife method. 

Our result indicates that the angular fluctuations of the Ly$\alpha$ forest transmission around LAEs should be $\lesssim 58\,\%$ at $10\,h^{-1}\rm Mpc$ relative to the global mean at $z\simeq4.9$. The physical interpretation of the result will be discussed in a companion paper (Kakiichi et al in prep).

\section{Ly$\alpha$ forest auto-correlation}\label{sec:auto}

\subsection{Estimating the Ly$\alpha$ forest auto-correlation}

\begin{figure*}
    \hspace*{-0.75cm}
	\includegraphics[width=1.1\columnwidth]{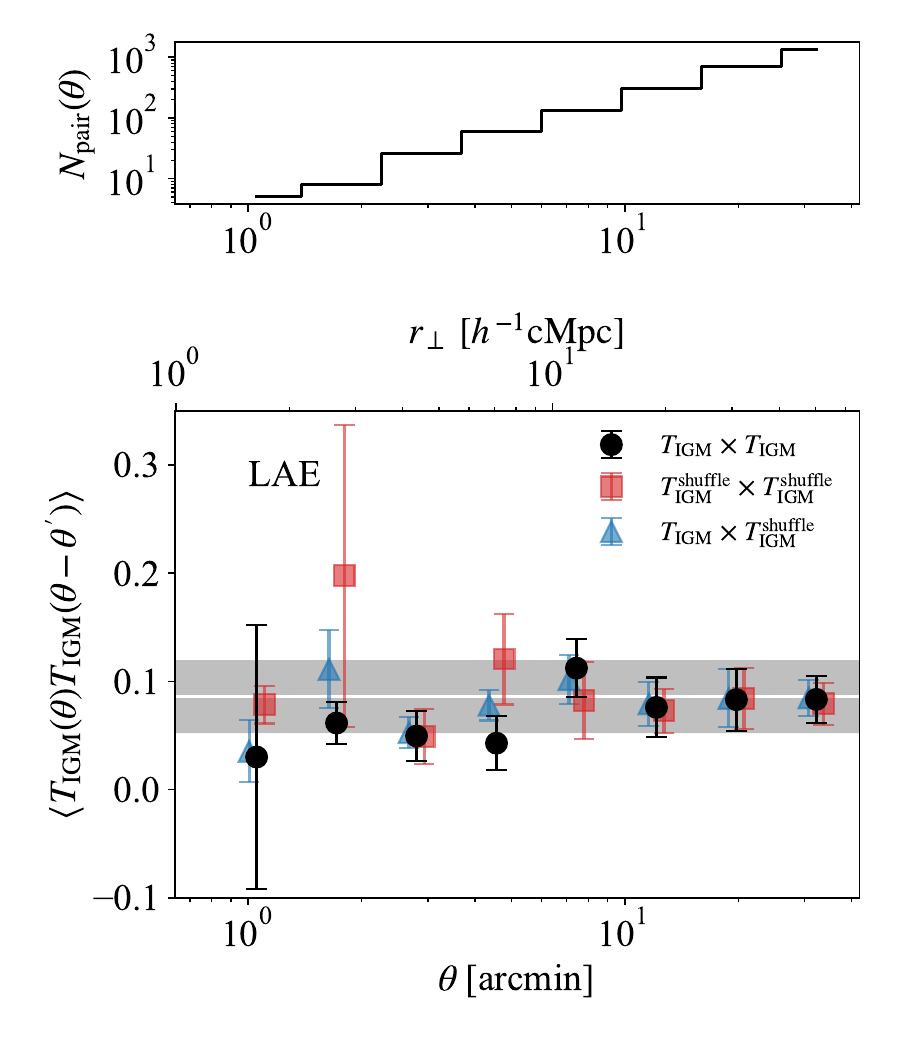}
    \hspace*{-0.5cm}
	\includegraphics[width=1.1\columnwidth]{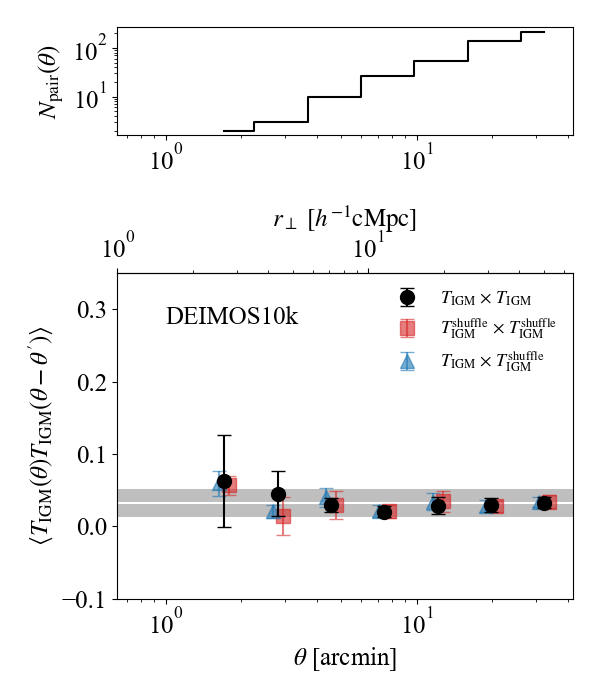}
	\vspace{-0.6cm}
	\caption{Ly$\alpha$ forest transmission auto-correlation function at $z=4.9$ using the $z=5.7$ LAE (left) and DEIMOS10k (right) samples. Values computed using only shuffled background sources (red squares) and the cross-correlation between data and shuffled samples (blue triangles) are shown offset $\pm0.02$ dex along the x-axis for clarity. The gray shaded region indicates the estimate in the case of no correlation based on the mean Ly$\alpha$ forest transmission and its $1\sigma$ error. Top panels indicate the number of sightline pairs for the LAE and DEIMOS10k samples.}
    \label{fig:TIGM_auto}
\end{figure*}

We now turn our attention to examine the spatial fluctuations of Ly$\alpha$ forest transmission. These are expected to spatially correlate across different sightlines due to large-scale fluctuations of the IGM. Unlike the measurement of the 3D Ly$\alpha$ forest auto-correlation from spectra \citep[e.g.][]{Slosar2011}, photometric IGM tomography measures the angular auto-correlation of Ly$\alpha$ forest transmission integrated over the line-of-sight width of the NB filter ($\simeq34\,h^{-1}\rm cMpc$). 

In order to estimate the Ly$\alpha$ forest angular auto-correlation function, using the pairs of NB-integrated Ly$\alpha$ forest transmission measurements, we first compute, at each angular bin, 
\begin{equation}
T_{\rm IGM}T_{\rm IGM}(\theta)=\frac{1}{N_{\rm pair}(\theta)}\sum_{i=1}^{N_{\rm bg}}\sum_{j>i}^{N_{\rm bg}}\TIGMi T_{{\rm IGM},j}\mathcal{I}(|\theta-\theta_{ij}|),
\end{equation}
where $N_{\rm pair}(\theta)=\sum_{i=1}^{N_{\rm bg}}\sum_{j>i}^{N_{\rm bg}}\mathcal{I}(|\theta-\theta_{ij}|)$ is the number of pairs in each angular bin.
For independent measurements of $\TIGMi$, the expectation value of the Ly$\alpha$ forest angular auto-correlation is given by
\begin{equation}
\langle\TIGM\TIGM(\theta)\rangle=\frac{1}{N_{\rm pair}(\theta)}\sum_{i=1}^{N_{\rm bg}}\sum_{j>i}^{N_{\rm bg}}\langle\TIGMi\rangle\langle T_{{\rm IGM},j}\rangle \mathcal{I}(|\theta-\theta_{ij}|).
\end{equation}
The error is computed from the Jackknife covariance matrix. The Ly$\alpha$ forest auto-correlation function is estimated by 
\begin{equation}
  \omega_{\rm\alpha\alpha}(\theta)=\langle\TIGM\TIGM(\theta)\rangle/\langle\overline{T}_{\rm IGM}\rangle^2-1.
\end{equation}

We can understand the scatter of $\langle\TIGM\TIGM(\theta)\rangle$ in the absence of any spatial correlation. As the Ly$\alpha$ forest auto-correlation has no complication from the window function or the survey geometry, $\langle\TIGM\TIGM(\theta)\rangle$ should be equal to $\langle\overline{T}_{\rm IGM}\rangle^2$ if there is no spatial correlation. To test this, we de-correlate the observed correlation by shuffling either one or both of the measured values of $\TIGM$ between the observed locations, i.e.
\begin{align}
&\langle\TIGM\TIGM^{\rm shuffle}(\theta)\rangle= \nonumber \\
&~~~~~~~~~~~
\frac{1}{N_{\rm pair}(\theta)}\sum_{i=1}^{N_{\rm bg}}\sum_{j>i}^{N_{\rm bg}}\langle\TIGMi\rangle\langle T_{{\rm IGM},j}^{\rm shuffle}(\theta)\rangle \mathcal{I}(|\theta-\theta_{ij}|),
\end{align}
or
\begin{align}
&\langle\TIGM^{\rm shuffle}(\theta)\TIGM^{\rm shuffle}(\theta)\rangle=\nonumber \\
&~~~~~~~~~~~
\frac{1}{N_{\rm pair}(\theta)}\sum_{i=1}^{N_{\rm bg}}\sum_{j>i}^{N_{\rm bg}}\langle\TIGMi^{\rm shuffle}(\theta)\rangle\langle T_{{\rm IGM},j}^{\rm shuffle}(\theta)\rangle \mathcal{I}(|\theta-\theta_{ij}|).
\end{align}
Using the real set of $\{T_{{\rm IGM},i}\}_{i=1,\dots,N_{\rm bg}}$, we generated a randomized set of $\TIGM$ values keeping the angular positions of the sightlines the same. This artificially de-correlates the possible correlation. If there is no systematic, this should approach $\simeq\langle\overline{T}_{\rm IGM}\rangle^2$.

\subsection{Result}

Figure \ref{fig:TIGM_auto} shows the observed auto-correlation function of the Ly$\alpha$ forest transmission at $z\simeq4.9$. 
The observed auto-correlation is consistent with the square of the mean and the shuffled results within $2\sigma$ error, indicating the observed signal is consistent with no auto-correlation. This null detection can be interpreted as the observed limit on the Ly$\alpha$ forest transmission fluctuations at $z\simeq4.9$. 

\subsubsection{Correcting for the contamination by low-redshift interlopers}

Similar to the angular mean Ly$\alpha$ forest transmission around LAEs, Figure \ref{fig:TIGM_auto} shows an offset between $\langle\TIGM\TIGM(\theta)\rangle$ measured using the background LAE and DEIMOS10k samples, which is likely caused by the low-redshift interlopers. The effect of the lower-redshift interlopers in $\langle\TIGM\TIGM(\theta)\rangle$ can be expressed as (see Appendix \ref{app:interloper})
\begin{align}
  &\langle\TIGM\TIGM(\theta)\rangle=(1-f_{\rm bg.int})^2\langle\TIGM\TIGM(\theta)\rangle^{\rm true}+\nonumber\\
  &2(1-f_{\rm bg.int})f_{\rm bg.int}\langle\TIGM\rangle^{\rm true}\langle\TIGM\rangle^{\rm bg.int}+(f_{\rm bg.int}\langle\TIGM\rangle^{\rm bg.int})^2.
\end{align}
The second and third terms indicate contaminations from cross-correlation between low-redshift interlopers and true background galaxies and the auto-correlation of low-redshift interlopers, assuming there is no spatial correlation. In terms of the Ly$\alpha$ forest angular auto-correlation function, the true auto-correlation function $\omega^{\rm true}_{\rm \alpha\alpha}(\theta)=\langle\TIGM\TIGM(\theta)\rangle^{\rm true}/(\langle\overline{T}_{\rm IGM}\rangle^{\rm true})^2-1$ is diluted by the interlopers such that
\begin{align}
\omega_{\rm \alpha\alpha}(\theta)=\left[\frac{(1-f_{\rm bg.int})\langle \overline{T}_{\rm IGM}\rangle^{\rm true}}{(1-f_{\rm bg.int})\langle \overline{T}_{\rm IGM}\rangle^{\rm true}+f_{\rm bg.int}\langle \overline{T}_{\rm IGM}\rangle^{\rm bg.int}}\right]^2\omega^{\rm true}_{\rm \alpha\alpha}(\theta).
\end{align}
Again, all factors can be determined {\it a posteriori} using spectroscopic follow-up of the background galaxy sample. Assuming $f_{\rm bg.int}=0.20$ and $\langle\overline{T}_{\rm IGM}\rangle^{\rm bg.int}=0.7$ for the background LAE sample and taking $\langle\overline{T}_{\rm IGM}\rangle^{\rm true}$ to be the value from the DEIMOS10k sample can explain the observed offset in $\langle\TIGM\TIGM(\theta)\rangle$. This corresponds to the damping of $\omega_{\alpha\alpha}(\theta)=0.39\,\omega_{\alpha\alpha}^{\rm true}(\theta)$ for the observed Ly$\alpha$ forest angular auto-correlation function from the LAE sample. There is no damping factor for the DEIMOS10k sample as the interloper contamination for the spectroscopically confirmed sample is zero. 

\section{IGM tomographic map}\label{sec:map}

\begin{figure*}
  \centering
    \includegraphics[width=\textwidth]{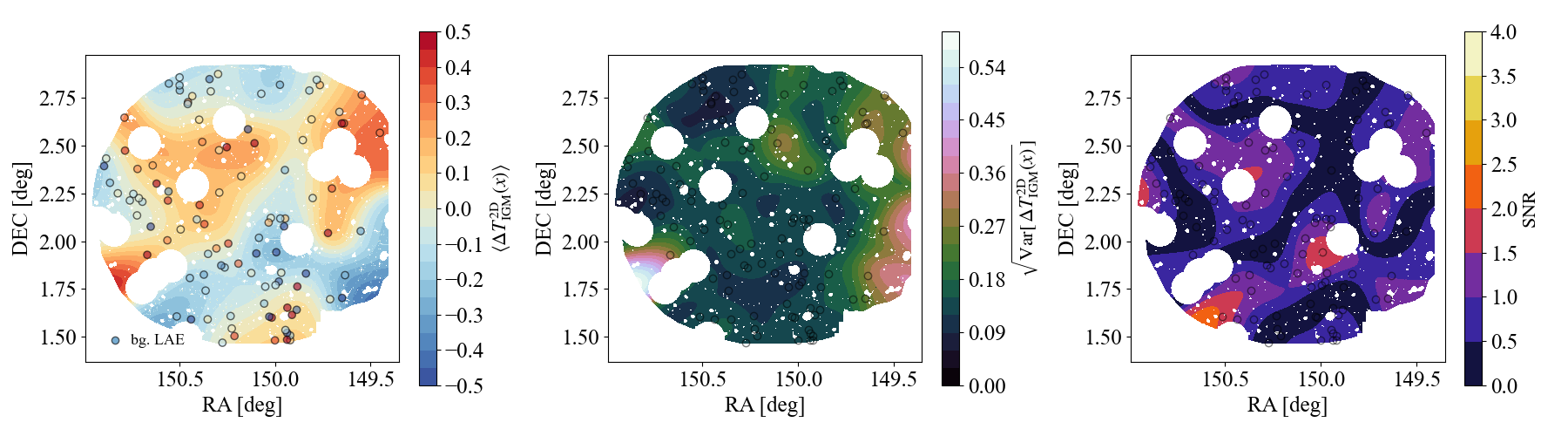}
      \caption{({\bf Left}): Reconstructed 2D tomographic map of the Ly$\alpha$ forest transmission of the IGM at $z\simeq4.9$ (colour map). The locations of the background ($z\simeq5.7$) LAEs are shown with coloured circles with the colours indicating the measured difference between the Ly$\alpha$ forest transmission along the sightline and the global mean. ({\bf Middle}): The standard deviation of the reconstructed map including the uncertainties from the photometric and continuum errors in the Bayesian SED fitting. The locations of the background LAE sightlines are shown with open circles. ({\bf Right}): The signal-to-noise ratio of the reconstructed map. The masked regions are left blank.}
      \label{fig:tomograhic_map}
\end{figure*}

\subsection{Reconstruction method}
Finally, we present a reconstructed tomographic map of the IGM. This is arguably the most unique aspect of photometric IGM tomography, since it enables us to directly visualise the large-scale structures of the IGM and galaxies in the same cosmic volume. To accomplish this we use the Nadaraya-Watson estimator for the 2D tomographic map of the IGM Ly$\alpha$ forest transmission fluctuations \citep{Kakiichi2022},
\begin{equation}
\Delta\TIGM^{\rm 2D}(\bm{\theta})=\frac{\sum^{N_{\rm bg}}_{\rm i=1}K_R(\bm{\theta}-\bm{\theta}_i)(\TIGMi-\langle\overline{T}_{\rm IGM}\rangle)}{\sum^{N_{\rm bg}}_{\rm i=1} K_R(\bm{\theta}-\bm{\theta}_i)},
\end{equation}
where $K_R(\bm{\theta})=(2\pi R^2)^{-1/2}\exp[-\bm{\theta}^2/(2R^2)]$ is a Gaussian kernel with a smoothing length $R$. In practice, we create a 2D map on pixelised map of $4096\times4096$ pixels. Following a similar argument as in previous sections, given the posterior of Ly$\alpha$ forest transmission,
the expectation value of the 2D tomographic map is given by
\begin{equation}
\langle\Delta\TIGM^{\rm 2D}(\bm{\theta})\rangle=
\frac{
  \sum_{i=1}^{N_{\rm bg}}
  \displaystyle
  K_R(\bm{\theta}-\bm{\theta}_i)(\langle \TIGMi\rangle-\langle\overline{T}_{\rm IGM}\rangle)
}
{
  \sum^{N_{\rm bg}}_{\rm i=1} K_R(\bm{\theta}-\bm{\theta}_i)
}.
\end{equation}
At each point $\bm{\theta}$, the estimator is simply the weighted sum of (independent) individual $\TIGMi$ measurements. Thus the variance can be computed as
\begin{align}
{\rm Var}[\Delta\TIGM^{\rm 2D}(\bm{\theta})]= \frac{
\sum_{i=1}^{N_{\rm bg}}\displaystyle
K_R(\bm{\theta}-\bm{\theta}_i)^2{\rm Var}\left[\TIGMi\right]
}
{
\left[\sum^{N_{\rm bg}}_{\rm i=1} K_R(\bm{\theta}-\bm{\theta}_i)\right]^2
}.
\end{align}
This includes errors from photometric noise and the UV continuum uncertainty.
We define the SNR map as the ratio between the observed Ly$\alpha$ forest transmission fluctuations and the standard deviation,
\begin{equation}
{\rm SNR}(\bm{\theta})=\frac{\left|\langle\Delta T^{\rm 2D}_{\rm IGM}(\bm{\theta})\rangle\right|}{\sqrt{{\rm Var}[\Delta T^{\rm 2D}_{\rm IGM}(\bm{\theta})]}}.
\end{equation}

\begin{figure*}
  \centering
    \includegraphics[width=\textwidth]{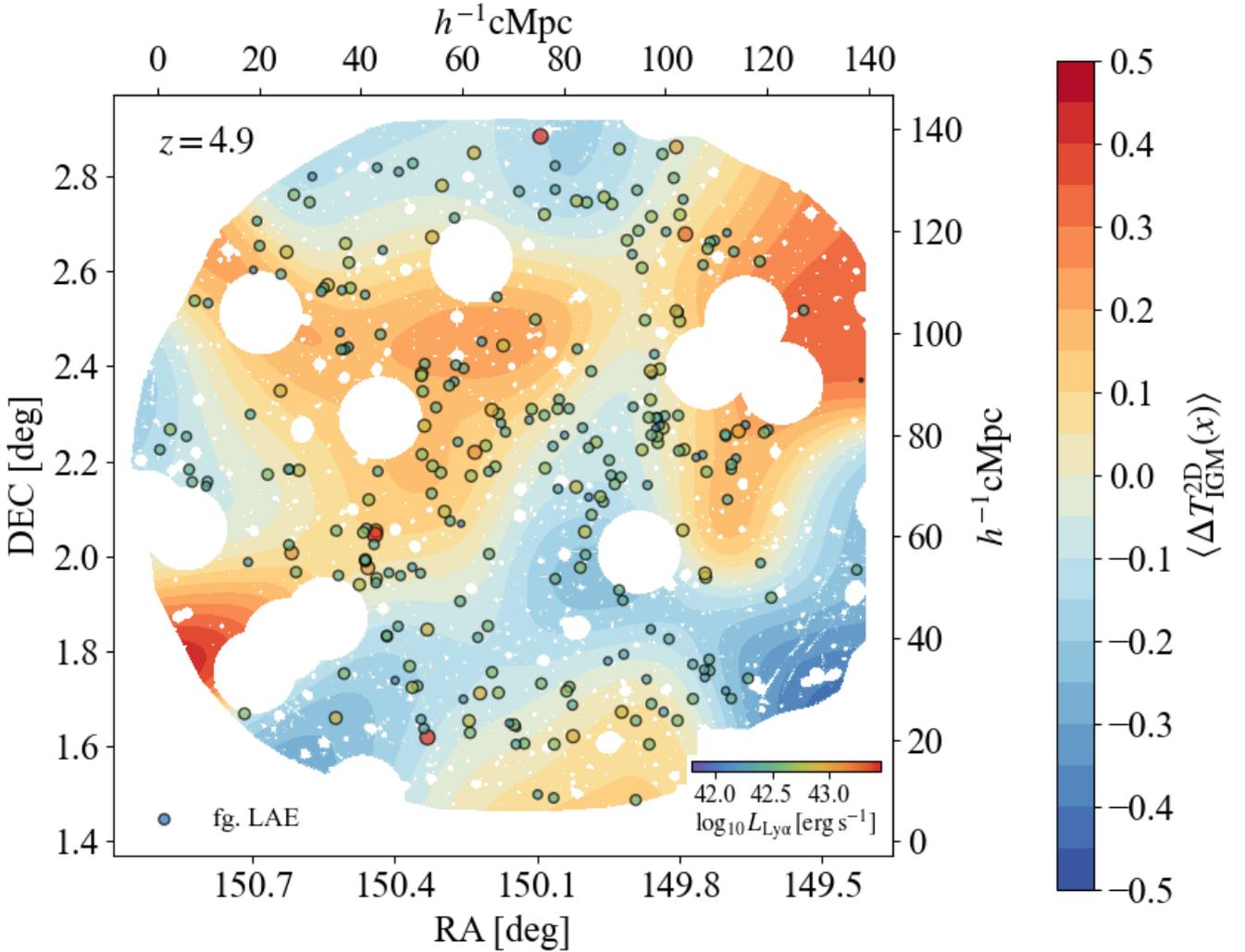}
    \vspace{-1cm}
      \caption{Reconstructed 2D tomographic map of the Ly$\alpha$ forest transmission of the IGM (coloured map) overlaid with the distribution of the $z=4.9$ LAEs (coloured circles). The angular resolution of the reconstructed map is $0.12\rm\,deg$ corresponding to $11\,h^{-1}\rm cMpc$. The colour of each circle indicates the Ly$\alpha$ luminosity of the LAE. Masked regions are left blank. Although the map is dominated by photometric noise, it represents the first IGM tomographic map co-spatial to the large-scale structure of LAEs in the same cosmic volume close to the end of cosmic reionization. }
      \label{fig:tomograhic_map_with_LAEs}
\end{figure*}

\begin{figure*}
  \centering
    \includegraphics[width=\textwidth]{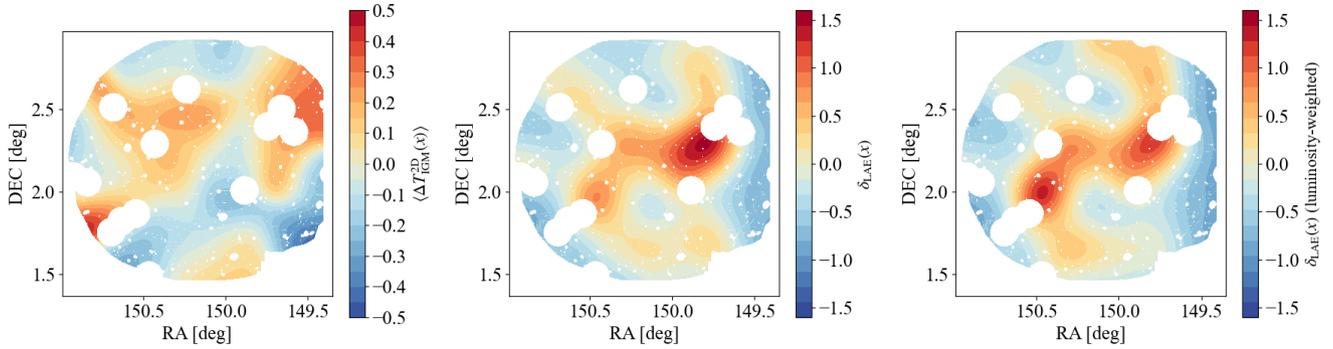}
    \vspace{-0.5cm}
    \caption{Fluctuations of the reconstructed 2D Ly$\alpha$ forest tomographic map (left), LAE number density field centre), and Ly$\alpha$ luminosity-weighted LAE number density field (right) with the angular resolution of $0.12\rm\,deg$ ($11\,h^{-1}\rm cMpc$) at $z\simeq4.9$.}
    \label{fig:map_comparison}
  \end{figure*}

\subsection{Result}

In Figure \ref{fig:tomograhic_map} we show the reconstructed 2D tomographic map of the Ly$\alpha$ forest transmission at $z\simeq4.9$. We only apply the map reconstruction to the background LAE sample because their distribution spans the entire field of view. The smoothing length is chosen as the mean inter-sightline separation, $R=1/\sqrt{\Sigma_{\rm LAE}}\simeq0.118\rm\,deg\,(7.1\arcmin)$ where $\Sigma_{\rm LAE}$ is the surface number density of the background LAEs. We can visually see large-scale fluctuations of the Ly$\alpha$ forest transmission with a median contrast of $\left|\langle\Delta T^{\rm 2D}_{\rm IGM}(\bm{x})\rangle\right|\sim0.1$. The typical standard deviation in the reconstructed map is $\sqrt{{\rm Var}[\Delta T^{\rm 2D}_{\rm IGM}(\bm{x})]}\sim0.14$.
We find the mean SNR of the reconstructed map as $\langle{\rm SNR}(\bm{x})\rangle=0.71$, indicating that our tomographic map is still noisy with contributions from photometric errors and the UV continuum uncertainty.

There are several tentative regions of transmissive and opaque transmission in the map located at ${\rm (RA,DEC)}\simeq(150.2^\circ,2.45^\circ)$ and ${\rm (RA,DEC)}\simeq(150.0^\circ,1.95^\circ)$ respectively, with the peak ${\rm SNR}\simeq1.5-2.0$. The Ly$\alpha$ forest transmission in these regions is reasonably coherent. We require deeper NB imaging data to secure the statistical significance. If confirmed, these opaque and transmissive regions of the IGM may represent a protocluster and a highly ionized region of the IGM by the enhanced UV background. 

Although there is a higher transmissive region towards the edge at ${\rm (RA,DEC)}\simeq(150.8^\circ,1.80^\circ)$, as there is no background galaxy here this is likely an artefact from the map reconstruction method. At the edge of the field of view, the reconstruction method is more affected by boundary effects. Although our estimator corrects for boundary effects by incorporating the sightline density, the estimator is more sensitive to the $\TIGM$ values of individual background galaxies, whereas at the centre of the field smoothing corrects outlier values of $\TIGM$.

Figure \ref{fig:tomograhic_map_with_LAEs} overlays the distribution of the $z=4.9$ LAEs on the reconstructed tomographic map of the Ly$\alpha$ forest transmission. This represents the highest redshift 2D tomographic map of the IGM with the galaxy distribution at the present time and demonstrates the potential of the NB tomographic technique to examine the galaxy-IGM connection closer to the reionization epoch.

We briefly examine the spatial correlation between the $z=4.9$ LAE distribution and the reconstructed Ly$\alpha$ forest transmission map. In order to apply the spatial correlation analysis at the map level, we first reconstruct LAE density map at the same smoothing scale using the Gaussian kernel density estimator with the boundary correction,
\begin{equation}
n_{\rm LAE}(\bm{\theta})=\frac{\sum^{N_{\rm bg}}_{\rm i=1} w_i K_R(\bm{\theta}-\bm{\theta}_i)}{\int m(\bm{\theta}')K_R(\bm{\theta}-\bm{\theta}')d\bm{\theta}'},
\end{equation}
where $m(\bm{\theta}')$ represents the mask and the denominator is the correction factor $C^{-1}=\int m(\bm{\theta}')K_R(\bm{\theta}-\bm{\theta}')d\bm{\theta}'$ for boundary effects including masked regions around bright stars. We introduce weights $w_i$, where $w_i=1$ for the ordinary LAE density field and $w_i=L_{\alpha,i}$ for the Ly$\alpha$ luminosity-weighted LAE density field. The galaxy density fluctuation map is then $\delta_{\rm LAE}(\bm{x})=n_{\rm LAE}/\bar{n}_{\rm LAE}-1$ where the mean density $\bar{n}_{\rm LAE}$ is computed by excluding the masked regions.

In Figure \ref{fig:map_comparison} we show the comparison of the 2D tomographic map with the LAE density and Ly$\alpha$ luminosity-weighted LAE density fields at the same smoothing length. Using all unmasked regions, the Pearson correlation coefficient $\rho$ indicates that there is a negligible correlation between the reconstructed 2D tomographic map and the (luminosity-weighted) LAE density map with $\rho=0.09$ ($0.06$). Within the precision of existing photometric data, it appears that $z\simeq4.9$ LAEs do not occupy extreme Ly$\alpha$ transmissive or opaque regions of the IGM. This is consistent with the two-point cross-correlation analysis between LAE and Ly$\alpha$ forest transmission. 

To avoid confusion from boundary effects, we have also limited the map-level analysis within the region with a low correction factor $C<2.0$. We then find that the Pearson correlation coefficient becomes $\rho=0.22$ for the LAE density field ($0.20$ for luminosity-weighted field), indicating a possible weak positive correlation between the LAE number density and the Ly$\alpha$ forest transmission map of the IGM. Although this weak correlation is intriguing, as noted above, our average SNR of the map is still low to conclude. 

\subsection{Systematics}
Unlike the cross- and auto-correlation functions between LAEs and Ly$\alpha$ forest transmission, the reconstruction of the 2D tomographic map demands a higher purity of the background galaxy sample. A fictitious Ly$\alpha$ forest transmission due to a low-redshift interloper would produce a fake transmissive IGM region. While having many background galaxies within a smoothing length of the reconstruction reduces the effect of interlopers, we have not found a way to statistically correct for this effect as is possible statistically using a spectroscopic subset in the case of the correlation functions. It is difficult to quantify the level of contamination in each transmissive or opaque region of the IGM identified with photometric IGM tomography without directly confirming the redshifts of all background galaxies spectroscopically. 
 
\section{Discussion}\label{sec:discussions}

\subsection{Improving photometric IGM tomography}

\subsubsection{Extremely-deep NB imaging and spectroscopic campaign}
As discussed in Section \ref{sec:individual_TIGM}, our estimate of $T_{\rm IGM}$ from individual background galaxies is dominated by photometric noise. This error propagates into our measurements of LAE-Ly$\alpha$ forest cross-correlation (Section \ref{sec:cross}) and Ly$\alpha$ forest auto-correlation functions (Section \ref{sec:auto})  and the reconstruction of the 2D tomographic map of the IGM (Section \ref{sec:map}). As the measured Ly$\alpha$ forest transmission depends on the contrast between the foreground NB flux and the BB flux of a background galaxy, the noise scales approximately as $\delta\TIGM\approx \delta f_{\rm NB}/f_{\rm BB}$. To improve the signal-to-noise ratio of photometric IGM tomography, we therefore require ({\it i}) deeper imaging in the foreground NB filter and/or ({\it ii}) enlarge the sample of bright (spectroscopically-confirmed) background galaxies for which we can more accurately measure $\TIGM$ given a great contrast between the NB and BB filters. 

While the current NB718 depth ($26.86\rm\, mag,\,3\sigma$) provides $3\sigma$ sensitivity to a typical mean value of the Ly$\alpha$ forest transmission when using $25.2\,\rm mag$ ($z$-band) background sources, the majority of our background galaxies are fainter. Secure ($>3\sigma$) detection of the Ly$\alpha$ forest transmitted flux along individual sightlines is currently only possible for rare bright background galaxies ($M_{\rm UV}\lesssim-21.4\rm\,mag$). Extremely-deep NB imaging reaching $27.6\rm\,mag$ ($28.2\rm\,mag$) at $3\sigma$ depth would allow us to detect typical Ly$\alpha$ forest transmitted fluxes to fainter $26.0-26.4\rm\,mag$ ($26.6-27.0\rm\,mag$) background sources at $2-3\sigma$ significance level which comprises $\approx50\,\%$ ($90\,\%$) of our background LAE+DEIMOS10k sample. Based on the existing depth of NB718 from CHORUS PDR1 after $t_{\rm exp}=7.7$ hour exposures \citep{Inoue2020}, and assuming a factor of $\propto t_{\rm exp}^{-1/2}$ reduction of photometric noise, such an extremely deep observation would require a total exposure of $\simeq28$ (100) hours in NB718. As the reconstructed tomographic map of the IGM is dominated by the photometric noise, the improvement in the NB image quality by longer integration will directly increase the signal-to-noise ratio of the IGM tomographic map. While this may seem a significant investment of the telescope time, given a large number of potential science applications of photometric IGM tomography as we will discuss later, such an investment would be of great interest.  

Concerning an increase in the number of bright spectroscopically-confirmed background sources, while we used a large compilation of  spectroscopic catalogues, previous surveys have focused primarily on the central part of the COSMOS field, providing only 36 background objects with suitable spectroscopic redshifts for our IGM tomography. According to the \citet{Bouwens2021} UV luminosity function, there should be numerous star-forming galaxies brighter than $m_{\rm UV}<25.5\rm\,mag$ ($M_{\rm UV}\lesssim-21$) in the appropriate redshift range ($4.98<z<5.89$) with the surface density of $\Sigma_{\rm g}\approx 726\rm\,deg^{-2}$. This corresponds to a total of $\approx1280$ sources that can be in principle accessed across the HSC's $1.76\,\rm deg^2$ field. Uncovering this population would provide a large boost in the number of background galaxies (cf. the surface density of our background LAEs of $\Sigma_{\rm LAE}\simeq71.8\rm\,deg^2$). The bright UV continua will ensure $\sim2-3\sigma$ detection of the Ly$\alpha$ forest transmission with the current NB718 depth. There are a number of photometric catalogues with dropout selection \citep[e.g. GOLDRUSH:][]{Harikane2022} and photometric redshifts \citep[COSMOS2020:][]{Weaver2022} in the COSMOS field. We expect at least $10-20\,\%$ of such UV continuum selected objects will show observable Ly$\alpha$ emission \citep{Stark2010,Stark2011,Mallery2012,Cassata2015,Arrabal-Haro2018,Kusakabe2020}. A wide-field multi-object spectroscopic (MOS) follow-up campaign in the ultra-deep HSC footprint of the COSMOS field can locate $\approx128-256$ background sources ($\Sigma_{\rm specz}\simeq72.6-145.2\rm\,deg^2$), providing $2-3\times$ increase in the total background sample (which is currently dominated by $z\simeq5.7$ LAEs with typical $\sim26.4\rm\,mag$ continua). As discussed in Section \ref{sec:systematics}, including a subset of the background LAEs in the spectroscopic follow-up campaign is also important as it allows us to statistically correct for the systematic bias in the correlation functions by low-redshift interlopers. A follow-up spectroscopic survey using wide-field MOS instruments such as Keck/DEIMOS and upcoming Subaru/Prime Focus Spectrograph (PFS) and VLT/MOONS is required to improve the significance and angular resolution of the photometric IGM tomography. 

Alternatively, it should also be possible to uncover large numbers of bright star-forming galaxies with secure redshifts using rest-frame optical emission line such as H$\beta+[\OIII]$ and H$\alpha$ using a wide-field redshift survey with the NIRCam wide-field slitless spectrograph (WFSS) on board {\it JWST}. Recently \citet{Sun2022a,Sun2022b} \citep[see also][]{Matthee2022b} suggest that $\sim88\,\%$ of bright $z\sim6$ star-forming galaxies show strong $\rm H\beta+[\OIII]$ and H$\alpha$ emission lines detectable with a shallow ($\sim20\rm\,min$) integration. If this holds true at $4.98<z<5.89$, the shallow wide-field NIRCam/WFSS survey tiling the HSC COSMOS field can uncover a factor of $\sim4-9$ larger number of background galaxies ($\Sigma_{\rm JWST}\simeq638\rm\,deg^{-2}$) than a ground-based wide-field MOS survey targeting Ly$\alpha$ lines.

\begin{figure}
  \centering
    \includegraphics[width=\columnwidth]{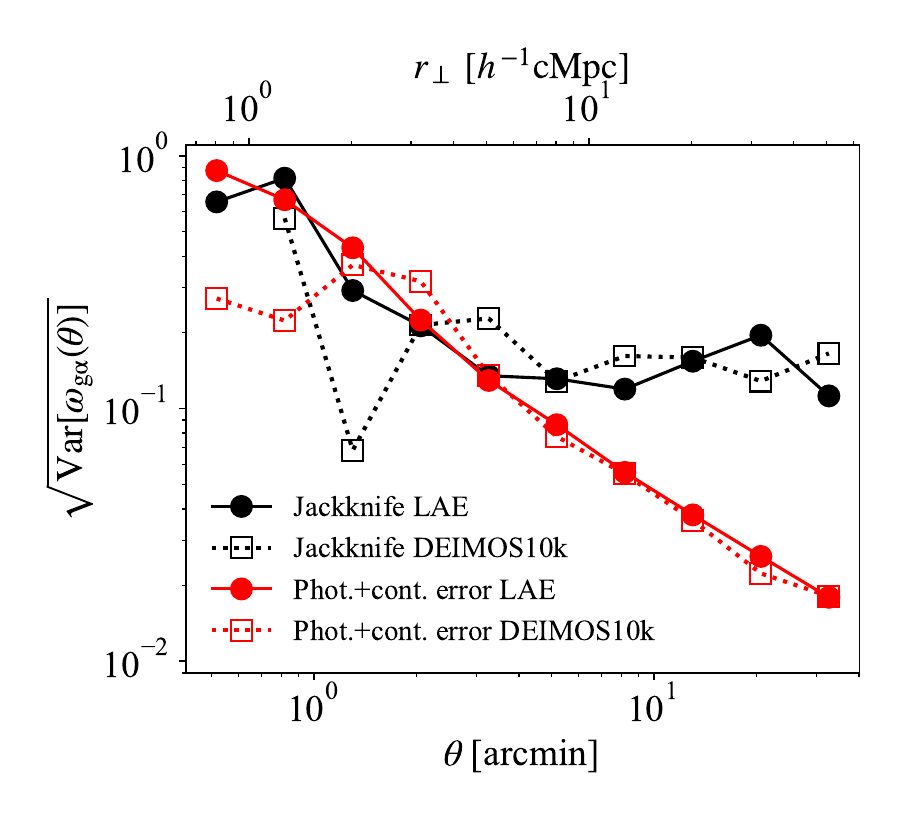}
    \vspace{-0.8cm}
      \caption{Comparison of the error budget in the cross-correlation estimated from the Jackknife covariance matrix (black) and analytic variance (red) for the LAE (filled circles) and DEIMOS10k (open squares) samples. The analytic variance includes the error from photometric noise and continuum slope uncertainty and follows the expected scaling $\sim(\mbox{(typical $\TIGM$ error)}/\langle \overline{T}_{\rm IGM}\rangle)/\sqrt{N_{\rm pairs}(\theta)}$ as the number of pairs per bin increases.}
      \label{fig:error_budget}
  \end{figure}

\subsubsection{Reducing the errors \& systematics in the statistical analysis}

For the statistical measurement of the correlation functions, there is patch-to-patch variance in addition to the photometric noise and UV continuum uncertainty. In Figure \ref{fig:error_budget} we compare the Jackknife variance with our propagated errors from photometric noise and UV continuum uncertainty (Equation \ref{eq:variance_xcorr}) from the Bayesian SED fitting framework in the angular mean Ly$\alpha$ forest transmission around LAEs $\langle T_{\rm IGM}(\theta)\rangle$. 

We find that the photometric noise is the dominant source of uncertainties at $\theta<3\rm\,arcmin$ ($<5\,h^{-1}\rm cMpc$). The Jackknife variance sometimes underestimates the error due to the small number of pairs in the inner angular bins. The photometric noise is comparable for both the LAE and DEIMOS10k samples because, while the individual error in $\TIGM$ is larger in the LAE sample, the larger sample size reduces the error in the cross-correlation function. 

At larger radii $\theta>3\rm\,arcmin$ ($>5\,h^{-1}\rm cMpc$), the Jackknife error becomes larger than the propagated error from photometric+continuum errors, indicating that the patch-to-patch variance in the field becomes the dominant uncertainty. While the photometric error is the dominant uncertainty in individual $\TIGM$, the error scale as $\propto\delta\TIGM/\sqrt{N_{\rm pair}(\theta)}$ where $\delta\TIGM$ is the typical photometric uncertainty in individual $\TIGM$'s. The patch-to-patch variance may arise from cosmic variance or systematics such as imperfect sky subtraction or a coherent error in the photometric colours across the field. We believe that the current dominant source of the patch-to-patch variance is systematics. We checked this by comparing the Jackknife variances between the LAE and DEIMOS10k samples. As the former is sampled from the larger area, if the large-scale patch-to-patch variance is due to cosmic variance, we expect the Jackknife variance to decrease. However, this is not the case. We have also compared the Jackknife variance using a smaller number of angular bins to reduce the photometric error. The photometric error decreases as expected. In the both cases, the large-scale Jackknife variance remains roughly constant, suggesting that the observed excess of the Jackknife variance compared to the photometric+continuum error is likely caused by systematics such as sky background subtraction or possibly reflected lights within the optical units during the NB imaging. An unaccounted coherent change in the photometric colours (e.g. ${\rm NB718}-z$ colour) across the field, e.g. due to imperfect PSF matching or Galactic dust extinction corrections, could also be a cause of the systematics. One possible way to quantify the systematics is to create an artificial IGM tomographic map using sky objects assuming constant Ly$\alpha$ forest transmission. While this artificial IGM map should be uniform across the field by construction, a coherent systematic change in the colours may introduce spatial fluctuations, which can be characterised by measuring the two-point correlation functions or similar statistics. More careful data reduction, background subtraction, and photometric calibration will be required to reduce the error in the large angular bins. We emphasise that these issues can be resolved with additional procedures during data reduction and calibration steps. 

 If the large-scale Jackknife variance is caused by the cosmic variance, we would benefit by enlarging the survey area to increase the sampling of the large-scale modes. Including other pointings such as the SXDS field would increase the constraining power on the large-scale correlation functions. Whether such an investment is worthwhile depends on the theoretically expected scale of the LAE-Ly$\alpha$ forest cross-correlation and Ly$\alpha$ forest auto-correlation functions (see companion paper, Kakiichi et al in prep). Further theoretical studies on photometric IGM tomography are necessary in order to understand the physical information contained on the different scales of the correlation functions.      

\subsubsection{Combining multi-wavelength data}

Our present analysis employed only two broad-band filters (HSC $z$ and $y$) to constrain the SEDs of background galaxies. The SED uncertainties (i.e continuum slope and SED template) can be mitigated by including multi-wavelength datasets available in the COSMOS field. Inclusion of near-infrared data such as UltraVISTA $JHK$ \citep{McCracken2012} will provide a longer baseline to better characterise the intrinsic SEDs. Constraining the dust and age of the background galaxies would eliminate the $\sim6-27\,\%$ error in the measured $\TIGM$ from SED fitting (Section \ref{sec:individual_TIGM}). The soon-available NIRCam imaging with F115W, F150W, F277W, F444W filters from COSMOS-Web (\citealt[ID 1727, PI:][]{COSMOS-Webb2021}, see also \citealt{Casey2022}) can precisely determine the rest UV-to-optical SEDs of background galaxies. 

\subsection{Science applications}

\subsubsection{LyC escape fraction and the nature of ionizing sources}

The statistical analysis of the LAE-Ly$\alpha$ forest angular cross-correlation from photometric IGM tomography provides a measure of the gas overdensity, temperature, and UV background fluctuations of the IGM around foreground LAEs. The angular cross-correlation is the NB-averaged version of the underlying galaxy-Ly$\alpha$ forest 3D cross-correlation that can be measured from the spectroscopic galaxy survey in quasar fields \citep{Kakiichi2018,Meyer2020}. We can thus adopt a similar approach to constrain the population-averaged LyC escape fraction from the LAE-Ly$\alpha$ forest cross-correlation. We will present the analysis in a companion paper (Kakiichi et al in prep).

The observed limit on the LAE-Ly$\alpha$ forest cross-correlation could also limit the contribution of bright LAEs to the UV background. \citet{Matthee2022,Naidu2022} claimed that bright LAEs with $L_{\rm Ly\alpha}\gtrsim10^{42}\rm\,erg\,s^{-1}$ could contribute significantly to the total ionizing budget at $z\gtrsim4$. Such bright LAEs may produce a large proximity zone which could be observed by photometric IGM tomography. The measurement of LAE-Ly$\alpha$ forest cross-correlation can be used to test this scenario. We will discuss the implication of our result on on the relative contribution of bright and faint galaxies to the total ionizing budget in a following paper (Kakiichi et al in prep).


\subsubsection{Search for ionized bubbles and protoclusters}
 
Besides the statistical analysis, the IGM tomography can also be used to search for ionized bubbles and/or protoclusters at high redshifts. While our present analysis focused on NB718 filter ($z\simeq4.9$), one can perform tomographic analyses at any redshift where NB filters are available, including NB527 ($z\simeq3.31$) and NB816 ($z\simeq5.72$) \citep[see][for the full list]{Kakiichi2022}, assuming adequate deep NB imaging and background galaxies are available. 

Ionized bubbles or the high UV background regions are expected to show transmissive Ly$\alpha$ forest with effective optical depth of $\tau_{\rm eff}\approx2$ at $z\simeq5.7$ \citep{Davies2018,Keating2020}. This corresponds to a NB-BB magnitude decrement of $m_{\rm NB816}-m_{\rm UV}=-2.5\log_{10} e^{-\tau_{\rm eff}}\approx2.2\rm\,mag$. With a NB816 depth of 27.7 mag, it is possible to search for such regions using 25.5 mag background sources at $5.81<z<6.86$. 

The present $2-3\sigma$ depth of $26.89-27.33\rm\,mag$ of the NB816 image already allows us to search for extreme transmissive IGM regions with $\tau_{\rm eff}\sim1-2$ ($m_{\rm NB816}-m_{\rm UV}\approx1.1-2.2\rm\,mag$). One such region ($\tau_{\rm eff}\simeq1.2$) has been already discovered serendipitously by \citet{Bosman2020} around a quasar using NB816 imaging. While current cosmological simulations \citep{Davies2018,Keating2020} do not predict such extreme values, they are performed under the assumption that reionization is driven by a large population of relatively faint galaxies. Highly transmissive regions may exist if the contribution from quasars \citep{Chardin2017} or luminous galaxies with high LyC escape fraction and ionizing photon production efficiency \citep[e.g.][]{Endsley2021,Topping2022,Marques-Chaves2022} are important. 

Similarly, one can conduct a search for protoclusters as coherently opaque regions of the IGM in the tomographic map \citep{Lee2016,Newman2020}. Indeed, \citet{Mawatari2017} used the photometric IGM tomographic technique to examine the protocluster region with NB497 ($z\simeq3.1$). The $10-40\,h^{-1}\rm cMpc$ scales of coherently strong Ly$\alpha$ absorption \citep{Cai2016} is shown to be associated with an overdensity at $z\sim2-3$ \citep{Cai2017,Shi2021} whose scale matches closely with the line-of-sight width of a NB filter. LAEs in the tomographic slice permit an immediately confirmation of whether the coherently opaque IGM region is associated with a galaxy overdensity without a separate spectroscopic or dedicated imaging campaign. This provides an opportunity to test the relation between galaxy overdensities and the IGM environments \citep[e.g.][]{Momose2021, Newman2022} at higher redshifts without a need of extremely-deep spectroscopy of background galaxies. 

\subsubsection{Quasar light-echoes and past AGN activity}

The relatively short variability timescale of a quasar $t_{\rm Q}\sim10^{6-8}\rm\,yr$  compared to the light crossing time of the IGM tomographic map means that the radiation from the quasar could leave an imprint on the ionization state of the IGM well after the non-thermal activity ceases \citep{Adelberger2004}. \citet{Schmidt2019,Kakiichi2022} examined the prospect of examining the lifetime/past AGN activity of an active quasar using this light echo signal. Our photometric IGM tomography demonstrates that searching for quasar light echoes is possible if an appropriate quasar field is targeted.

In principle, the search for quasar light-echoes is not limited to the region around a luminous quasar, but can be applied to any region around a massive galaxy where quasar activity might have occurred during its previous $\sim10^8\rm\,yr$. Searches for fossil light-echoes around LAEs, or LBGs at redshift within the region of influence of the NB tomographic slice, could be used to constrain the past luminous ionizing activity of a source as a function of the travel time between the source and a position of the IGM. \citet{Bosman2020} used the photometric detection of the Ly$\alpha$ forest transmission in the NB816 filter located slightly foreground of $z\simeq5.8$ quasar to show that the quasar was active for at least $\sim 2\times10^4 \rm\,yr$ in the past. \citet{Bowler2015,Bowler2020,Ono2018,Harikane2022} suggest that the double-power law luminosity function at $z\gtrsim4$ may be a sign of inefficient quenching by AGN feedback acting on luminous galaxies. Finding a lack of highly transmissive regions around luminous galaxies would imply that these systems could not release a significant ionizing radiation by quasar activity during their history. 

\subsubsection{Correlation with direct LyC, $\HeII$, and $\CIV$ emitters}

Multiple HSC NB imaging available in the COSMOS field provides a NB photometric measure of LyC leakage and the identification of strong $\HeII$ and $\CIV$ emission lines for $z\simeq4.9$ LAEs via the CHORUS survey \citep{Inoue2020}. The correlation of these populations with photometric IGM tomographic map will have a number of applications.

For example, direct search for LyC emission along the line-of-sight of galaxies is increasingly difficult at $z\gtrsim3.5$ as the IGM becomes increasingly opaque on average. \citet{Bassett2021,Bassett2022} showed that any LyC detection could be biased towards the rare transmissive regions of the IGM and the bias introduced by assuming an average IGM transmission is severe at $3<z<5$. While LyC leaking candidates have reported at this redshift range 
\citep[e.g.][]{Shapley2016,Vanzella2018,Ji2020,Mestric2020,Prichard2022,Rivera-Thorsen2022,Marques-Chaves2022}, 
the uncertain IGM transmission value against LyC photons make it difficult to convert the observed LyC flux to an absolute LyC escape fraction and sometimes results in unphysical values $f_{\rm esc}>1$. \citet{Fletcher2019} also noted the possibility of spatial variations in the inferred LyC escape fraction due to the uncertain IGM transmission. Photometric IGM tomography provides an useful independent measure of the IGM transmission and could be used to alleviate the issue of uncertain IGM LyC opacities along the lines-of-sight to high-redshift galaxies. 

Combining the direct detection of LyC leakage and the indirect measurement from galaxy-Ly$\alpha$ forest cross-correlation, one can test the relative contributions of luminous and faint galaxies to the ionising budget. The former measures the ionising contribution of galaxies above the detection limit, while the latter provides a population-averaged LyC escape fraction including those below the detection limit.
NB measurements of $\CIV$ and $\HeII$ from LAEs are also valuable for examining hard ionising sources like AGN and X-ray binaries. The IGM tomographic map enables us to connect the properties of these populations with their large-scale IGM environments. 

\section{Conclusions}\label{sec:conclusion}

We present a novel technique called photometric IGM tomography to map the large-scale structure of the IGM at $z\sim5$ in the COSMOS field. The technique utilizes ultra-deep NB718 imaging to detect the Ly$\alpha$ forest transmission along sightlines to various background galaxies including spectroscopically-confirmed DEIMOS10k sources and NB816-selected LAE catalogue from SILVERRUSH. In this paper, we describe a science verification of this new technique using public HSC data including HSC-SSP DR3 and CHORUS PDR1. 

We have developed a Bayesian SED fitting framework to measure the Ly$\alpha$ forest transmission along background galaxies. This allows us to accurately propagate the error from photometric noise and uncertainty from the assumed SED template into the final measurement of the Ly$\alpha$ forest transmission. At the current imaging depths, photometric noise from NB718 imaging dominates the total error in the estimated Ly$\alpha$ forest transmission. Uncertainties from the UV continuum slopes and SED templates are subdominant in the present analysis. 

Using a total sample of 140 background sources, we have photometrically measured the NB718-integrated mean Ly$\alpha$ forest transmission at $z\simeq4.9$. We find that our result is consistent with the previous measurement using quasar spectra, demonstrating that an accurate photometric NB measurement of the Ly$\alpha$ forest transmission is practical. We argue that the most likely systematic is contamination from low-redshift interlopers in the NB-selected LAE sample, which if not taken into account, would cause a fictitious Ly$\alpha$ forest transmission. This may explain an offset we see in the measured values between spectroscopically-confirmed background sources (DEIMOS10k) and NB-selected background sources (SILVERRUSH $z\simeq5.7$ LAEs). Fortunately, this can be corrected statistically provided that the low-redshift interloper fraction is known e.g. from spectroscopic follow-up of a subset of the population. 

We developed a method for measuring the angular LAE-Ly$\alpha$ forest cross-correlation and the Ly$\alpha$ forest auto-correlation functions. Our method incorporates the  individual posteriors from the Bayesian SED fitting framework to measure the angular correlation functions consistent with the propagated error including photometric noise and SED uncertainties. Low-redshift interlopers in our foreground ($z\simeq4.9$) and background ($z\simeq5.7$) LAE samples are again a main systematic uncertainty which can also be corrected statistically using a partial spectroscopic follow-up of the parent LAE samples. Applying the technique to the present data, we did not detect any angular LAE-Ly$\alpha$ forest cross-correlation and auto-correlation of the Ly$\alpha$ forest at $z\simeq4.9$. Our result is consistent with no Ly$\alpha$ forest fluctuations around LAEs and should be below $58\,\%$ at $10\,h^{-1}\rm cMpc$ compared to the global mean transmission at $z\simeq4.9$. We will discuss the physical implications of this limit in a companion paper (Kakiichi et al in prep). 

Finally, we presented a reconstructed 2D tomographic map of the IGM at $z\simeq4.9$, co-spatial with the distribution of foreground LAEs in the same cosmic volume in the COSMOS field. The map embraces a field $140\,h^{-1}\rm cMpc$ in diameter with a transverse spatial resolution of $\simeq11\,h^{-1}\rm cMpc$. While the current map is still dominated by photometric noise, it represents the most detailed tomographic map close to the end of cosmic reionization. The ability of photometric IGM tomography to map both the large-scale structures of galaxies and the IGM across a large region of the sky is extremely appealing, allowing us to apply the technique for many science cases including constraints on LyC escape fractions, the nature of ionizing sources through the UV background and thermal fluctuations of the IGM, and the searches for ionized bubbles, protoclusters, and quasar light-echoes.     

Photometric IGM tomography can be embedded in traditional NB imaging and wide-field spectrscopic surveys and is applicable at all redshifts from $z\sim2$ to 6 where NB filters are available, thus making it possible to examine the co-evolution of galaxies and the cosmic web during the first few billion years of cosmic history. We argue that the technique can be improved through extremely-deep NB imaging and large spectroscopic follow-up campaigns. Although this would require a large, but nonetheless practical, investment of telescope time with existing 8-10 m telescopes, we make the case that such an investment is worthwhile. Combining multi-wavelength dataset including the UltraVISTA JHK and {\it Spitzer}/IRAC imaging and soon available {\it JWST}/NIRCam imaging in the COSMOS field will enable us to better control the systematic uncertainty arising from the intrinsic SEDs of background galaxies. Future Subaru/PFS surveys will greatly increase the number of spectroscopically-confirmed background galaxies. Furthermore, a large-scale JWST spectroscopic survey tiling the COSMOS field would push the redshift frontier closer to the reionization epoch. The wide-field capability of the photometric IGM tomography is highly complementary to surveys based on using extremely-deep spectra with ELT/MOSAICS and TMT/WFOS as their small fields of view require interesting target regions to be pre-selected, e.g. from photometric IGM tomographic maps. As such, photometric IGM tomography has great potential to uncover the physics of galaxy-cosmic web connection in the early Universe in the coming decade.

\section*{Acknowledgements}

KK thanks Harley Katz, Fred Davies, and K-G Lee for discussions and constructive comments. We thank the referee for carefully reading the manuscript. This project has received funding from the European Research Council (ERC) under the European Union's Horizon 2020 research and innovation programme (grant agreement No 885301). RSE acknowledges financial support from ERC Advanced Grant FP7/669253. RAM acknowledges support from the ERC Advanced Grant 740246 (Cosmic Gas). This work is supported by the World Premier International Research Center Initiative
(WPI Initiative), MEXT, Japan, as well as KAKENHI Grant-in-Aid for Scientific Research (A) (20H00180 and 21H04467)
through the Japan Society for the Promotion of Science (JSPS).
This work was supported by the joint research program of the
Institute for Cosmic Ray Research (ICRR), University of
Tokyo.

This paper is based on data collected at the Subaru Telescope and retrieved from the HSC data archive system, which is operated by the Subaru Telescope and Astronomy Data Center (ADC) at NAOJ. Data analysis was in part carried out with the cooperation of Center for Computational Astrophysics (CfCA), NAOJ. We are honored and grateful for the opportunity of observing the Universe from Maunakea, which has the cultural, historical and natural significance in Hawaii. The Hyper Suprime-Cam (HSC) collaboration includes the astronomical communities of Japan and Taiwan, and Princeton University. The HSC instrumentation and software were developed by the National Astronomical Observatory of Japan (NAOJ), the Kavli Institute for the Physics and Mathematics of the Universe (Kavli IPMU), the University of Tokyo, the High Energy Accelerator Research Organization (KEK), the Academia Sinica Institute for Astronomy and Astrophysics in Taiwan (ASIAA), and Princeton University. Funding was contributed by the FIRST program from the Japanese Cabinet Office, the Ministry of Education, Culture, Sports, Science and Technology (MEXT), the Japan Society for the Promotion of Science (JSPS), Japan Science and Technology Agency (JST), the Toray Science Foundation, NAOJ, Kavli IPMU, KEK, ASIAA, and Princeton University. This paper makes use of software developed for Vera C. Rubin Observatory. We thank the Rubin Observatory for making their code available as free software at \url{http://pipelines.lsst.io/}.

\section*{Data Availability}

All the original HSC data and spectroscopic catalogues including DEIMOS10k are available online through the HSC-SSP website and NASA/IPAC IRSA (see the links listed in this paper). The SILVERRUSH catalogue is available at \url{http://cos.icrr.u-tokyo.ac.jp/rush.html}. The table of the measured Ly$
\alpha$ forest transmission along our background galaxies is available through online supplementary material. 



\bibliographystyle{mnras}
\bibliography{reference} 




\appendix

\section{Derivation of the estimators}\label{app:derivation}

For clarity, we explicitly show the derivation of the estimator and the variance. The mean Ly$\alpha$ forest transmission is given by $\langle \overline{T}_{\rm IGM}\rangle=\int \overline{T}_{\rm IGM} P(\overline{T}_{\rm IGM}|\{f_{{\rm NB},i}^{\rm obs}, \boldsymbol{f}_{{\rm BB},i}^{\rm obs}\}_{i=1,\dots,N_{\rm bg}})d\overline{T}_{\rm IGM}$. By substituting Equation \ref{eq:full_posterior}, we have 
\begin{align}
\langle\bar{T}_{\rm IGM}\rangle
&=\int \left(\frac{1}{N_{\rm bg}}\sum_{i=1}^{N_{\rm bg}}T_{{\rm IGM},i}\right)P(T_{{\rm IGM},i}|f_{{\rm NB},i}^{\rm obs}, \boldsymbol{f}_{{\rm BB},i}^{\rm obs})\prod_{i=1}^{N_{\rm bg}} dT_{{\rm IGM},i}, \nonumber \\
&=\frac{1}{N_{\rm bg}}\sum_{i=1}^{N_{\rm bg}} \int T_{{\rm IGM},i}P(T_{{\rm IGM},i}|f_{{\rm NB},i}^{\rm obs}, \boldsymbol{f}_{{\rm BB},i}^{\rm obs})dT_{{\rm IGM},i},
\nonumber \\
&=\frac{1}{N_{\rm bg}}\sum_{i=1}^{N_{\rm bg}} \langle \TIGMi\rangle. \nonumber
\end{align}

The variance of the mean Ly$\alpha$ forest transmission is given by 
$\sigma^2_{\overline{T}_{\rm IGM}}=\int (\overline{T}_{\rm IGM}-\langle\overline{T}_{\rm IGM}\rangle)^2 P(\overline{T}_{\rm IGM}|\{f_{{\rm NB},i}^{\rm obs}, \boldsymbol{f}_{{\rm BB},i}^{\rm obs}\}_{i=1,\dots,N_{\rm bg}})d\overline{T}_{\rm IGM}$. By substituting Equation \ref{eq:full_posterior} and the above result, we have
\begin{align}
&\sigma^2_{\overline{T}_{\rm IGM}}= \nonumber \\
&\frac{1}{N^2_{\rm bg}} \int \left(\sum_{i=1}^{N_{\rm bg}}T_{{\rm IGM},i}-\langle T_{{\rm IGM},i}\rangle\right)^2P(T_{{\rm IGM},i}|f_{{\rm NB},i}^{\rm obs}, \boldsymbol{f}_{{\rm BB},i}^{\rm obs})\prod_{i=1}^{N_{\rm bg}} dT_{{\rm IGM},i}. \nonumber 
\end{align}
Since all the individual measurements are uncorrelated, the cross-terms are zeros, i.e.
\begin{align}   
\iint dT_{{\rm IGM}, i}dT_{{\rm IGM},j} P(T_{{\rm IGM}, i}|f_{{\rm NB}, i}^{\rm obs}, \boldsymbol{f}_{{\rm BB}, i}^{\rm obs}) P(T_{{\rm IGM}, j}|f_{{\rm NB}, j}^{\rm obs}, \boldsymbol{f}_{{\rm BB},j}^{\rm obs}) \nonumber \\
\times \left(T_{{\rm IGM}, i}-\langle T_{{\rm IGM}, i}\rangle\right)\left(T_{{\rm IGM},j}-\langle T_{{\rm IGM},j}\rangle\right)=0. \nonumber
\end{align}
Thus, we obtain
\begin{align}
&\sigma^2_{\overline{T}_{\rm IGM}}= \nonumber \\
&\frac{1}{N_{\rm bg}^2}\sum_{i=1}^{N_{\rm bg}} \int \left(T_{{\rm IGM},i}-\langle T_{{\rm IGM},i}\rangle\right)^2P(T_{{\rm IGM},i}|f_{{\rm NB},i}^{\rm obs}, \boldsymbol{f}_{{\rm BB},i}^{\rm obs})dT_{{\rm IGM},i}.\nonumber
\end{align}

The calculation for the angular averaged Ly$\alpha$ forest transmission around galaxies is identical to that for the estimator and the variance for the mean Ly$\alpha$ forest transmission for each angular bin. Thus, we obtain the estimator,
\begin{align}
\langle\overline{T}_{\rm IGM}(\theta)\rangle=&\frac{1}{N_{\rm pair}(\theta)}\sum_{j=1}^{N_{\rm fg}}\sum_{i=1}^{N_{\rm bg}}\mathcal{I}(|\theta-\theta_{ij}|)\nonumber \\
&~~~~~~~~~~~~~\times\int \TIGMi P(\TIGMi|f_{{\rm NB},i}^{\rm obs}, \boldsymbol{f}_{{\rm BB},i}^{\rm obs})d\TIGMi, \nonumber
\end{align}
and the variance, 
\begin{align}
&{\rm Var}[\overline{T}_{\rm IGM}(\theta)]=
\frac{1}{N_{\rm pair}(\theta)^2}\sum_{j=1}^{N_{\rm fg}}\sum_{i=1}^{N_{\rm bg}}\mathcal{I}(|\theta-\theta_{ij}|) \nonumber \\
&~~~~~~~~~~~~\times\int (\TIGMi-\langle\TIGMi\rangle )^2 P(\TIGMi|f_{{\rm NB},i}^{\rm obs}, \boldsymbol{f}_{{\rm BB},i}^{\rm obs})d\TIGMi. \nonumber
 \end{align}


\begin{table*}
\centering
\caption{The measured values of the Ly$\alpha$ forest transmission $\TIGM$, UV magnitude $\Muv$, and the UV continuum slope $\beta$ from the Bayesian SED fitting framework for all background sources in the DEIMOS10k sample. The full table and the machine-readable file are available as online supplementary material.}\label{table:A1}
\begin{tabular}{rlllllll}
\hline
   ID & Object name           & RA           & DEC           & $\TIGM$                    & $\Muv$                   & $\beta$                 & Note   \\
\hline
    1 & DEIMOS\_2018\_L222036 & 10h02m20.94s & +01d37m06.78s & $+0.049_{-0.194}^{+0.196}$ & $-20.65_{-0.13}^{+0.13}$ & $-1.56_{-1.67}^{+1.65}$ &        \\
    2 & DEIMOS\_2018\_L416105 & 10h02m45.66s & +01d55m35.91s & $+0.531_{-0.176}^{+0.178}$ & $-21.18_{-0.12}^{+0.12}$ & $-1.31_{-1.07}^{+1.07}$ &        \\
    3 & DEIMOS\_2018\_L244697 & 10h01m59.64s & +01d39m16.81s & $+0.045_{-0.064}^{+0.064}$ & $-21.45_{-0.02}^{+0.02}$ & $-1.69_{-0.70}^{+0.70}$ &        \\
    4 & DEIMOS\_2018\_L254463 & 10h01m58.94s & +01d40m10.67s & $+0.098_{-0.049}^{+0.049}$ & $-21.62_{-0.02}^{+0.02}$ & $-2.39_{-0.62}^{+0.63}$ &        \\
    5 & DEIMOS\_2018\_L263567 & 10h01m35.65s & +01d41m08.01s & $+0.070_{-0.141}^{+0.139}$ & $-20.44_{-0.18}^{+0.18}$ & $-2.81_{-1.49}^{+1.47}$ &        \\
     & $\dots$ &  &  &  &  & & \\
\hline
\end{tabular}
\end{table*}

\begin{table*}
    \centering
    \caption{Same as Table \ref{table:A1} but for background sources ($z\simeq5.7$ LAEs) from the SIVLERRUSH catalogue. The full table and the machine-readable file are available as online supplementary material.}\label{table:A2}
    \begin{tabular}{rlllllll}
    \hline
       ID & Object name             & RA (J2000)   & DEC (J2000)   & $\TIGM$                    & $\Muv$                   & $\beta$                 & Note   \\
    \hline
        1 & SILVERRUSH\_2021\_15477 & 09h57m44.50s & +02d16m39.00s & $+3.004_{-1.648}^{+1.693}$ & $-19.47_{-0.32}^{+0.32}$ & $-1.33_{-2.32}^{+2.27}$ & $94\%$ outlier \\
        2 & SILVERRUSH\_2021\_14862 & 09h57m47.81s & +02d16m18.53s & $+1.431_{-0.798}^{+0.826}$ & $-19.83_{-0.27}^{+0.27}$ & $-0.87_{-2.16}^{+2.04}$ & $63\%$ outlier \\
        3 & SILVERRUSH\_2021\_5731  & 09h57m48.23s & +02d33m56.51s & $+0.604_{-0.381}^{+0.401}$ & $-20.57_{-0.26}^{+0.27}$ & $-1.15_{-2.17}^{+2.12}$ &        \\
        4 & SILVERRUSH\_2021\_18336 & 09h58m11.04s & +02d45m50.21s & $+0.621_{-0.455}^{+0.481}$ & $-20.33_{-0.28}^{+0.28}$ & $-1.47_{-2.22}^{+2.21}$ &        \\
        5 & SILVERRUSH\_2021\_10859 & 09h58m12.36s & +02d03m09.22s & $-0.006_{-0.368}^{+0.372}$ & $-20.31_{-0.21}^{+0.22}$ & $-0.52_{-1.82}^{+1.77}$ &        \\
         & $\dots$ &  &  &  &  & & \\
    \hline
    \end{tabular}
\end{table*}

\section{Interloper contamination}\label{app:interloper}

The contamination by low-redshift interlopers in the foreground and background LAE samples dilutes the angular mean Ly$\alpha$ forest transmission around the foreground LAEs. When we measure the observed angular mean Ly$\alpha$ forest transmission around LAEs using the background LAE sample, one can be decomposed the estimator into four different contributions: (1) true foreground LAE - true background LAE pairs (fg-bg pairs), (2) foreground interloper - true background LAE pairs (fg.int-bg pairs), (3) true foreground LAE - background interloper pairs (fg-bg.int pairs), and (4) foreground interloper - background interloper pairs (fg.int-bg.int pairs),
\begin{align}
&\overline{T}_{\rm IGM}(\theta)= \frac{1}{N_{\rm pair}(\theta)}\times \nonumber \\
&~~
\left(
\mathop{\sum^{N^{\rm true}_{\rm fg}}\sum^{N^{\rm true}_{\rm bg}}}_{\substack{\rm fg-bg\,pairs}}
\TIGMi^{\rm true}\mathcal{I}(|\theta-\theta_{ij}|)
+
\!\!
\mathop{\sum^{N^{\rm int}_{\rm fg}}\sum^{N^{\rm true}_{\rm bg}}}_{\substack{\rm fg.int-bg\,pairs}}
\TIGMi^{\rm true}\mathcal{I}(|\theta-\theta_{ij}|)\right.
\nonumber \\
&\left.+
\!\!
\mathop{\sum^{N^{\rm true}_{\rm fg}}\sum^{N^{\rm int}_{\rm bg}}}_{\substack{\rm fg-bg.int\,pairs}}
\!\!
\TIGMi^{\rm bg.int}\mathcal{I}(|\theta-\theta_{ij}|)
+
\!\!\!\!
\mathop{\sum^{N^{\rm int}_{\rm fg}}\sum^{N^{\rm int}_{\rm bg}}}_{\substack{\rm fg.int-bg.int\,pairs}}
\!\!\!\!
\TIGMi^{\rm bg.int}\mathcal{I}(|\theta-\theta_{ij}|)
\right), \nonumber
\end{align}
where  $N_{\rm pairs}(\theta)$ is the total number of observed pairs per angular bin $N_{\rm pairs}(\theta)$, $N_{\rm fg}^{\rm true}$ and $N_{\rm bg}^{\rm true}$ are the number of true foreground and background LAEs, $N_{\rm fg}^{\rm int}$ and $N_{\rm bg}^{\rm int}$ are the number of foreground and background interlopers, and $\TIGMi^{\rm true}$ and $\TIGMi^{\rm bg.int}$ are the measured Ly$\alpha$ forest transmission along true background LAEs and background LAE interlopers. The expectation value of the angular averaged Ly$\alpha$ forest transmission around LAEs is then given by
\begin{align}
\langle\overline{T}_{\rm IGM}(\theta)\rangle&= \frac{N_{\rm pair}^{\rm fg-bg\,pairs}(\theta)}{N_{\rm pair}(\theta)}
\langle\overline{T}_{\rm IGM}^{\rm true}(\theta)\rangle \nonumber \\
&
+
\frac{N_{\rm pair}^{\rm fg.int-bg\,pairs}(\theta)}{N_{\rm pair}(\theta)}
\langle\overline{T}_{\rm IGM}^{\rm true}(\theta)\rangle_{\rm fg.int-bg\,pairs} \nonumber \\
&
+
\frac{N_{\rm pair}^{\rm fg-bg.int\,pairs}(\theta)}{N_{\rm pair}(\theta)}
\langle\overline{T}_{\rm IGM}^{\rm bg.int}(\theta)\rangle_{\rm fg-bg,int\,pairs} \nonumber \\
&
+
\frac{N_{\rm pair}^{\rm fg.int-bg.int\,pairs}(\theta)}{N_{\rm pair}(\theta)}
\langle\overline{T}_{\rm IGM}^{\rm bg.int}(\theta)\rangle_{\rm fg.int-bg.int\,pairs}, \nonumber
\end{align}
where $N_{\rm pairs}^{\rm fg-bg\,pairs}(\theta)$, $N_{\rm pairs}^{\rm fg.int-bg\,pairs}(\theta)$, $N^{\rm fg-bg.int\,pairs}_{\rm pairs}(\theta)$, and $N^{\rm fg.int-bg.int\,pairs}_{\rm pairs}(\theta)$ are the number of foreground LAE - background LAE, foreground LAE interloper - background LAE, foreground LAE - background LAE interloper, and foreground LAE interloper - background LAE interloper pairs per bin.

The total number of observed pairs depends on the angular correlation function of the observed populations $\omega_{\rm fg-bg}^{\rm obs}(\theta)$, 
\begin{equation}
  N_{\rm pairs}(\theta)=N_{\rm fg}^{\rm obs}N_{\rm bg}^{\rm obs}\left[1+\omega_{\rm fg-bg}^{\rm obs}(\theta)\right]\frac{2\pi\theta\Delta\theta}{\Omega_{\rm survey}}
\end{equation}
where $\Omega_{\rm survey}$ is the survey area and $\Delta\theta$ is the angular bin size. Similarly, the number of true foreground LAE - true background LAE pairs etc is given by
\begin{align}
  &
  N_{\rm pairs}^{\rm fg-bg\,pairs}(\theta)=N_{\rm fg}^{\rm true}N_{\rm bg}^{\rm true}\left[1+\omega_{\rm fg-bg}(\theta)\right]\frac{2\pi\theta\Delta\theta}{\Omega_{\rm survey}}, \nonumber \\ 
  &
  N_{\rm pairs}^{\rm fg.int-bg\,pairs}(\theta)=N_{\rm fg}^{\rm int}N_{\rm bg}^{\rm true}\left[1+\omega_{\rm fg.int-bg}(\theta)\right]\frac{2\pi\theta\Delta\theta}{\Omega_{\rm survey}}, \nonumber \\
  &
  N_{\rm pairs}^{\rm fg-bg.int\,pairs}(\theta)=N_{\rm fg}^{\rm true}N_{\rm bg}^{\rm int}\left[1+\omega_{\rm fg-bg.int}(\theta)\right]\frac{2\pi\theta\Delta\theta}{\Omega_{\rm survey}}, \nonumber \\
  &
  N_{\rm pairs}^{\rm fg.int-bg.int\,pairs}(\theta)=N_{\rm fg}^{\rm int}N_{\rm bg}^{\rm int}\left[1+\omega_{\rm fg.int-bg.int}(\theta)\right]\frac{2\pi\theta\Delta\theta}{\Omega_{\rm survey}} \nonumber. 
\end{align}

Defining the interloper fractions in the observed foreground and background LAE samples to be $f_{\rm fg.int}=N_{\rm fg}^{\rm int}/N^{\rm obs}_{\rm fg}$ and $f_{\rm bg.int}=N_{\rm bg}^{\rm int}/N^{\rm obs}_{\rm bg}$, the observed angular cross-correlation function from the foreground and background LAE samples has contributions from true and interloper pairs,
\begin{align}
  \omega_{\rm fg-bg}^{\rm obs}(\theta)
  & =(1-f_{\rm fg.int})(1-f_{\rm bg.int})\omega_{\rm fg-bg}(\theta) \nonumber \\
  &  +f_{\rm fg.int}(1-f_{\rm bg.int})\omega_{\rm fg.int-bg}(\theta) \nonumber \\
  & +f_{\rm bg.int}(1-f_{\rm fg.int})\omega_{\rm fg-bg.int}(\theta) \nonumber \\
  &  +f_{\rm bg.int}f_{\rm fg.int}\omega_{\rm fg.int-bg.int}(\theta). \nonumber
\end{align}
Since both interlopers and true LAEs in the two different NB-selected samples are located at different redshifts, all these angular correlation functions should be zeros. The objects in foreground and background LAE samples are not spatially correlated to each other. Thus,
\begin{align}
&  \langle\overline{T}_{\rm IGM}^{\rm true}(\theta)\rangle_{\rm fg.int-bg\,pairs}
  \approx\langle\overline{T}_{\rm IGM}^{\rm true}\rangle, \nonumber \\
&  \langle\overline{T}_{\rm IGM}^{\rm bg.int}(\theta)\rangle_{\rm fg-bg,int\,pairs}
  \approx\langle\overline{T}_{\rm IGM}^{\rm bg.int}\rangle, \nonumber \\
&  \langle\overline{T}_{\rm IGM}^{\rm bg.int}(\theta)\rangle_{\rm fg.int-bg.int\,pairs}
  \approx\langle\overline{T}_{\rm IGM}^{\rm bg.int}\rangle. \nonumber 
\end{align}

Therefore, we find that the effect of the interloper contamination on the observed angular-averaged Ly$\alpha$ forest transmission profile is expressed as
\begin{align}
\langle\overline{T}_{\rm IGM}(\theta)\rangle&=
(1-f_{\rm fg.int})(1-f_{\rm bg.int})\langle\overline{T}_{\rm IGM}^{\rm true}(\theta)\rangle \nonumber \\
&+f_{\rm fg.int}(1-f_{\rm bg.int})\langle\overline{T}_{\rm IGM}^{\rm true}\rangle \nonumber \\
&+f_{\rm bg.int}(1-f_{\rm fg.int})\langle\overline{T}_{\rm IGM}^{\rm bg.int}\rangle \nonumber \\
&+f_{\rm fg.int}f_{\rm bg.int}\langle\overline{T}_{\rm IGM}^{\rm bg.int}\rangle.  \nonumber
\end{align}

As the observed LAE-Ly$\alpha$ forest cross-correlation is defined with respect to the observed mean Ly$\alpha$ forest transmission which is also affected by the interlopers in the background LAE sample, i.e. $\langle\overline{T}_{\rm IGM}^{\rm obs}\rangle=(1-f_{\rm bg.int})\langle\overline{T}_{\rm IGM}\rangle^{\rm true}+f_{\rm bg.int}\langle\overline{T}_{\rm IGM}\rangle^{\rm bg.int}$, we find
\begin{align}
\omega^{\rm obs}_{\rm g\alpha}(\theta)&=\frac{\langle\overline{T}^{\rm obs}_{\rm IGM}(\theta)\rangle}{\langle\overline{T}^{\rm obs}_{\rm IGM}\rangle}-1 \nonumber \\
&=\frac{(1-f_{\rm fg.int})(1-f_{\rm bg.int})\langle \overline{T}_{\rm IGM}^{\rm true}\rangle}{(1-f_{\rm bg.int})\langle \overline{T}_{\rm IGM}^{\rm true}\rangle+f_{\rm bg.int}\langle \overline{T}^{\rm bg.int}_{\rm IGM}\rangle}\omega^{\rm true}_{\rm g\alpha}(\theta), \nonumber
\end{align}
where $\omega_{\rm g\alpha}^{\rm true}(\theta)=\langle\overline{T}_{\rm IGM}^{\rm true}(\theta)\rangle/\langle\overline{T}_{\rm IGM}^{\rm true}\rangle-1$.

\section{Tables of the estimated Ly$\alpha$ forest transmission}

Tables \ref{table:A1} and \ref{table:A2} show the first five rows of the tables of the estimated Ly$\alpha$ forest transmission along all the background sources used in this paper. The full tables and the corresponding machine-readable files are available as online supplementary materials.

\bsp	
\label{lastpage}
\end{document}